\documentclass[12pt]{article}

\usepackage{acro}
\usepackage{amsmath}
\usepackage{amssymb}
\usepackage{amsthm}
\usepackage{mathrsfs}
\usepackage{color}
\usepackage{fullpage}
\usepackage{hyperref}
\usepackage{cleveref}
\usepackage{natbib}
\usepackage{url}
\usepackage[table]{xcolor}
\usepackage{diagbox}
\usepackage{nicefrac}
\usepackage{nicematrix}
\usepackage{color, colortbl}
\definecolor{LightCyan}{rgb}{0.88,1,1}
\usepackage{caption}
\usepackage[table]{xcolor} 
\usepackage{graphicx}
\usepackage{subcaption}
\definecolor{LightCyan}{rgb}{0.88, 1, 1} 
\usepackage{float}
\usepackage{tabularray}

\usepackage{array}
\usepackage{makecell, multirow, booktabs}
\usepackage{siunitx} 
\usepackage{algpseudocode}
\usepackage{algorithm}
\usepackage{scalerel}
\UseTblrLibrary{booktabs}

\newtheorem{definition}{Definition}
\newtheorem{theorem}{Theorem}
\newtheorem{corollary}{Corollary}
\newtheorem{lemma}{Lemma}
\newtheorem{assumption}{Standing Assumption}
\newtheorem{example}{Example}
\newtheorem{remark}{Remark}

\DeclareAcronym{pdf}{short = {pdf}, long  = {probability density function}}
\DeclareAcronym{mse}{short = {MSE}, long  = {mean square error}}
\DeclareAcronym{smc}{short = {SMC}, long  = {sequential Monte Carlo}}
\DeclareAcronym{ti}{short = {TI}, long  = {thermodynamic integration}}
\DeclareAcronym{it}{short = {IT}, long  = {importance tempering}}
\DeclareAcronym{ess}{short = {ESS}, long  = {effective sample size}}
\DeclareAcronym{gp}{short = {GP}, long  = {Gaussian process}}
\DeclareAcronym{ode}{short = {ODE}, long  = {ordinary differential equation}}
\DeclareAcronym{lvm}{short = {LVM}, long  = {Lotka--Volterra model}}
\DeclareAcronym{mrna}{short = {mRNA}, long  = {messenger ribonucleic acid}}
\DeclareAcronym{name}{short = {ELATE}, long  = {ExtrapoLAting Tempered Expectations}}

\algnewcommand{\Inputs}[1]{%
  \State \textbf{Inputs:}
  \Statex \hspace*{\algorithmicindent}\parbox[t]{.96\linewidth}{\raggedright #1}
}
\algnewcommand{\Initialize}[1]{%
  \State \textbf{Initialize:}
  \Statex \hspace*{\algorithmicindent}\parbox[t]{.96\linewidth}{\raggedright #1}
}

\begin{document}

\title{Extrapolation of Tempered Posteriors}
\author{Mengxin Xi$^1$, Zheyang Shen$^2$, Marina Riabiz$^1$ \\
Nicolas Chopin$^3$, Chris J. Oates$^{2,4}$ \\
\small $^1$ King's College London, UK \\
\small $^2$ Newcastle University, UK \\
\small $^3$ ENSAE, Institut Polytechnique de Paris, France \\
\small $^4$ Alan Turing Institute, UK }
\maketitle

\begin{abstract}
    Tempering is a popular tool in Bayesian computation, being used to transform a posterior distribution $p_1$ into a reference distribution $p_0$ that is more easily approximated.
    Several algorithms exist that start by approximating $p_0$ and proceed through a sequence of intermediate distributions $p_t$ until an approximation to $p_1$ is obtained. 
    Our contribution reveals that high-quality approximation of terms up to $p_1$ is not essential, as knowledge of the intermediate distributions enables posterior quantities of interest to be extrapolated.
    Specifically, we establish conditions under which posterior expectations are determined by their associated tempered expectations on any non-empty $t$ interval.
    Harnessing this result, we propose novel methodology for approximating posterior expectations based on extrapolation and smoothing of tempered expectations, which we implement as a post-processing variance-reduction tool for sequential Monte Carlo.    
\end{abstract}


\section{Introduction}
\label{sec: intro}

This paper focuses on sampling methods rooted in \emph{tempering}, a computational device that has been exploited also for optimisation tasks \citep{kirkpatrick1983optimization} and in diverse application domains such as chemistry \citep{khachaturyan1979statistical}, physics \citep{swendsen1986replica}, and operational research \citep{pincus1970monte}.
The main idea of tempering is to construct a smoothly-varying sequence $(p_t)_{0 \leq t \leq 1}$, with $p_0$ representing a simple or tractable distribution and $p_1$ representing the distribution of interest.
Approximation of the simpler distributions in this sequence can be leveraged for approximation of more complicated distributions.
Note that in our convention the index $t$ can be interpreted as an \emph{inverse} temperature, so that tempering refers to the process of transforming $p_1$ into $p_0$, while \emph{annealing} refers to the process of transforming $p_0$ into $p_1$, and this terminology is also widely used.

In Bayesian settings where tempering is used, it is typical (but not essential) for $p_0$ to be the prior and $p_1$ to be the posterior distribution of interest.
Several different approaches to posterior approximation exploit tempering, whether this be in the form of importance sampling \citep{neal2001annealed}, parallel tempering \citep{swendsen1986replica,marinari1992simulated,geyer1995annealing}, sequential Monte Carlo \citep{chopin2002sequential,chopin2023connection}, piecewise deterministic Markov processes \citep{sutton2022continuously}, or gradient flows \citep{nusken2024stein,maurais2024sampling}.
Several methods also exploit tempering for computation of the marginal likelihood \citep{gelman1998simulating,friel2008marginal}.

All existing tempering-based methods for posterior approximation, to the best of our knowledge, attempt to approximate $p_t$ across a range of values for $t$ that span the interval $[0,1]$.
The precise values of $t$ that are considered may be specified at the outset or at run-time, and need not be uniformly spaced, but in all cases some computational resources are devoted to direct approximation of the posterior itself, corresponding to $t = 1$.
Since the posterior is, by construction, usually the most complicated distribution being approximated, this complexity is the principal determinant of the total computational resources required.
However, our contribution reveals that accurate approximation of $p_1$ may be unnecessary, as posterior quantities of interest can in principle be extrapolated based on their tempered equivalents with $t < 1$.

In mathematics, the property of being able to extrapolate a function is expressed as \emph{analyticity}; a (real) analytic function is defined as having a convergent power series at any point in its domain, and it can be shown that an analytic function is fully determined by knowledge of that function in any open set.
On the theoretical side, our main contribution is to establish weak sufficient conditions on the prior, the likelihood, and the functional of interest $f : \mathbb{R}^d \rightarrow \mathbb{R}$, under which the map $t \mapsto \mathbb{E}_t[f]$ (where the subscript $t$ denoted expectation with respect to $p_t$) is analytic on $t \in [0,1]$.
This implies that the tempered expectations in any non-empty interval $(0,t)$ fully determine the posterior expectation of interest.
On the practical side, although we cannot directly access tempered expectations to exploit analyticity, we take inspiration from these theoretical results and endow tempering \ac{smc} methods \citep{chopin2002sequential} with novel extrapolation and smoothing functionality, which we call \ac{name}. 
Roughly speaking, \ac{name} can be thought of as a novel variance-reduction tool capable of leveraging the approximations produced at inverse temperatures $t < 1$ to better approximate expectations at $t = 1$. 
Similar functionalities could in principle be applied to any of the aforementioned methods based on tempering, but to promote the use of these methods we focus on the state-of-the-art \emph{waste-free} \ac{smc} \citep{dau2022waste}, providing \ac{name} as an add-on to the \texttt{particles} package of \citet{chopin2020introduction}.

The idea of leveraging samples from tempered posteriors for the purpose of approximating posterior expectations is not itself novel:
\citet{moller1993discussion} noted that importance reweighting can be applied to adjust for the bias that is incurred due to sampling from a tempered version of the target, and several subsequent authors developed this importance sampling idea in the context of various tempering-based algorithms, including \citet{neal1996sampling,neal2001annealed,neal2005estimating,gramacy2010importance,nguyen2014improving,zanella2019scalable,li2023importance,karamanis2024persistent}.
Our approach is distinct from these works, in that we formulate a regression task that implicitly prioritises predictive \ac{mse}, as opposed to relying on importance sampling where unbiased estimation is prioritised.
In fact, we demonstrate how \ac{name} can be combined with the 
\emph{\acl{it}}
method of \citep{gramacy2010importance}, 
which provides a variance reduction technique for combining several importance sampling estimators, which in our case arise from \ac{smc}. 
The result is a potential `double' accuracy boost for tempering \ac{smc} in the form of a pure post-processing method. 

The remainder of the paper proceeds as follows:
Our setting, notation, and statements of our main theoretical results are contained in \Cref{sec: theory}.
These set the scene for presenting \ac{name} in \Cref{sec: methods}.
The empirical performance of \ac{name} is assessed in \Cref{sec: applications}, and a discussion of our findings is contained in \Cref{sec: discuss}.

\section{Theoretical Foundations for Extrapolation of Tempered Posteriors}
\label{sec: theory}

This section presents our main theoretical results, which serve as motivation for developing \ac{name} in \Cref{sec: methods}.
To simplify presentation, we consider only distributions on $\mathbb{R}^d$ for some $d \in \mathbb{N}$, but we note that our arguments do not depend strongly on the domain and could in principle be generalised (c.f. \Cref{rem: non euclidean}).

Our central object of study is a \emph{tempered posterior}
\begin{align}
p_t(x) = \frac{p_0(x) L(x)^t}{Z_t} ,  \label{eq: tempered posterior}
\end{align}
where for the purposes of this paper the prior \ac{pdf} $p_0 : \mathbb{R}^d \rightarrow [0,\infty)$ (with respect to the Lebesgue measure on $\mathbb{R}^d$) is assumed to exist, $L : \mathbb{R}^d \rightarrow (0,\infty)$ is the likelihood, $t\in [0,1]$ is the inverse temperature, and $Z_t$ is the appropriate normalisation constant.
A standing assumption is made in this work which ensures tempered posteriors are well-defined:  

\begin{assumption} \label{ass: stand}
    $L : \mathbb{R}^d \rightarrow (0,\infty)$ is bounded.
\end{assumption}

\noindent \Cref{ass: stand} implies that the normalising constants exist, i.e. $Z_t \in (0,\infty)$, and thus the \acp{pdf} $p_t$ also exist;  moreover, each $p_t$ can be bounded by a multiple of $p_0$; see \Cref{lem: temp moment} in \Cref{app: proof main}. 

Let $f : \mathbb{R}^d \rightarrow \mathbb{R}$ be a $p_0$ integrable function whose posterior expectation is of interest.
The tempered posterior expectations, which we will denote throughout as 
$$
g(t) := \mathbb{E}_t[f] := \mathbb{E}_{X_t \sim p_t}[f(X_t)] ,
$$
exist as a consequence of $p_0$ integrability and \Cref{ass: stand}. 
The main mathematical question that we pose and solve is ``how smooth is the function $g$?''.
Before presenting our results, it is instructive to consider an example where tempered expectations can be explicitly computed:

\begin{example}[Gaussian location model]
\label{ex: gauss}
    For the Gaussian location model 
    $$
    y_1,\dots,y_n \stackrel{\mathrm{iid}}{\sim} \mathcal{N}(x , \sigma^2), 
    $$
    with $\sigma > 0$ fixed, consider a conjugate prior $p_0(x) = \mathcal{N}(x ; \mu_0,\sigma_0^2)$, so that the tempered posterior is again Gaussian with

    \begin{center}
    \begin{tabular}{lll}
    $p_t(x) = \mathcal{N}(x;\mu_t,\sigma_t^2)$, \quad where \quad & 
    $\mu_t := \sigma_t^2 (\sigma_0^{-2}\mu_0+tn\sigma^{-2}\bar{y})$ & \; \; and \; \;  $\bar{y} := \frac{1}{n} \sum_{i=1}^n y_i$. \\
    & $\sigma_t^2 := (\sigma_0^{-2}+ tn \sigma^{-2})^{-1}$ & 
    \end{tabular}
    \end{center}
    
    \noindent Considering the identity function $f(x) = x$, corresponding to the first moment, we have
    \begin{align}
    g(t) = \mathbb{E}_t[f] = \frac{ (\sigma_0^{-2} \mu_0) + t ( n \sigma^{-2} \bar{y} ) }{ (\sigma_0^{-2}) + t (n \sigma^{-2}) }, \label{eq:Gauss location 1d}
    \end{align}
    which is a rational function of $t$ with positive denominator, and is therefore analytic on $t \in [0,1]$.
\end{example}

\noindent Although for this example it was straightforward to compute the tempered expectations and to deduce that the map $t \mapsto g(t)$ is analytic, this calculation will not be possible in general.
Our main theoretical contribution is a set of weak and easily-verifiable conditions under which analyticity can be deduced.
To this end, we first present a general result on the existence of higher-order derivatives of $g(t)$ in \Cref{thm: main}, and then use this to obtain explicit sufficient conditions in \Cref{cor: sufficient}.
To state our first result, let $\ell(x) := \log L(x)$, which is well-defined from \Cref{ass: stand}.

\begin{theorem}[Regularity of tempered expectations] \label{thm: main}
If the expectations $\mathbb{E}_0[|f \ell^i|]$ and $\mathbb{E}_0[|\ell^i|]$ exist for $i = 0,\dots,k$, then $g^{(k)}(t)$ is well-defined for all $t \in [0,1]$.
Further, if for some $t \in [0,1]$ it holds that 
\begin{align}
\sum_{k=0}^\infty \left| \frac{\mathbb{E}_t[f \ell^k]}{k!} \right| < \infty , \label{eq: f assum}
\end{align}
and, for some $\epsilon > 0$,
\begin{align}
\mathbb{E}_t[\exp\{ (1+\epsilon) |\ell| \}] < \infty , \label{eq: sub exp}
\end{align}
then $g$ is analytic on $[0,1]$.
\end{theorem}

\noindent The proof of \Cref{thm: main} is provided in \Cref{app: proof main}, and we highlight that it is `elementary' and self-contained.
Despite being elementary, the ideas used to obtain this result are perhaps surprising in their diversity; repeated application of the product rule for differentiation introduces a recursive relationship among derivatives that we represent using \emph{lag polynomials}, a concept borrowed from the time series literature \citep{hamilton2020time}.
These lag polynomials are in turn related, via a ring isomorphism, to complex power series, and properties of complex analytic functions are used to obtain the final result.

It is worth emphasising the strength of the second part of \Cref{thm: main}; being (real) analytic is a much stronger property than simply being smooth (i.e. having derivatives of all orders).
Indeed, if the function $g$ is analytic on $[0,1]$, then knowledge of $g$ on any non-empty interval $(0,t)$ is sufficient to perfectly extrapolate $g$ to the whole of the interval \citep[this fact follows from the \emph{identity theorem} for real analytic functions; see][Corollary 1.2.7]{krantz2002primer}.
This would not be true for smooth functions in general, as the existence of \emph{mollifiers} such as
\begin{align*}
    t \mapsto \left\{ \begin{array}{ll} 0, & t \in [0,1/2), \\ \exp\{ - 1 / ( 1 - 4 |t-1|^2 ) \}, & t \in [1/2,1],  \end{array}  \right.
\end{align*}
demonstrates that a smooth function can vanish everywhere in an interval and yet take non-zero values outside of that interval.
Analytic functions are thus especially well-suited to being extrapolated.
Strikingly, the map $t \mapsto g(t)$ is analytic quite generally in the tempering context.
Indeed, \Cref{cor: sufficient} below establishes weak and easily-verifiable sufficient conditions on the prior $p_0$, likelihood $L$, and function $f$ of interest that imply conditions \eqref{eq: f assum} and \eqref{eq: sub exp} of \Cref{thm: main} hold, and in doing so unlocks the potential for novel methodology designed to exploit the regularity of tempered posteriors.
These sufficient conditions will now be discussed:

\begin{definition}[Informative prior condition on $p_0$ and $L$]
\label{def: info}
    If, for some $\epsilon > 0$, we have $\int p_0(x) L(x)^{-\epsilon} \; \mathrm{d}x < \infty$, then we say that an \emph{informative prior} condition on $p_0$ and $L$ is satisfied.
\end{definition}

\noindent The intuition for \Cref{def: info} is that the likelihood $L$ cannot vanish too quickly relative to the prior $p_0$.
This condition is satisfied, for example, when $p_0$ is Gaussian and $L$ is continuous with a Gaussian tail.
On the other hand, this condition would preclude, for example, the situation where $p_0$ is a Laplace distribution and $L$ is a Gaussian likelihood. 
It would also preclude an improper prior, but note that we already assumed $p_0$ is a proper probability distribution at the outset, to ensure \eqref{eq: tempered posterior} is well-defined (a work-around to handle improper priors is outlined in \Cref{rem: weak prior}).

\begin{definition}[Growth condition on $f$]
\label{def: growth}
    If, for some $C_1, C_2 \in (0,\infty)$ and some $m \in \mathbb{N}$, it holds that $|f| \leq C_1 + C_2 |\ell|^m$, then we say that a \emph{growth condition} on $f$ is satisfied.
\end{definition}

\noindent To build intuition for \Cref{def: growth} note that, for a Gaussian likelihood, $|\ell(x)| \asymp x^2$ as $\|x\| \rightarrow \infty$, and thus this growth condition allows $f$ to grow at most polynomially fast.

\begin{corollary}[Sufficient conditions on $f$, $p_0$ and $L$]
\label{cor: sufficient}
    Suppose that an informative prior condition on $p_0$ and $L$, and a growth condition on $f$, are satisfied.
    Then the conditions of \Cref{thm: main} are satisfied and $g$ is analytic on $[0,1]$.
\end{corollary}

\noindent An elementary proof is contained in \Cref{app: proof corollary}.
Notice that no continuity-type regularity of $f$, $p_0$ or $L$ was assumed.
This result demonstrates that analyticity of $g$ is quite general, and moreover the conditions in \Cref{def: info,def: growth} can typically be validated if the tail behaviour of $p_0$, $L$ and $f$ is known.

Before going on to develop computational methodology that exploits extrapolation in \Cref{sec: methods}, we first make a few remarks to explain why our theoretical analysis may be of more general interest: 

\begin{remark}[Improper priors]
    \label{rem: weak prior}
    Improper priors $p_0$ can be handled by considering an alternative sequence of tempered distributions $q_t(x) \propto q_0(x) [p_0(x) L(x) / q_0(x)]^t$ for a suitable \ac{pdf} $q_0 : \mathbb{R}^d \rightarrow [0,\infty)$.
    Our theoretical results can then be directly applied, substituting $p_0$ with $q_0$ and substituting $L$ with $p_0 L / q_0$.
\end{remark}

\begin{remark}[Tempered generalised posteriors]
    Our analysis does not require the interpretation of $L$ as a likelihood, and one could consider any loss function \emph{in lieu} of $- \log L$, for example arising in generalised Bayesian inference \citep{bissiri2016general,knoblauch2022optimization}.
\end{remark}

\begin{remark}[Tempering beyond Bayesian statistics]
Though we have couched our results in terms of Bayesian statistics, they can be applied to general geometric tempering of the form $p_t = p_0^{1-t} p_1^t$ by setting $L := p_1 / p_0$. 
\end{remark}

\begin{remark}[Generalisation to other domains]
\label{rem: non euclidean}
Inspection of the proofs of \Cref{thm: main} and \Cref{cor: sufficient} reveals that we do not use the mathematical structure of $\mathbb{R}^d$, except for a technical result on the interchange of derivative (with respect to $t$) and integral (with respect to $x$) in \Cref{lem: interchange}.  
Thus we anticipate our analysis can be extended to other smooth spaces for which such an interchange is allowed.
\end{remark}

\begin{remark}[A non-elementary proof]
It is possible to shorten the proof of the second part of \Cref{thm: main} by appealing to the fact that complex differentiability implies complex analyticity; details of this `non-elementary' approach are provided in \Cref{app: shortened proof}.
Nevertheless, the elementary proof is presented here because we also require the intermediate calculations from \Cref{lem: binom} in \Cref{sec: methods}.  
\end{remark}

\paragraph{Numerical Analytic Continuation}

In practice we will not have exact access to the tempered expectations $g(t)$, but a plethora of numerical methods are available that can produce approximations $\hat{g}(t)$.
A theorem of \citet{landau1986extrapolating} states, roughly speaking, that if $\hat{g}(t) = g(t) + O(\epsilon)$ are provided up to some horizon $t_{\max}$, then \emph{numerical analytic continuation} is possible up to a horizon $t_{\max} + O(- \log \epsilon)$ \citep[see also][]{trefethen2023numerical}.
Our results therefore motivate leveraging the approximations $\hat{g}(t)$ as `data' with $t < 1$, when estimating the posterior expectation $g(1)$.
However, existing methods for numerical analytic continuation require data to be gathered at specific designs $\{t_i\}_{i=1}^n$, which is in conflict with the use of state-of-the-art sampling methods where $\{t_i\}_{i=1}^n$ are adaptively selected.
They are also not designed to integrate derivative data, which we will demonstrate is available in our context.
To bridge this gap, a novel regression method for numerical analytic continuation is required, and this is the focus of \Cref{sec: methods}.

\section{\acf{name}}
\label{sec: methods}

This section introduces novel methodology to harness extrapolation of tempered expectations in the context of \ac{smc}. 
Intuitively, we will use a regression model to `smooth out' the errors of the estimators of the tempered expectations arising from \ac{smc}, and evaluate the fitted regression model at $t = 1$ to obtain an improved estimator for the posterior expectation of interest.
\Cref{subsec: tempered smc} recalls the main ingredients of a modern \ac{smc} method. \Cref{subsec: extrap as reg} then casts extrapolation from the tempered \ac{smc} output as a regression task and proposes a suitable regression model, that underlies \ac{name}. The idea is illustrated in \Cref{subsec: illust}.

\subsection{Tempered Sequential Monte Carlo}
\label{subsec: tempered smc}

At a high level, at each inverse temperature $t_i$, a tempering \ac{smc} method constructs a collection of  \emph{particles} $\{x_j^{(i)}\}_{j=1}^N \subset \mathbb{R}^d$ and \emph{weights} $\{w_j^{(i)}\}_{j=1}^N$, such that the tempered distribution $p_{t_i}$ is approximated by the empirical distribution
\begin{align}
Q_N^{(i)} := \sum_{j=1}^N w_j^{(i)} \delta_{x_j^{(i)}}, \label{eq:empirical_approximation_pti}
\end{align}
where the number of particles is $N \in \mathbb{N}$. 
The inverse temperature is initialised at $t_0 = 0$, the particles are initialised as independent samples from $p_0$, and the weights are initialised at~${1}/{N}$.
The inverse temperature is then gradually increased following a schedule $\{t_i\}_{i=1}^n$ that can be pre-specified or determined at run-time \citep{chopin2002sequential,chopin2023connection}.
As the inverse temperature is increased, the weights and particles are updated to better reflect the tempered distribution being approximated; see \Cref{subsection:tempered_smc} for more details. 
A mature literature on \ac{smc} provides a range of options for how the weights and particles are evolved \citep{chopin2020introduction}. 
For a suitable tempered \ac{smc} algorithm, at each inverse temperature~$t_i$ we can read off a consistent and asymptotically normal estimator of $g(t_i)$
\begin{align}
    \hat{g}(t_i) := Q_N^{(i)}[f] = \sum_{j=1}^N w_j^{(i)} f(x_j^{(i)}), \qquad \sqrt{N}\left( \hat{g}(t_i) - g(t_i) \right) \stackrel{d}{\rightarrow} \mathcal{N}(0,\sigma_i^2 ), \label{eq:AN_SMC}
\end{align}
where $\sigma^2_i$ denotes the asymptotic variance, and with convergence occurring in the limit $N \rightarrow \infty$.
Further, for some modern \ac{smc} algorithms such as the \emph{waste-free} \ac{smc} algorithm of \citet[][summarised in \Cref{subsection:WF_SMC}]{dau2022waste}, the asymptotic variance can be automatically estimated, say by $\hat{\sigma}_i^2$, from a single run of \ac{smc} \citep[see Section 4.3 of][]{dau2022waste}.
Access to such variance estimates is a prerequisite for the \ac{name} methodology that we introduce next; where needed, in the sequel we will use the notation $\sigma_i^2[f]$ and $\hat{\sigma}_i^2[f]$ to make the dependence on $f$ explicit.

\subsection{Extrapolation of Tempered Posteriors as a Regression Task}
\label{subsec: extrap as reg}

The analysis of \Cref{sec: theory} motivates practical regression methodology that leverages approximate tempered expectations to more accurately approximate posterior expectations of interest.
This section explains how we approached this regression task.
There are two types of data that we exploit; direct approximation of function values $g(t_i)$, and indirect approximation of the gradients $g'(t_i)$.
Our approach to regression involves heuristic use of a heteroscedastic Gaussian error model, similar in spirit to \emph{least squares Monte Carlo} \citep{carriere1996valuation}.

\paragraph{Function Value Data}

From \ac{smc} 
we have access to point estimates $\hat{g}(t_i)$ for each $g(t_i)$, together with the variance estimates $\hat{\sigma}_i^2$.
Asymptotic normality motivates the heuristic use of a Gaussian heteroscedastic error model
\begin{align}
    \hat{g}(t_i) \stackrel{\approx}{\sim} \mathcal{N} ( g(t_i) , \hat{\sigma}_i^2 ), \label{eq: gauss like}
\end{align}
where the symbol $\stackrel{\approx}{\sim}$ can be read as `approximately distributed as', and where, for simplicity, we treat these data as independent \citep[in practice, such dependence is mitigated by the \emph{propagation of chaos} effect in waste-free \ac{smc};][]{dau2022waste}.

\paragraph{Gradient Data}

From the recurrence relation established in \Cref{lem: binom} of \Cref{app: proof main}, we can express the derivatives of $g$ in terms of expectations, and these can thus also be approximated using \ac{smc}.
Taking the first-order derivative, we have
\begin{align*}
    g'(t_i) & = \mathbb{E}_{t_i}[f\ell] - \mathbb{E}_{t_i}[f] \mathbb{E}_{t_i}[\ell],
\end{align*}
which can be estimated using
\begin{align}
    \hat{g}'(t_i) := Q_N^{(i)}[f\ell] - Q_N^{(i)}[f] Q_N^{(i)}[\ell] .  \label{eq: derivative estimator}
\end{align}
Heuristically, via an assumption of independent errors and the delta method, the asymptotic variance of \eqref{eq: derivative estimator} can be estimated as
\begin{align*}
    \hat{\gamma}_i^2 := \hat{\sigma}_i^2[f \ell] + Q_N^{(i)}[f]^2  \hat{\sigma}_i^2[\ell]  + Q_N^{(i)}[\ell]^2 \hat{\sigma}_i^2[f], 
\end{align*}
which motivates augmenting the Gaussian heteroscedastic error model \eqref{eq: gauss like} with additional gradient data using
\begin{align}
    \hat{g}'(t_i) \stackrel{\approx}{\sim} \mathcal{N}( g'(t_i) , \hat{\gamma}_i^2 ) .  \label{eq: gauss like 2}
\end{align}
Consideration of higher-order derivative data is possible in principle but would introduce further heuristic approximations, so we stop at first-order gradient data and explore whether there is a benefit from inclusion of such data through the empirical assessment in \Cref{sec: applications}.



\paragraph{Regression Likelihood}
In summary, by accounting for both function value and gradient data, we arrive at a heuristic log-likelihood
\begin{align}
\mathfrak{L}(g) := - \sum_{i=0}^h \frac{ ( g(t_i) - \hat{g}(t_i) )^2 }{  2 \hat{\sigma}_i^2 } - \sum_{i=0}^h \frac{ ( g'(t_i) - \hat{g}'(t_i) )^2 }{ 2  \hat{\gamma}_i^2 }, \label{eq: heur log like}
\end{align}
where $h\leq n$ denotes the number of initial temperatures for which we evaluate the data. Therefore, $h<n$ corresponds to \emph{extrapolation} of data from $t < 1$ to the target temperature $t=1$, while when $h=n$ corresponds to \emph{smoothing} of data that spans the interval $t \in [0,1]$.

\paragraph{Regression Model}

Having described the heuristic likelihood \eqref{eq: heur log like}, attention now turns to selecting a prior distribution for the function $g$, that is the regression model.
Motivated by conjugacy, we considered a \ac{gp} regression model $g \sim \mathcal{GP}( m_\theta , k_\phi )$, where $m_\theta : [0,1] \rightarrow \mathbb{R}$ is a prior mean function parametrised by $\theta$, and $k_\phi : [0,1] \times [0,1] \rightarrow \mathbb{R}$ is a prior covariance function parametrised by $\phi$.
For the prior covariance we set $k_\phi(t,t') = \lambda^2 \exp( - \ell^{-2} (t-t')^2 )$ where $\phi = \{\lambda , \ell\}$ are positive parameters to be specified; this ensures that samples from the \ac{gp} are analytic \citep[][p406]{handcock1993bayesian}, informed by \Cref{thm: main}.
To arrive at a suitable prior mean function we consider again the Gaussian location model from \Cref{ex: gauss}, noting that the 
moments of a Bayesian posterior distribution are often important quantities of interest:

\begin{example}[Gaussian location model, continued]
\label{ex: gauss 2}
    For the Gaussian location model in \Cref{ex: gauss}, consider monomial functions of interest $f(x) = x^k$.
    In this case the tempered posterior expectation is a rational function whose numerator and denominator are polynomials of order $k$.
    If we were to evaluate $g$ at $2(k+1)$ distinct inputs, then we could uniquely determine the coefficients $\theta$ of a rational function
    \begin{align}
    g_\theta(t) = \frac{a_0 + a_1 t + \dots + a_r t^r}{b_0 + b_1 t + \dots + b_s t^s} , \qquad \theta = (a_0,\dots,a_r,b_0,\dots,b_s), \label{eq: rat fn}
\end{align}
with $(r,s) = (k,k)$ and extrapolate to exactly recover $g(1) = g_\theta(1)$.
\end{example}

\noindent Of course, we cannot expect exactness under rational function extrapolation to hold for general posteriors and general functionals of interest, but we can still employ rational functions as in \eqref{eq: rat fn} as a prior mean $m_\theta$, motivated by the ubiquity of Gaussian-like posteriors due to the Bernstein--von Mises theorem \citep[][Section 10.2]{van2000asymptotic}.
Fortuitously, rational function approximation is also considered state-of-the-art for numerical analytic continuation \citep[][Section 7]{trefethen2023numerical}.
In practice, we maximize the GP marginal likelihood to jointly select the numerator and denominator degrees ($r$ and $s$) of the rational prior mean function, as well as estimate its parameters $\theta$ and the prior covariance function parameters $\phi$. 
Full details on the GP conditional distributions based on sample tempered expectations and their gradients, and the expression of the marginal likelihood are contained in \Cref{ap: gp computation}.

\paragraph{\ac{name} Output}
We define the estimate of $g(1)$ as the mean of the \ac{gp} conditional distribution at $t=1$, leveraging the analyticity of $g(t)$ while inherently minimizing the predictive \ac{mse}. 
Although we display the \ac{gp} posterior covariance in our plots for completeness, our quantitative assessment of \ac{name} relies solely on the posterior mean. This choice reflects a practical consideration: the underlying estimators of tempered expectations exhibit cross-temperature dependencies and heteroscedastic Monte Carlo variances that are themselves noisy. While such variance estimation instability can lead the analytic \ac{gp} posterior variance to become unreliable or overconfident, the conditional mean remains well regularized by the \ac{gp} prior smoothing.

\paragraph{Computational Cost}
The  cost associated with fitting the \ac{gp} is dominated by the parameter search, where evaluation of the marginal likelihood incurs a $O(n^3)$ cost. 
However, across our numerical experiments, the number of temperatures 
$n$ visited never exceeds one hundred, making this $O(n^3)$ cost  negligible relative to the cost of generating the underlying \ac{smc} output.
Importantly, because \ac{name} regression is one-dimensional in the  variable $t$, its computational overhead is  independent of the dimension of the model parameter~space.

\subsection{Illustration of \ac{name}}
\label{subsec: illust}

\begin{figure}[t!]
    \centering
        \includegraphics[width=\textwidth]{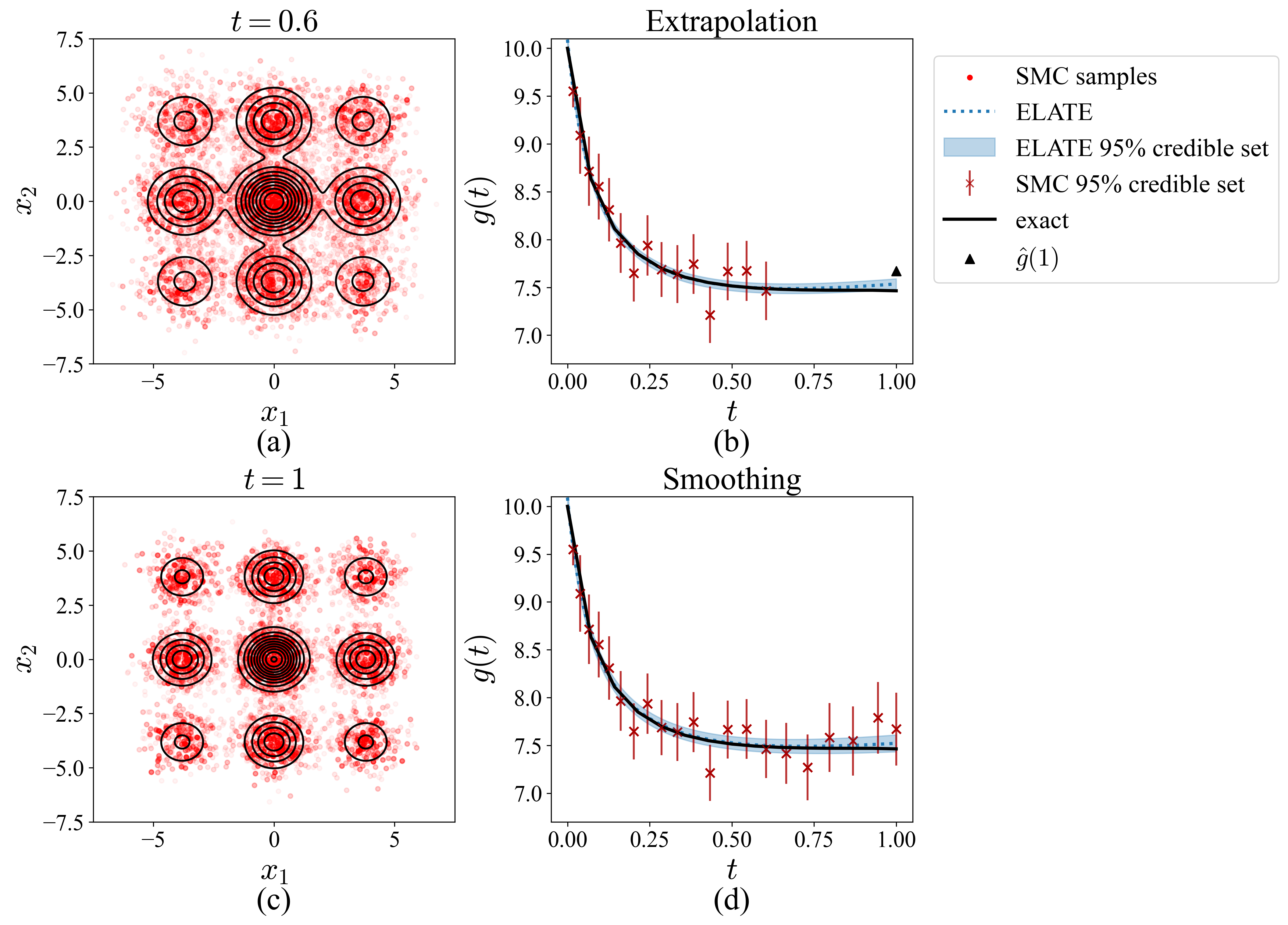}
    \caption{
    Illustration of \ac{name}.
    \Ac{smc} is used to sample from tempered versions of a \textbf{Gaussian mixture} target, shown here for (a) $t = 0.6$ and (c) $t = 1$.
    In panels (b) and (d) the exact tempered expectation $g(t) = \mathbb{E}_t[f]$, for $f(x) = x_1^2$, is shown in solid black line. Red crosses denote the  tempered expectations estimated by \ac{smc}, and the red error bars indicate the corresponding estimator variances. The blue dashed line indicates the \ac{gp} posterior predictive mean, while the blue shaded interval denotes the corresponding 95\% predictive credible set. 
    Panel (b) differs to panel (d) in that only training data with $t<0.6$ are used and the black triangle visualizes the reference  \ac{smc} estimator $\hat{g}(1)$; this illustrates the information content already present in these data, while further improvement is achieved in~(d) when using the full training dataset. 
   }

\label{fig:illustration:ELATE_SMC}
\end{figure}

To illustrate \ac{name}, in  \Cref{fig:illustration:ELATE_SMC} we consider  a two-dimensional Gaussian mixture model  and $f(x) = x_1^2$ as the function of interest; in this case, the conditions for analyticity of $g(t)$ are satisfied (c.f. \Cref{app: illus} for details).
Panels (a) and (c) show contour plots of tempered distributions $p_t$ alongside the \ac{smc} samples at two different temperatures.
Panels~(b) and~(d) display the function value data $\hat{g}(t_i)$, along with their associated  $\pm 1.96\hat{\sigma}_i$ error bars. 
As illustrated, the temperatures $\{t_i\}_{i=1}^n$ adaptively selected by the \ac{smc} sampler are more sparsely distributed near $t = 1$, where the variance of the estimated tempered expectations is  larger. This behavior suggests that more stable approximations obtained at lower inverse temperatures can be leveraged to refine our ultimate estimate of the target posterior expectation. Remarkably, panel (b) reveals that training the \ac{gp} solely on data within the restricted range $t < 0.6$ produces a more accurate approximation of the target expectation than the raw  \ac{smc} output at $t = 1$. In panel (d) all data are utilized, and the \ac{gp} predictive mean consistently outperforms the \ac{smc} baseline; in this setting, the corresponding \ac{gp} predictive uncertainty provides a more faithful representation of the true $g(t)$ trajectory than what is achieved in panel~(b).


\section{Empirical Assessment}
\label{sec: applications}

This section presents a detailed empirical evaluation of \ac{name}, focusing on standard quantities of interest such as the first and second moment of a posterior distribution. Although \Cref{sec: methods} introduced the regression-based extrapolation framework using estimators derived directly from \ac{smc}, the methodology is readily adaptable to alternative tempered expectation frameworks. In particular, \ac{name} can be applied to \acl{it} (\acs{it}) estimators \citep{gramacy2010importance}, which we review in \Cref{app: IT}. 
In our approach, the \ac{it} variances are quantified via bootstrapping, and the Gaussian heuristic likelihood in \eqref{eq: heur log like} is less justified for \ac{it} than for the \ac{smc} estimators. Nevertheless, to ensure a comprehensive evaluation, we establish both the ``vanilla" \ac{smc} estimator $\hat{g}_{\mathrm{SMC}}(1)$ and the \ac{it} estimator $\hat{g}_{\mathrm{IT}}(1)$ as baselines against which to assess the performance gains of \ac{name}.
In our simulations, both  \ac{smc} and \ac{it} estimators are constructed from the  same underlying simulation output. Tempered posterior samples are generated using the waste-free \ac{smc} algorithm of \citet{dau2022waste}, detailed in \Cref{app: SMC}. This setup uses a total of $N$ samples composed of $M$ particles, each evolved for $P$ steps under a $p_t$-invariant Markov kernel, with an adaptive temperature ladder governed by the minimum effective sample size threshold $\text{ESS}_{\text{min}}$.

When optimizing the \ac{gp} parameters,  
empirical results indicate that allowing numerator and denominator of the rational prior mean function to be high-degree polynomials does not yield systematic improvements in predictive performance; instead it significantly increases  instability during parameter optimization. Consequently, the degrees $(r, s)$ are restricted to the small discrete set $\mathcal{R} = \{0,1,2\}^2$. Additionally, the fitted rational prior mean function occasionally presents poles within the interval $[0, 1]$, which can severely degrade extrapolation performance. To mitigate this issue, a penalty term is incorporated into the fitting process, as explained in more detail in  \Cref{ap: gp computation}, where we also provide pseudocode to detail the practical implementation of \ac{name}.

In \Cref{subsec: gmm} we report additional findings for the 
simulated setting of the Gaussian mixture model presented above, 
followed by an application to Bayesian analysis for \acp{ode} in \Cref{subsec: first}. Finally, in \Cref{subsec: thermo} we  investigate the use of ELATE 
in thermodynamic integration, 
for estimating the marginal likelihood. Python code to reproduce these results is available at \texttt{https://github.com/K211Mengxin/ELATE}.

\subsection{Gaussian Mixture Model}
\label{subsec: gmm}
Comprehensive simulation results for the Gaussian mixture model introduced in \Cref{subsec: illust} are presented in \Cref{app: illus}. Our findings shows that the performance of \ac{name} is largely governed by the variance profile of the underlying estimators. When based on \ac{smc} data, both \ac{name} smoothing and extrapolation enhance the accuracy of  estimates in high variability settings, and they  do not degrade it in low variability settings, where smoothing works better than extrapolation. 
On the other hand, because  \ac{it} achieves a substantial variance reduction compared to plain \ac{smc} in this two-dimensional example, further post-processing the \ac{it} output via \ac{name} yields only marginal benefits. 
Finally, incorporating gradient information ensures better performance across experiments, supporting the inclusion of both terms in the heuristic
likelihood~\eqref{eq: heur log like}.

Guided by these observations, we focus in the remainder of this paper on assessing the performance of \ac{name} smoothing when using noisy observations of tempered \ac{smc} and \ac{it} expectations and their derivatives. 

\subsection{Parameter Inference for ODEs}
\label{subsec: first}

\begin{figure}[t!]
    \centering
        \includegraphics[width=\textwidth]{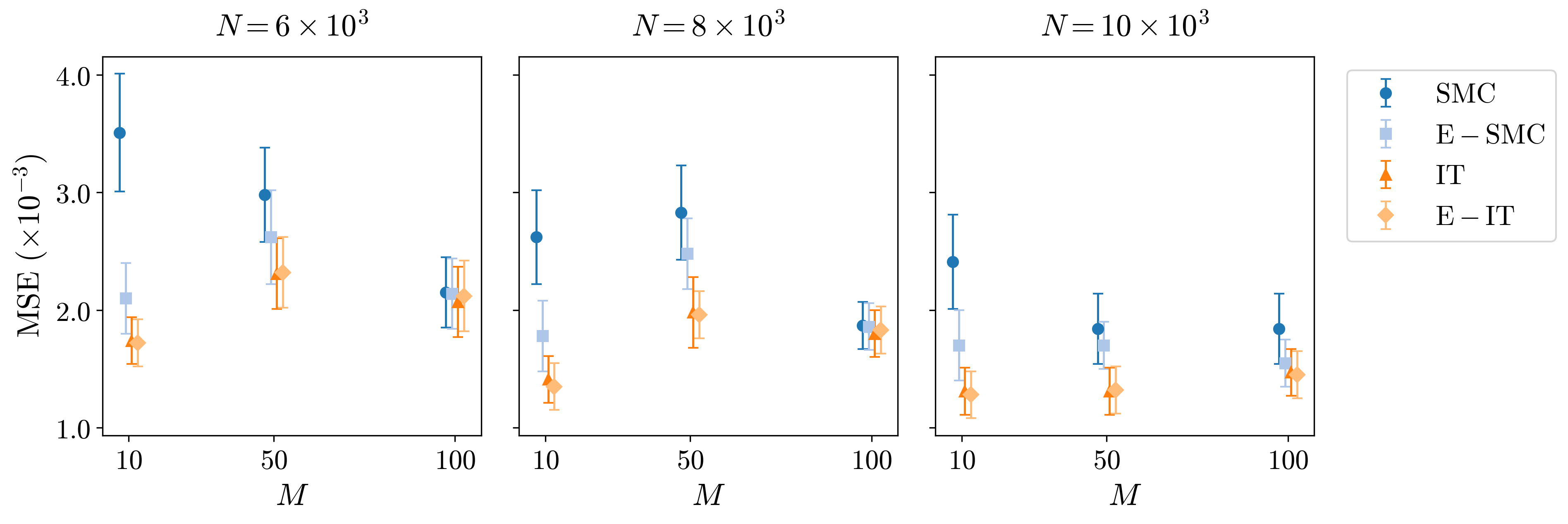}
    \caption{Parameter estimation for \acp{ode}. Estimator \ac{mse} and associated standard error for the \textbf{mRNA} model, and  the test function 
$f(\vartheta) = \delta$, computed over 100 independent realizations of SMC. Blue circle markers indicate the plain \ac{smc} estimators, and light-blue squares their \ac{name}-enhanced counterparts; orange triangles indicate the plain \ac{it} estimators, and light-orange diamond markers their \ac{name}-enhanced counterparts. Panels display increasing particle sizes $N$ from left to right, with standalone markers displaying performance across distinct resample sizes $M$. \ac{mse} values are presented in units of $10^{-3}$.  
   }
\label{fig: lvm mse}
\end{figure}

Motivated by settings where the effectiveness of sampling-based methods is limited due to the computational cost associated with evaluation of the likelihood, 
here we consider
parameter inference for \acp{ode}.
Our test-bed is the \ac{mrna} model of \cite{leonhardt2014single}, a prototypical system of coupled \acp{ode} describing the dynamics of mRNA delivery and protein expression in cells following transfection.  
The  model comprises four parameters, 
$\vartheta = \{ \psi, \delta, \beta, \tau_0 \}$, all of which are inferred. 
Following \cite{ballnus2017comprehensive, surjanovic2022parallel} the prior distributions are taken to be uniform over a finite interval, thus satisfying the  informative prior condition (c.f. \Cref{def: info}.)
Data are generated from the \ac{mrna} \ac{ode} under an additive Gaussian noise model, giving rise to a Gaussian likelihood. Two of the parameters are exchangeable, leading to a bimodal posterior distribution, and the computational challenge is to accurately approximate the mean and second moment of each posterior marginal.
Full details of the test-bed can be found in \Cref{appendix:mrna}. 

\begin{figure}[t!]
    \centering
        \includegraphics[width=0.75\textwidth]{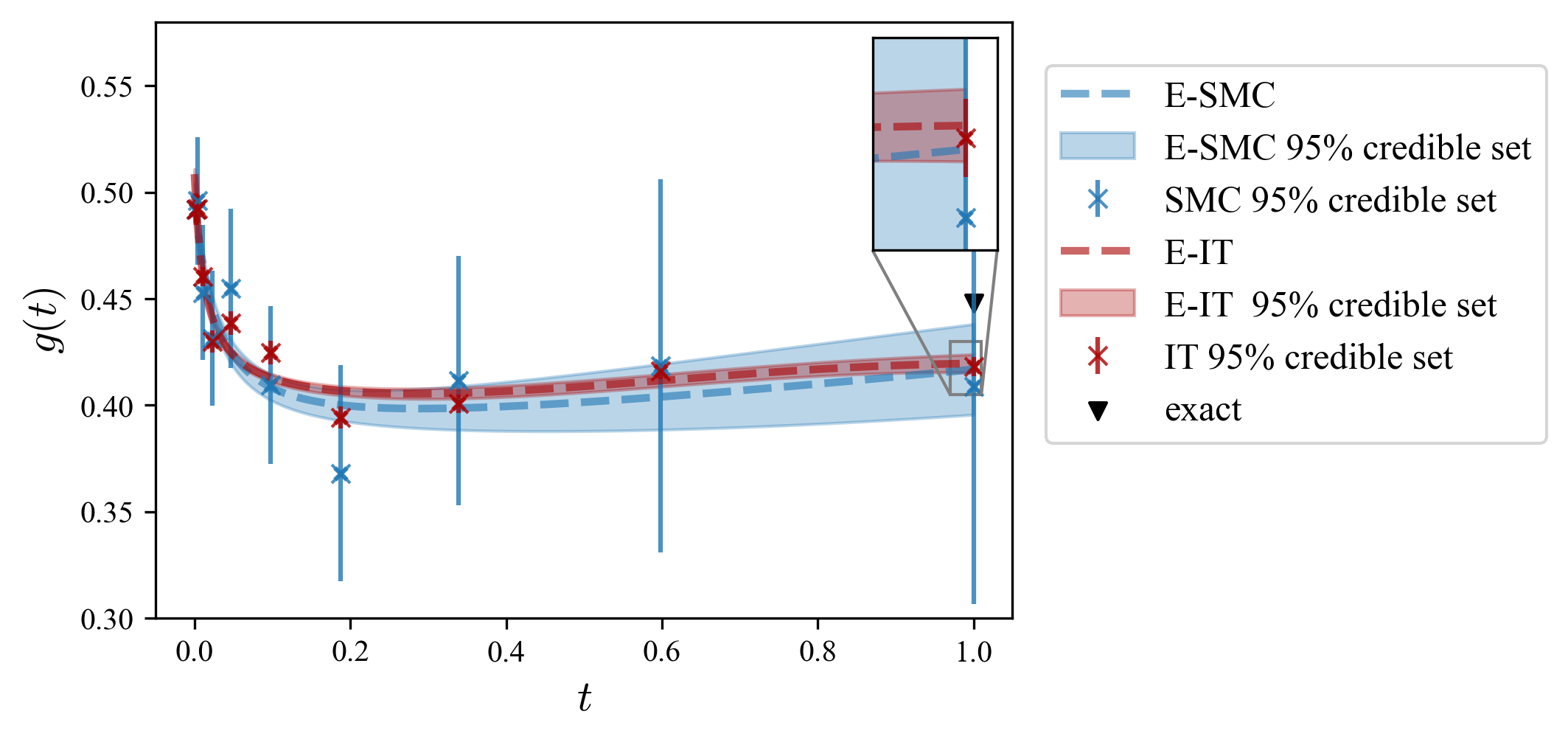}
    \caption{Illustration of \ac{name}, for the \textbf{\ac{mrna}} \ac{ode} model, using~\ac{smc} data (blue crosses and  vertical lines denote the \ac{smc} 95\% credible interval) and  the corresponding~\ac{it} data (red crosses and  vertical lines denote the \ac{it} 95\% credible interval). The blue and  dashed lines represent  the corresponding \ac{gp}  posterior means, and the blue and red shaded areas indicate the corresponding predictive credible intervals.
}
    \label{fig:visual_table1_mrna}   
\end{figure}

Performance is measured in terms of the \ac{mse} relative to a gold standard obtained averaging 100 brute-force extended \ac{smc} runs, each with $M=200$ resampled particles, 
 chain length $P=2500$
 for a total $N=500\times10^3$ total particles,
and  $\text{ESS}_{\text{min}} = 0.7$.
\Cref{fig: lvm mse} 
compares the baseline estimators (\ac{smc} and \ac{it}) with their \ac{name}-enhanced smoothing counterparts (E-SMC and E-IT), obtained using both function and gradient data, reporting the average \ac{mse} across 100 experiments when estimating the posterior mean of $\delta$. 
Applying \ac{name} either improves the corresponding baseline estimators
or it results in only negligible degradation. 
Overall, \ac{name} provides the largest benefits in the highly variable variable regimes, while remaining competitive when the baseline estimators are already accurate, consistent with findings for  the synthetic Gaussian mixture model results in \Cref{subsec: gmm}. 
Specifically, E-SMC  yields lower \ac{mse} than \ac{smc} across all values of 
$N$~and~$M$, whereas the gains of E-IT  upon \ac{it} are modest, and mainly observed at smaller resample size $M$.  
\Cref{fig:visual_table1_mrna} visually contrasts the four methods for one  realization of the \ac{smc}~data, with $M=10$ and $N =10 \times 10^3$.
To evaluate the sensitivity of our conclusions to the choice of integrand, we repeat the analysis estimating the means and second moments of all four parameters of the \ac{mrna} model for a fixed value of $M = 50$, while varying $N$. The results, presented in \Cref{app: LVM vary integrand}, corroborate the finding  that applying \ac{name} to baseline estimators can improve estimation accuracy, particularly when the baseline estimates exhibit high variability.

    Finally, we seek to illustrate potential failure modes for \ac{name} by considering $(a)$ a weakly informative Cauchy prior for which \Cref{def: info} does not hold;  $(b)$ a highly irregular function of interest $f(\vartheta) =\sin(100\delta)$; and $(c)$ a combination of both.
The results, presented in \Cref{app: failure modes}, indicate that under these conditions, the performance of \ac{name} depends critically on the accuracy of the underlying vanilla \ac{smc} estimates.

\subsection{Thermodynamic Integration}
\label{subsec: thermo}

Path sampling and the closely-related technique of \emph{\acl{ti}} (\textup{\acs{ti}}) emerged from the physics community as a computational approach to compute normalising constants, and are now a popular tool for computing marginal likelihood \citep{gelman1998simulating}. 
Empirical investigations have revealed \ac{ti} to be among the most promising approach to estimation of model evidence \citep{friel2012estimating,llorente2023marginal}, though other promising approaches compatible with tempered \ac{smc} have been developed \citep[e.g.][]{zhou2016toward,syed2024optimised}.
Focusing on \ac{ti}, in this section we investigate whether approximation accuracy can be improved using \ac{name}.

\paragraph{\ac{name} for \ac{ti}} 
The standard thermodynamic identity, in our notation, is
\begin{align*}
    \log Z_1 = \int_0^1 g(t) \; \mathrm{d}t , \qquad g(t) := \mathbb{E}_t[\ell] ,
\end{align*}
and various quadrature-based approximations to the integral have been developed \citep{friel2008marginal,calderhead2009estimating,oates2016controlled}.
These approximations take the form 
\begin{align}
\log Z_1 \approx \sum_{i=1}^n \alpha_i \hat{g}(t_i), \label{eq: approx thermo} 
\end{align}
where $\{\alpha_i\}_{i=1}^n$ are quadrature weights, $\{t_i\}_{i=1}^n$ are quadrature nodes, and $\{\hat{g}(t_i)\}_{i=1}^n$ are sample-based approximations to the integrand.
The overall error in \eqref{eq: approx thermo} can be decomposed into quadrature error and Monte Carlo error.
Our main observation here is that \emph{quadrature error will decrease exponentially fast in $n$ if we have an analytic integrand}, provided an appropriate quadrature rule is used \citep{gotz2001optimal}. 
This suggests that, \emph{if} $g(t)$ is analytic, only a small number of quadrature nodes may be needed in \eqref{eq: approx thermo}, which in turn would usefully control the computational cost since there would be fewer tempered expectations that need to be approximated.
However, to the best of our knowledge earlier works focused on using a trapezoidal rule, or a Simpson's rule, which does not take full advantage of the regularity of the integrand.
Our main result in \Cref{thm: main} is a set of sufficient conditions under which functions of the form $t \mapsto \mathbb{E}_t[f]$ are analytic, and thus our result can be applied to \ac{ti} in the specific case where $f = \ell$.
In this case, the growth condition on $f$ is automatically satisfied and we can present a particularly simple sufficient condition:

\begin{corollary} \label{cor: thermo}
    If an informative prior condition on $p_0$ and $L$ is satisfied then $t \mapsto \mathbb{E}_t[\ell]$ is analytic on $[0,1]$.
\end{corollary}

\noindent
There is a literature on the quadrature of analytic functions, but it tends to focus on specific sets of quadrature nodes, such as the roots of orthogonal polynomials \citep[e.g.][]{irrgeher2015integration,kuo2017multivariate}.
At the same time, the literature on \ac{ti} has considered selection of non-uniform quadrature nodes to account for the increased Monte Carlo variance often associated values of $t$ closer to 1, including methods that are both non-adaptive \citep{calderhead2009estimating,oates2016controlled} and adaptive \citep{miasojedow2013adaptive,behrens2012tuning,zhou2016toward,hug2016adaptive}.
These existing approaches are not immediately compatible with \ac{name}, since our quadrature nodes are automatically determined by waste-free \ac{smc}.
To proceed, we instead employ \emph{Bayesian quadrature} \citep{larkin1972gaussian}, since this  allows for arbitrary quadrature nodes, it provides epistemic uncertainty quantification consistent with the \ac{gp} model in \ac{name}, and it enables us to directly exploit the analyticity of the integrand \citep{karvonen2021integration}.
In our setting, Bayesian quadrature amounts to using the posterior \ac{gp} from \ac{name} as a surrogate for the exact integrand; full details are reserved for \Cref{app: BQ}.

\paragraph{\ac{name}-v2 for \ac{ti}} Given that traditional \ac{ti} is known to incur quadrature error, in addition to \ac{name} as just described, we note that $\tilde{g}(t) := \log Z_t$ satisfies $\tilde{g}'(t) = \mathbb{E}_t[\ell]$.
It follows that: 
\begin{corollary} \label{cor: thermo_v2}
 Under the same conditions as \Cref{cor: thermo}, the function $t\rightarrow \log Z_t$ is analytic on $[0, 1]$. 
\end{corollary}
\noindent
We therefore also consider the extrapolation of $\log Z_t$ by applying  \ac{name}  to the estimates of the tempered log-marginal likelihood and its variance, provided by \ac{smc} \citep[][Proposition~2]{dau2022waste}.  
We refer to this as \ac{name}-v2.

\begin{figure}[t!]
    \centering
        \includegraphics[width=\textwidth]{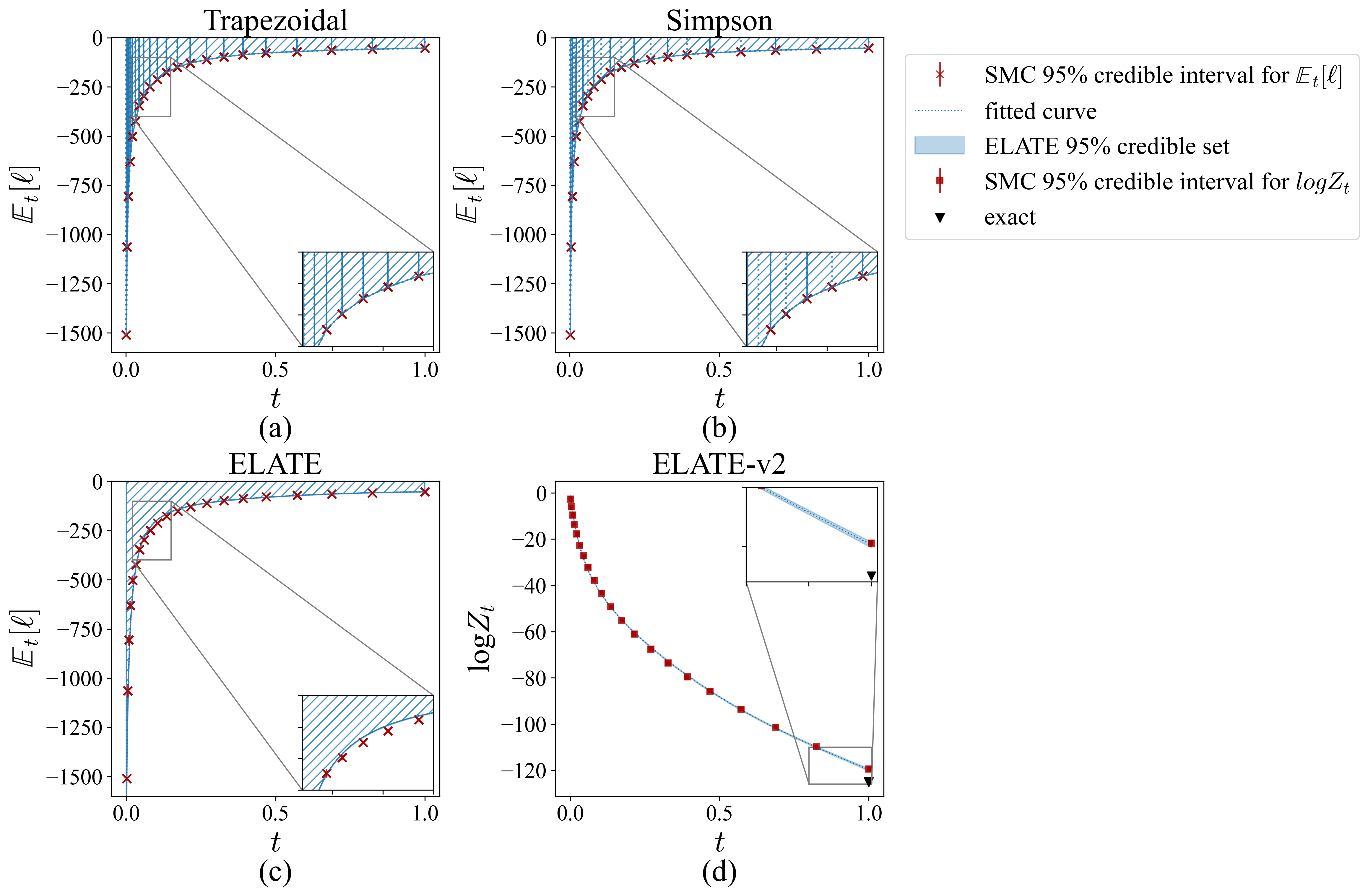}
    \caption{Visualization  of \ac{ti} for the \textbf{Sonar} dataset. Panel~(a) illustrates the trapezoidal rule,  Panel~(b)  Simpson’s rule, and Panel~(c) Bayesian quadrature based on \ac{name}. In Panels~(a), (b), and (c), the red crosses represent the \ac{smc} estimators of \(   \mathbb{E}_t[\ell] \), the solid blue vertical lines indicate the integration intervals, while the shaded regions represent the area under the curve being integrated. In panel (c) we also plot \ac{smc} error bars, non visible because they have small values, {and the dashed blue line represents the posterior mean of the \ac{gp} fitted to the estimators of $\mathbb{E}_t[\ell]$.}
    Panel~(d) illustrates \ac{name}-v2; the red sqaures denote the estimates of \(  \log Z_t \), with (non visible) error bars indicating the associated estimator variance.
    The  dashed blue line corresponds to the posterior mean of the \ac{gp} fitted to the estimators of  $\log Z_t$, and the blue shaded area indicates the  corresponding predictive credible interval.
    }
\label{fig:illustration:nl_sonar}
\end{figure}

We examine a challenging high-dimensional example,
where we fit a logistic regression model to the \textit{Sonar} data from \citep{Dua:2019} detailed in \Cref{app: sonar_model}, where, for completeness, we also report estimates of certain moments of the parameter vector. 
\Cref{fig:illustration:nl_sonar} illustrates the baseline trapezoidal rule \citep{friel2008marginal}, and Simpson's rule \citep{hug2016adaptive}, compared to \ac{name}, based on estimates of  $\mathbb{E}_{t}[\ell]$ and \ac{name}-v2, based on estimates of $\log Z_t$. 
Whilst the trapezoidal and Simpson's rules numerically integrate the discrete estimates of $\mathbb{E}_{t}[\ell]$ across the predefined tempering grid, \ac{name} fits a smooth surrogate function to these expectations and evaluates the integral via the resulting posterior mean within a Bayesian quadrature framework. In all cases, \ac{smc} provides reliable point estimates, particularly concentrated in the region \( t < 0.2 \), where more temperatures are automatically selected. As \( t \) increases beyond 0.2, the curvature of \( \mathbb{E}_{t}[\ell] \) flattens, where larger spaced temperatures have less impact on integration accuracy.
The same figure also shows that \ac{name}-v2 provides high-quality fits of \( \log Z_t \),  thereby supporting the use of ELATE-v2 to estimate the marginal log-likelihood.

\begin{figure}[t!]
    \centering
    \includegraphics[width=0.75\textwidth]{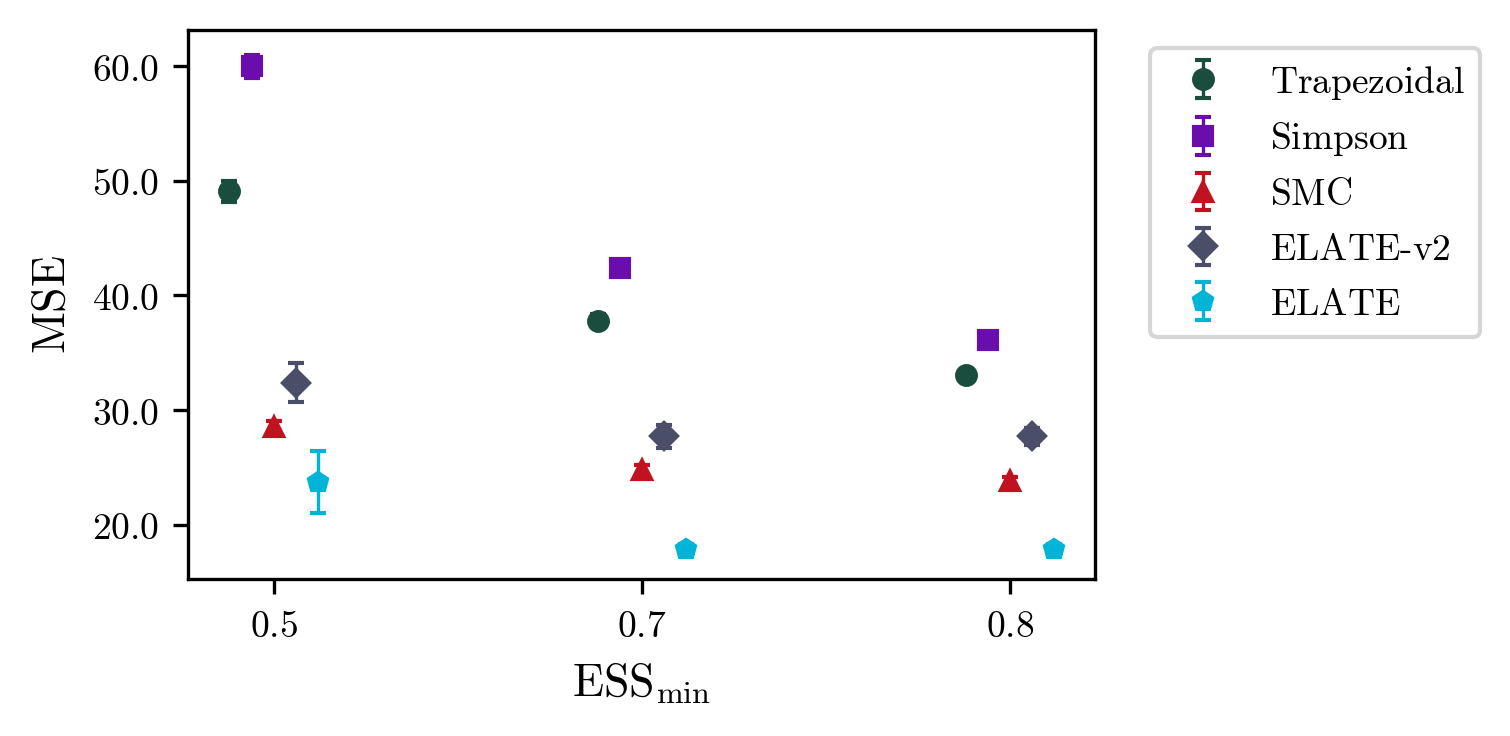}
    \caption{Thermodynamic integration.
Estimator \ac{mse} and associated standard error for the marginal log-likelihood of  the logistic regression on the \textbf{Sonar} dataset, computed over 100 independent realizations of \ac{smc}, varying the $\text{ESS}_{\text{min}}$ threshold.
We compare standard thermodynamic integration  based on trapezoidal and Simpson's quadratures (deep-green circle and deep-purple square markers, respectively), the estimate of the normalising constant estimates produced by \ac{smc} (red triangle markers), the Bayesian quadrature approach of \ac{name} (teal pentagon markers), and extrapolating $\log Z_t$ estimates produced by \ac{smc} using \ac{name}-v2 (deep-gray diamond markers).
}
\label{fig: thermo}
\end{figure}

To quantify the effectiveness of the evaluated methods in estimating the log-marginal likelihood, we perform 100 independent replications and compare the \ac{mse} of the respective estimators, as illustrated in \Cref{fig: thermo}. 
The gold standard is obtained using Simpson’s rule, with 130 equally spaced inverse temperatures, at which high-precision Monte
Carlo samples were generated.
To control for confounding, in the empirical assessment of \ac{name} for \ac{ti} we compare estimates of the log-marginal likelihood computed from the same \ac{smc} output, with $M=50$ and $P=400$. Furthermore, because these estimation methods are sensitive to the temperature ladder, the sequence of inverse temperatures $n$ is set adaptively according to the minimum effective sample size threshold $\text{ESS}_{\text{min}}$. 
Our simulations consistently demonstrate that \ac{name} outperforms alternative approaches in estimating the log-marginal likelihood. Notably, the \ac{mse} of the \ac{name} estimator remains identical for  $\text{ESS}_{\text{min}} =0.7$ and $\text{ESS}_{\text{min}}=0.8$, whereas the performance of classical quadrature rules and vanilla \ac{smc} continues to fluctuate. This validates the qualitative behavior observed in \Cref{fig:illustration:nl_sonar}: \ac{name} delivers stable, robust estimation with fewer quadrature nodes by leveraging the analyticity of $\mathbb{E}_t[\ell]$ through the integration of the GP posterior mean. This conclusion
applies to the other test beds studied in the paper and reported in  \Cref{app:TI_additional_results}, establishing a compelling use case for \ac{name}.


    



\section{Discussion}
\label{sec: discuss}

Our contribution was motivated by the strong regularity properties of tempering, and we used this observation to increase the accuracy of posterior expectations computed using tempered \ac{smc}. Our empirical results demonstrate that applying \ac{name} substantially improves estimation for multimodal distributions whose modes are inadequately represented by the raw \ac{smc} samples, while introducing no numerical degradation when the baseline performance is already highly precise, or when the estimated \ac{smc} variance is low.
Importantly, this additional accuracy was achieved at negligible computational overhead compared to running \ac{smc}.
This was made possible by conceiving \ac{name} as a post-processing method, but in principle the nodes $\{t_i\}_{i=1}^n$ could be chosen in a goal-driven manner to optimise the performance of \ac{name}; we leave this active node selection as an promising direction for future work.

Perhaps the main limitation of this work is that it considered only scalar posterior quantities of interest.
As future work it would be interesting not just to extrapolate expectations, but probability distributions themselves, making use of the individual samples generated by \ac{smc}.
The main difficulty with this approach, as we see it, is that it would require additional mathematical analysis to generalise the concept of analyticity to functions that are distribution-valued.

Finally, we note that there are a plethora of other computational techniques that exploit tempering, and we expect that in many of these cases some form of extrapolation can also be performed. In this work, we explored applying \ac{name} to \ac{it} estimators obtained from \ac{smc}, for which the baseline Gaussian likelihood model is quite heuristic. 
More broadly, the design of effective transformations from one distribution to another is an active research topic, particularly in machine learning, and emerging alternatives to tempering such as convolutions \citep{song2019generative} and dilation \citep{chehab2024practical} may also confer sufficient regularity to enable expectations to be extrapolated.


\paragraph{Acknowledgments}

The authors are grateful for discussions with Ben Adcock, Fran\c{c}ois-Xavier Briol, Jon Cockayne, Toni Karvonen, Anna Korba, Francesca Romana Crucinio and Gareth Roberts.
ZS and CJO were supported by EPSRC (EP/W019590/1).
CJO was supported by the Leverhulme Trust (PLP-2023-004).
This research was supported by the Heilbronn Institute for Mathematical Research, through the UKRI/EPSRC Additional Funding Programme for Mathematical Sciences.
MX was supported by the China Scholarship Council. Computation was performed using \citet{CREATE2025} at King's College London, UK.

\bibliographystyle{abbrvnat}
\bibliography{bibliography}

\clearpage
\appendix

\section{Proofs}

This appendix contains full proofs for the theoretical results stated in the main text.
Preliminary technical lemmas are contained in \Cref{app: technical}.
The main result, \Cref{thm: main}, is proven in \Cref{app: proof main}, while \Cref{cor: sufficient} is proven in \Cref{app: proof corollary}. Finally \Cref{app: shortened proof} provides a more compact, but non-elementary, approach for characterizing the analyticity of tempered expectations. 
Note that measurability is implicitly assumed throughout.

\subsection{Technical Lemmas}
\label{app: technical}

This section contains several technical lemmas that will be called upon in the proof of \Cref{thm: main} and \Cref{cor: sufficient}.
The key technical idea is to represent higher-order derivatives of tempered expectations using a tool called \emph{lag polynomials} from the time-series literature \citep[][Chapter 2]{hamilton2020time}.
Lag polynomials are introduced in \Cref{subsubsec: lag poly}, (complex) power series in \Cref{app: power series}, and an isomorphism between the two in \Cref{app: isomo}.
This isomorphism enables arguments that are more natural and straight-forward in one setting to be transferred to the other setting.
To this end, we prove a technical lemma on complex analytic functions (\Cref{app: complex anal}) that will have consequences for lag polynomials via the isomorphism that we described.
Further, we require technical lemmas on the interchange of limits (\Cref{app: interchange}) and moment generating functions (\Cref{app: moment}).
To state these lemmas, several pieces of notation will be introduced as they are required.

\subsubsection{Lag Polynomials}
\label{subsubsec: lag poly}

Let $\ell_1 := \{ x \in \mathbb{R}^{\mathbb{N}_0} : \| x \|_1 := \sum_{n=0}^\infty |x_n| < \infty \}$ be the set of infinite sequences whose sum is absolutely convergent.
As a convention, for $x \in \ell_1$ define $x_{-1} = x_{-2} = \dots = 0$.
The \emph{lag operator} $\mathcal{L} : \ell_1 \rightarrow \ell_1$ acts on elements $x \in \ell_1$ via $[\mathcal{L}(x)]_i = x_{i-1}$ for each $i \in \mathbb{N}_0$.
Following the time-series literature, we can consider polynomials constructed using powers of the lag operator \citep[][Chapter 2]{hamilton2020time}.
The set of \emph{lag polynomials} absolutely convergent in a ball of radius $R \geq 0$ is denoted $\mathcal{R}_{1,R} = \{ h(\mathcal{L}) = \sum_{n=0}^\infty a_n \mathcal{L}^n : \sum_{n=0}^\infty |a_n R^n | < \infty \}$.
A lag polynomial $h(\mathcal{L}) \in \mathcal{R}_{1,R}$ acts on vectors $x \in \ell_1$ as 
$$
[h(\mathcal{L}) (x)]_i := \left[ \left( \sum_{n=0}^\infty a_n \mathcal{L}^n \right) (x) \right]_i = \sum_{n=0}^\infty a_n x_{i-n}
$$
and $h(\mathcal{L})(x) \in \ell_1$ follows, provided that $R \geq 1$, from the following \Cref{lem: discrete conv}:

\begin{lemma}[Absolute convergence of Cauchy product] \label{lem: discrete conv}
    Let $x,y \in \ell_1$.
    Then $z \in \mathbb{R}^{\mathbb{N}_0}$ with $z_n := \sum_{i=0}^n x_i y_{n-i}$ satisfies $z \in \ell_1$.
\end{lemma}
\begin{proof}
Since $\|y\|_1 < \infty$ then
$$
\sum_{n = 0}^\infty \sum_{k = 0}^\infty  | x_n y_{k-n} | 
= \sum_{n = 0}^\infty |x_n| \sum_{k = 0}^\infty  | y_{k-n} |
= \| x \|_1 \| y \|_1 < \infty
$$
and we are allowed to rearrange the order of summation to conclude that
$$
\sum_{k = 0}^\infty \sum_{n = 0}^\infty  | x_n y_{k-n} |  = \sum_{n = 0}^\infty \sum_{k = 0}^\infty  |x_n y_{k-n} | 
$$
and thus
$$
\|z\|_1 = \sum_{k = 0}^\infty \left| \sum_{n = 0}^\infty x_n y_{k-n} \right| \leq \sum_{k = 0}^\infty \sum_{n = 0}^\infty | x_n y_{k-n} |  < \infty 
$$
so that $z \in \ell_1$, as claimed.
\end{proof}

\begin{lemma}[Lag polynomials as a ring] \label{lem: lag poly ring}
    The set $\mathcal{R}_{1,R}$ is a ring when equipped with addition $(\sum_{n=0}^\infty a_n \mathcal{L}^n) + (\sum_{n=0}^\infty b_n \mathcal{L}^n) = \sum_{n=0}^\infty (a_n + b_n) \mathcal{L}^n$ and multiplication $(\sum_{n=0}^\infty a_n \mathcal{L}^n) \cdot (\sum_{n=0}^\infty b_n \mathcal{L}^n) = \sum_{n=0}^\infty c_n \mathcal{L}^n$ with $c_n = \sum_{i=0}^n a_i b_{n-i}$.
\end{lemma}
\begin{proof}
The properties of a ring are trivially verified once we are satisfied that the addition and multiplication operations are well-defined.
Further, it is trivial that addition is well-defined, so the remaining task is to establish that multiplication is well-defined.
To this end, notice that
$$
c_n R^n = \sum_{i = 0}^n (a_i R^i) (b_{n-i} R^{n-i})
$$
and we are interested in the absolute convergence of $\sum_{n=0}^\infty c_n R^n $, to deduce whether or not this series is an element of $\mathcal{R}_{1,R}$.
This series has the form of a Cauchy product of the series $\sum_{n=0}^\infty a_n R^n$ and $\sum_{n=0}^\infty b_n R^n$.
\Cref{lem: discrete conv} shows that the Cauchy product of two absolutely convergent series is absolutely convergent, hence we establish that $\sum_{n=0}^\infty |c_n R^n| < \infty$ and thus we have closure under multiplication.
\end{proof}

\subsubsection{Complex Power Series}
\label{app: power series}

The set of complex power series that are absolutely convergent in a ball of radius $R \geq 0$ is denoted $\mathcal{R}_{2,R} := \{ \sum_{n=0}^\infty a_n z^n , \sum_{n=0}^\infty |a_n R^n| < \infty \}$. 

\begin{lemma}[Complex power series as a ring]
The set $\mathcal{R}_{2,R}$ is a ring when equipped with addition $(\sum_{n=0}^\infty a_n z^n) + (\sum_{n=0}^\infty b_n z^n) = \sum_{n=0}^\infty (a_n + b_n) z^n$ and multiplication $(\sum_{n=0}^\infty a_n z^n) \cdot (\sum_{n=0}^\infty b_n z^n) = \sum_{n=0}^\infty c_n z^n$ with $c_n = \sum_{i=0}^n a_i b_{n-i}$.
\end{lemma}
\begin{proof}
Entirely analogous to the proof of \Cref{lem: lag poly ring}.
\end{proof}

\subsubsection{Ring Isomorphism}
\label{app: isomo}

To transfer arguments for complex analytic functions into properties of lag polynomials, we use the natural isomorphism between the rings $\mathcal{R}_{1,R}$ and $\mathcal{R}_{2,R}$.
The proof of the following result is trivial:

\begin{lemma}[Ring isomorphism] \label{lem: iso}
    The map $\iota : \mathcal{R}_{1,R} \rightarrow \mathcal{R}_{2,R}$ which sends $\sum_{n=0}^\infty a_n \mathcal{L}^n$ to $\sum_{n=0}^\infty a_n z^n$ is an isomorphism of the rings $\mathcal{R}_{1,R}$ and $\mathcal{R}_{2,R}$.
\end{lemma}

\noindent The main technical motivation for considering complex power series instead of lag polynomials is that we will require a multiplicative inverse for a lag polynomial in the proof of \Cref{thm: main}, but deducing the existence of a well-defined multiplicative inverse to a lag polynomial appears somewhat difficult.
In contrast, for complex analytic functions, such arguments are quite natural.
Indeed, to identify a multiplicative inverse to a lag polynomial $h(\mathcal{L})$, from the ring isomorphism we can find a multiplicative inverse to the equivalent complex power series $h(z)$, say $h^{-1}(z)$, and deduce that the equivalent lag polynomial $h^{-1}(\mathcal{L})$ is a multiplicative inverse to $h(\mathcal{L})$.

\subsubsection{Complex Analytic Functions}
\label{app: complex anal}

Let $B_R(0) := \{ z \in \mathbb{C} : |z| < R \}$ denote the open ball of radius $R \geq 0$ centred at the origin in $\mathbb{C}$.
A complex power series $\sum_{n=0}^\infty a_n z^n$ from $\mathcal{R}_{2,R}$ defines a complex analytic function $h : B_R(0) \rightarrow \mathbb{C}$ via $h(z) = \sum_{n=0}^\infty a_n z^n$.

\begin{lemma}[Inversion of complex analytic functions] \label{lem: inversion}
Let $h$ be complex analytic and non-zero on $B_R(0)$ for some $R>1$.
Then $h(z)^{-1} = \sum_{n = 0}^\infty \psi_n z^n$ exists and is complex analytic on $B_R(0)$, for some coefficients $\psi_n$ with $\| \psi \|_1 = \sum_{n = 0}^\infty | \psi_n | < \infty$.
\end{lemma}
\begin{proof}
Since the complex analytic function $h : B_r(0) \rightarrow \mathbb{C}$ is non-zero on the open set $B_R(0)$, its reciprocal is well-defined and analytic on that same set.
Thus, since $0 \in B_R(0)$ and $h$ is complex analytic, we can write $h(z)^{-1} = \sum_{n=0}^\infty \psi_n z^n$ for some coefficients $\psi_n$ and all $|z| \leq R$.
(Recall that the power series of a complex analytic function has radius of convergence equal to the distance between the origin of the Taylor series and the edge of the domain on which it is complex analytic; in this case $R$.)
Since $h^{-1}$ is well-defined at some $z$ with $r := |z| > 1$, the terms in the series $h^{-1}(z) = \sum_{n=0}^\infty \psi_n z^n$ must satisfy $|\psi_n z^n| = |\psi_n| r^n \rightarrow 0 $ whence $\|\psi\|_1 < \infty$.
\end{proof}

\subsubsection{Interchange of Limits}
\label{app: interchange}

To calculate the derivative of a tempered expectation it will be necessary to commute the partial derivative $\partial_t$ with the expectation, and this interchange must be justified.
The following conditions are sufficient for the interchange of derivative and integral:

\begin{lemma}[Interchange of derivative and integral] \label{lem: interchange}
    Consider a collection of integrable functions $h_t : \mathbb{R}^d \rightarrow \mathbb{R}$ indexed by $t \in [0,1]$, such that the partial derivatives $\partial_t h_t : \mathbb{R}^d \rightarrow \mathbb{R}$ exists for almost all $x \in \mathbb{R}^d$, and such that $\sup_{t \in [0,1]} | \partial_t h_t(x) | \leq b(x)$ for some integrable function $b$ on $\mathbb{R}^d$.
    Then $\partial_t \int h_t(x) \; \mathrm{d}x = \int \partial_t h_t(x) \; \mathrm{d}x$.
\end{lemma}
\begin{proof}
    See \citet{folland1999real}.
\end{proof}

\subsubsection{Moment Generating Functions}
\label{app: moment}

Finally, for the proof of \Cref{cor: sufficient}, a basic result about the existence of moment generating functions will be required:

\begin{lemma}[Existence of moment generating function] \label{lem: mgfs}
    Let $s > 0$ and let $X$ be a real-valued random variable with $\mathbb{E}[\exp\{s|X|\}] < \infty$.
    Then the moments $\mathbb{E}[|X|^k]$ exist for all $k \in \mathbb{N}$.
    Further, the moment generating function $m(s) := \mathbb{E}[\exp\{sX\}]$ exists and admits the power series
    $$
    m(s) = \sum_{k=0}^\infty \frac{\mathbb{E}[X^k]}{k!} s^k ,
    $$
    with this series being absolutely convergent.
\end{lemma}
\begin{proof}
    From the power series representation of the exponential function
    $$
    \mathbb{E}[\exp\{s|X|\}] = \mathbb{E} \left[ \sum_{k=0}^\infty \frac{|X|^k}{k!} s^k \right]
    = \sum_{k=0}^\infty \frac{\mathbb{E}[|X|^k]}{k!} s^k
    $$
    where the interchange of expectation and sum is justified by the monotone convergence theorem.
    From this it follows that $\mathbb{E}[|X|^k] < \infty$ for all $k \in \mathbb{N}$.
    Finally, the dominated convergence theorem gives the final part.
\end{proof}

\subsection{Proof of \Cref{thm: main}}
\label{app: proof main}

From \Cref{ass: stand} we have $L_{\sup} := \sup_{x \in \mathbb{R}^d} L(x) < \infty$.
Further, since the tempered distributions $p_t$ defined by \eqref{eq: tempered posterior} are invariant to multiplication of $L$ by an arbitrary positive constant, we can without loss of generality suppose that $L_{\sup} < 1$.
This will be assumed in the sequel.

\subsubsection{Sufficient Conditions for Differentiability}

The calculations that we wish to perform involve tempered moments, and it is necessary to first establish conditions under which such moments are well-defined:

\begin{lemma}[Existence of tempered moments] \label{lem: temp moment}
    \Cref{ass: stand} implies that the tempered densities $p_t$ are well-defined, and satisfy $\sup_{t \in [0,1]} p_t(\cdot) \leq C p_0(\cdot)$ for a finite constant $C \in (0,\infty)$.
    In particular, if $h : \mathbb{R}^d \rightarrow [0,\infty)$ satisfies $\mathbb{E}_0[h] < \infty$, then $\mathbb{E}_t[h] \leq C \mathbb{E}_0[h]$ for all $t \in [0,1]$.
\end{lemma}
\begin{proof}
    Since $L_{\sup} < 1$,
    \begin{align*}
        Z_{\inf} := \inf_{t \in (0,1)} \int p_0(x) L(x)^t \; \mathrm{d}x
        & \geq \int p_0(x) \inf_{t \in (0,1)} L(x)^t \; \mathrm{d}x \\
        & = \int p_0(x) \min\{ 1 , L(x) \} \; \mathrm{d}x 
        = \int p_0(x) L(x) \; \mathrm{d}x > 0 .
    \end{align*}
    It follows that
    \begin{align*}
        \sup_{t \in (0,1)} p_t(x)
        = \sup_{t \in (0,1)} \frac{p_0(x) L(x)^t}{Z_t} 
        & \leq \frac{ p_0(x) }{Z_{\inf}}
    \end{align*}
    and, letting $C := Z_{\inf}^{-1}$,
    \begin{align*}
        \sup_{t \in [0,1]} \mathbb{E}_t[h] = \sup_{t \in [0,1]} \int h(x) p_t(x) \; \mathrm{d}x
        \leq \int h(x) \sup_{t \in [0,1]} p_t(x) \; \mathrm{d}x
        \leq C \int h(x) p_0(x) \; \mathrm{d}x = C \mathbb{E}_0[h] ,
    \end{align*}
    as claimed.
\end{proof}

As a warm-up, we first consider in detail how to take the first derivative of a tempered \ac{pdf}.
The result will also be useful for the subsequent development.

\begin{lemma}[Derivative of tempered posterior pdf] \label{lem: diff pt}
Assume that $\mathbb{E}_0[|\ell|]$ exists.
Then $\partial_t p_t(x) = \{ \ell(x) - \mathbb{E}_t[\ell] \} p_t(x)$.
\end{lemma}
\begin{proof}
Let $h_t(x) = p_0(x) L(x)^t$.
Then $\partial_t h_t(x) = p_0(x) \ell(x) L(x)^t$, and
\begin{align*}
    0 \leq \sup_{t \in (0,1)} | \partial_t h_t(x) | = p_0(x) |\ell(x)| \max\{ 1 , L(x) \} & \leq p_0(x) |\ell(x)| + p_0(x) |\ell(x)| L(x) \\
    & = p_0(x) |\ell(x)| + Z_1 p_1(x) |\ell(x)| ,
\end{align*}
which is integrable over $\mathbb{R}^d$ since we assumed $\mathbb{E}_0[ | \ell | ]$ exists, and the existence of $\mathbb{E}_1[|\ell|]$ follows from \Cref{lem: temp moment}.
Thus we may apply \Cref{lem: interchange} to $h_t(x)$ to justify the interchange of derivative and integral,
$$
\partial_t Z_t = \partial_t \int p_0(x) L(x)^t \; \mathrm{d}x = \int p_0(x) \ell(x) L(x)^t \; \mathrm{d}x = \int \ell(x) p_t(x) \; \mathrm{d}x \times Z_t = \mathbb{E}_t[\ell] Z_t .
$$
From the quotient rule for differentiation,
\begin{align*}
\partial_t p_t(x) & = \frac{ Z_t \partial_t [p_0(t) L(x)^t] - [p_0(x) L(x)^t ] \partial_t Z_t }{ Z_t^2 } \\
& = \frac{p_0(t) \ell(x) L(x)^t}{Z_t} - \frac{p_0(x) L(x)^t \mathbb{E}_t[\ell]}{Z_t} 
=  \{ \ell(x) - \mathbb{E}_t[\ell] \} p_t(x) 
\end{align*}
as claimed.
\end{proof}

Armed with a formula for first derivative of the tempered \ac{pdf}, we next derive a recurrence relation for derivatives of the tempered expectations which will be key to the proof of \Cref{thm: main}:

\begin{lemma}[Recurrence relation for $g^{(k)}$] \label{lem: binom}
Assume that the moments $\mathbb{E}_0[|f \ell^n|]$ and $\mathbb{E}_0[|\ell^n|]$ exist for all $n \in \{0,1, \dots , k\}$ and a fixed $k \in \mathbb{N}$.
Then
$$
\mathbb{E}_t[f \ell^k] = \sum_{n=0}^k \binom{k}{n} g^{(k-n)}(t) \mathbb{E}_t[\ell^n] .
$$
\end{lemma}
\begin{proof}
The proof is by induction with base case $k=1$.
For the base case, let $h_t(x) = f(x) p_t(x)$.
From \Cref {lem: diff pt}, $\partial_t h_t(x) = f(x) \{ \ell(x) - \mathbb{E}_t[\ell] \} p_t(x)$, and from \Cref{lem: temp moment},
\begin{align*}
    0 \leq \sup_{t \in [0,1]} |\partial_t h_t(x)| & \leq |f(x)| \left[ |\ell(x)| + \sup_{t \in [0,1]} |\mathbb{E}_t[\ell]| \right] p_t(x)
    \leq |f(x)| [ |\ell(x)| + C \mathbb{E}_0[\ell] ] C p_0(x)
\end{align*}
which is integrable since we assumed $\mathbb{E}_0[|f|]$, $\mathbb{E}_0[|\ell|]$, and $\mathbb{E}_0[|f \ell|]$ exist.
Thus from \Cref{lem: interchange} we may interchange derivative and integral to obtain
\begin{align}
\partial_t \mathbb{E}_t[f] = \partial_t \int f(x) p_t(x) \; \mathrm{d}x 
= \int f(x) \partial_t p_t(x) \; \mathrm{d}x 
& = \int f(x) \{ \ell(x) - \mathbb{E}_t[\ell] \} p_t(x) \; \mathrm{d}x  \nonumber \\
& = \mathbb{E}_t[f \ell] - \mathbb{E}_t[f] \mathbb{E}_t[\ell] = \mathbb{C}_t[f,\ell] . \label{eq: diff 1}
\end{align}
Recognising $g(t) = \mathbb{E}_t[f]$ and $g'(t) = \partial_t \mathbb{E}_t[f]$, we can rearrange \eqref{eq: diff 1} to obtain $\mathbb{E}_t[f \ell] = g'(t) + g(t) \mathbb{E}_t[\ell]$, so that the base case is established.
Now for the inductive step, with starting point
\begin{align}
\mathbb{E}_t[f \ell^{k-1}] = \sum_{n=0}^{k-1} \binom{k-1}{n} g^{(k-1-n)}(t) \mathbb{E}_t[\ell^n] \label{eq: induction}
\end{align}
for some $k \geq 1$.
Differentiating both sides, for which we can conveniently re-use the argument from \eqref{eq: diff 1} with $f \mapsto f \ell^{k-1}$ and $f \mapsto \ell^n$, justified by the corresponding assumptions that $\mathbb{E}_0[|f \ell^{k-1}|]$, $\mathbb{E}_0[|f \ell^k|]$, and $\mathbb{E}_0[|\ell^k|]$ exist, gives that
\begin{align*}
\mathbb{C}_t[f \ell^{k-1} , \ell] = \sum_{n=0}^{k-1} \binom{k-1}{n} \left\{ g^{(k-n)}(t) \mathbb{E}_t[\ell^n] + g^{(k-1-n)}(t) \mathbb{C}_t[\ell^n , \ell] \right\}
\end{align*}
which implies that, again using the inductive assumption \eqref{eq: induction},
\begin{align*}
\mathbb{E}_t[f \ell^k] & = \mathbb{E}_t[f \ell^{k-1}] \mathbb{E}_t[\ell] \\
& \qquad + \sum_{n = 0}^{k-1} \binom{k-1}{n} \left\{ g^{(k-n)}(t) \mathbb{E}_t[\ell^n] + g^{(k-1-n)}(t) \mathbb{E}_t[\ell^{n+1}] - g^{(k-1-n)}(t) \mathbb{E}_t[\ell^n] \mathbb{E}_t[\ell] \right\} \\
& = \mathbb{E}_t[\ell] \left\{ \sum_{n=0}^{k-1} \binom{k-1}{n} g^{(k-1-n)}(t) \mathbb{E}_t[\ell^n] \right\} \\
& \qquad + \sum_{n=0}^{k-1} \binom{k-1}{n} \left\{ g^{(k-n)}(t) \mathbb{E}_t[\ell^n] + g^{(k-1-n)}(t) \mathbb{E}_t[\ell^{n+1}] - g^{(k-1-n)}(t) \mathbb{E}_t[\ell^n] \mathbb{E}[\ell] \right\} .
\end{align*}
The coefficient of $g^{(k)}(t)$ in this expression is $1$ and the coefficient of $g^{(k-n)}(t)$ for $n > 0$ in this expression is 
\begin{align*}
& \mathbb{E}_t[\ell] \binom{k-1}{n-1} \mathbb{E}_t[\ell^{n-1}] + \binom{k-1}{n} \mathbb{E}_t[\ell^n] + \binom{k-1}{n-1} \mathbb{E}_t[\ell^n] - \binom{k-1}{n-1} \mathbb{E}_t[\ell^{n-1}] \mathbb{E}_t[\ell] \\
& = \left\{ \binom{k-1}{n-1} + \binom{k-1}{n} \right\} \mathbb{E}_t[\ell^n] 
= \binom{k}{n} \mathbb{E}_t[\ell^n] ,
\end{align*}
from which the inductive step is established.
\end{proof}

\subsubsection{Proof of \Cref{thm: main}}

Now we are ready to present the proof of \Cref{thm: main}:

\begin{proof}[Proof of \Cref{thm: main}]
Fix $t \in [0,1]$.
The preconditions of \Cref{lem: binom} are satisfied for moments up to order $k$, and thus
\begin{align*}
\frac{g^{(k)}(t)}{k!} & = \frac{\mathbb{E}_t[f \ell^k]}{k!} - \sum_{n=1}^k \frac{\mathbb{E}_t[\ell^n]}{n!}  \frac{g^{(k-n)}(t)}{(k-n)!} 
= \frac{\mathbb{E}_t[f \ell^k]}{k!} - \sum_{n=0}^{k-1} \frac{\mathbb{E}_t[\ell^{k-n}]}{(k-n)!}  \frac{g^{(n)}(t)}{n!} ,
\end{align*}
which shows that $g^{(k)}$ is well-defined.
For the second part of the theorem, we use \eqref{eq: sub exp} and \Cref{lem: mgfs} applied to $\ell$ to deduce both the existence of the moments $\mathbb{E}_t[|\ell|^k]$ for all $k \in \mathbb{N}$, and that the power series 
\begin{align}
    m_2(z) := \sum_{k=0}^\infty \frac{\mathbb{E}_t[\ell^k]}{k!} z^k 
    \label{eq: series initial}
\end{align}
is absolutely convergent for all $|z| \leq 1 + \epsilon$, i.e. $m_2 \in \mathcal{R}_{2,1 + \epsilon}$.
Since moments of all orders exist, we can leverage \Cref{lem: binom} for arbitrary order $k$, enabling us to cast $x_k := g^{(k)}(t)/k!$ as the solution of an infinite order autoregressive process
\begin{align*}
x_k = b_k + a_1 x_{k-1} + \dots + a_k x_0 , \qquad a_n := - \frac{ \mathbb{E}_t[\ell^n] }{ n! } , \qquad b_n := \frac{ \mathbb{E}_t[f \ell^n] }{ n! } .
\end{align*}
Recalling the lag operator $\mathcal{L}$ from \Cref{subsubsec: lag poly}, we can write this autoregressive process as $x_k = b_k + (a_1 \mathcal{L} + \dots + a_k \mathcal{L}^k) x_k$, so that in terms of a lag polynomial, 
\begin{align*}
m_1(\mathcal{L}) x = b, \qquad m_1(\mathcal{L}) := I - \sum_{n=1}^\infty a_n \mathcal{L}^n = \sum_{n=0}^\infty \frac{\mathbb{E}_t[\ell^n]}{n!} \mathcal{L}^n 
\end{align*}
where $x = (x_0,x_1,\dots)$ and $b = (b_0,b_1,\dots)$.
To see that this series is well-defined as an element of $\mathcal{R}_{1,1 + \epsilon}$, we observe that the equivalent complex power series $m_2$ in \eqref{eq: series initial} satisfies $m_2 \in \mathcal{R}_{2,1 + \epsilon}$ and use the ring isomorphism in \Cref{lem: iso}.
Further, from \Cref{lem: inversion}, since $m_2$ exists on $B_{1 + \epsilon}(0)$ and does not have complex roots (the complex exponential has no roots), we can write
$$
m_2(z)^{-1} = \sum_{n = 0}^\infty \psi_n z^n, \qquad \| \psi \|_1 = \sum_{n = 0}^\infty |\psi_n| < \infty
$$
so that $m_2^{-1} \in \mathcal{R}_{2,1+\epsilon}$, and therefore $m_2^{-1} \in \mathcal{R}_{2,1}$, where radius of convergence 1 is of interest because tempered expectations $g$ are supported on $[0,1]$. 
Returning to the lag polynomial domain using \Cref{lem: iso}, we have shown that $m_1^{-1} \in \mathcal{R}_{1,1}$ with
$$
m_1(\mathcal{L})^{-1} = \sum_{n = 0}^\infty \psi_n \mathcal{L}^n, \qquad
x_k = \sum_{n = 0}^\infty \psi_n b_{k - n} .
$$
If in addition $\| b \|_1 < \infty$, as requested in \Cref{eq: f assum}, then $\| x \|_1 < \infty$ by \Cref{lem: discrete conv}.
It follows that $g$ is analytic on $[0,1]$ with the convergent series expansion
$$
g(s) = \sum_{k=0}^\infty \frac{g^{(k)}(t)}{k!} (s-t)^k , \qquad \sum_{k=0}^\infty \left| \frac{g^{(k)}(t)}{k!} (s-t)^k \right| \leq \sum_{k=0}^\infty \left| \frac{g^{(k)}(t)}{k!} \right| = \| x \|_1 < \infty 
$$
holding for $s \in [0,1]$, as claimed.
\end{proof}

\subsection{Proof of \Cref{cor: sufficient}}
\label{app: proof corollary}

This section is dedicated to the proof of \Cref{cor: sufficient}.
Before presenting this argument, two preliminary lemmas are required:

\begin{lemma}[Tail condition implies exponential moment] \label{lem: finite info}
    If the informative prior condition is satisfied, then for some $\epsilon > 0$ we have $\mathbb{E}_1[\exp\{(1 + \epsilon)|\ell|\}] < \infty$.
\end{lemma}
\begin{proof}
    Since $L_{\sup} < 1$, it follows that $|\ell| = - \ell$, and
\begin{align*}
    \mathbb{E}_t[\exp\{(1 + \epsilon) |\ell| \}] =  \mathbb{E}_t[\exp\{-(1 + \epsilon) \ell \}] 
    = \mathbb{E}_t[L^{-(1 + \epsilon)}]
    =  \frac{1}{Z_t} \int p_0(x) L(x)^{t - 1 - \epsilon} \; \mathrm{d}x.
\end{align*}
Taking $t = 1$, our finite information hypothesis ensures the final integral exists, and completes the proof.
\end{proof}

To state the next lemma, let $f_+(x) := \max\{f(x),0\}$ and $f_-(x) := \min\{f(x),0\}$, so that $f(x) = f_+(x) + f_-(x)$.

\begin{lemma}[Log-likelihood bounded growth constraint] \label{lem: f growth}
    Suppose that for some $\epsilon > 0$ we have $\mathbb{E}_1[\exp\{(1 + \epsilon)|\ell|\}] < \infty$.
    If in addition $|f| \leq C |\ell|^m$ for some $C \in (0,\infty)$ and $m \in \mathbb{N}$ then \eqref{eq: f assum} is satisfied.
\end{lemma}
\begin{proof}
Since $L_{\sup} < 1$, also $\ell \leq \ell_{\sup} < 0$.  
Our assumption implies that $|f_+ \ell^{-m}| \leq C$ and $|f_- \ell^{-m}| \leq C$.
Since $f_+ \ell^{-m}$ and $\ell^{k-m}$ have constant sign, $|\mathbb{E}_t[(f_+ \ell^{-m}) (\ell^{k+m})]| \leq C |\mathbb{E}_t[\ell^{k+m}]|$ holds for all $t \in [0,1]$ and $k \in \mathbb{N}_0$, with an analogous bound holding for $f_-$ as well.
This leads to a bound
\begin{align}
|\mathbb{E}_t[f \ell^k]| & \leq |\mathbb{E}_t[f_+ \ell^k]| + |\mathbb{E}_t[f_- \ell^k]| \nonumber \\
& = |\mathbb{E}_t[f_+ \ell^{-m} \ell^{k+m}]| + |\mathbb{E}_t[f_- \ell^{-m} \ell^{k+m}]| 
\leq 2 C |\mathbb{E}_t[\ell^{k+m}]| . \label{eq: flk bound}
\end{align}
Now, let $K \in \mathbb{N}$ be large enough that $m \log (k + m) / (k + m) \leq \log( 1 + \epsilon)$ for all $k > K$.
Then, for all $k > K$, $(k + m)^m \leq (1 + \epsilon)^{k+m}$ and
$$
\frac{1}{k!} \leq \frac{(1 + \epsilon)^{k+m}}{(k+m)!} .
$$
Then from \eqref{eq: flk bound} with $t = 1$,
\begin{align*}
\frac{1}{2C} \sum_{k=0}^\infty \left| \frac{\mathbb{E}_1[f \ell^k]}{k!} \right|
\leq \sum_{k=0}^\infty \left| \frac{\mathbb{E}_1[\ell^{k+m}]}{k!} \right|
& = \sum_{k=0}^K \left| \frac{\mathbb{E}_1[\ell^{k+m}]}{k!} \right| + \sum_{k=K+1}^\infty \left| \frac{\mathbb{E}_1[\ell^{k+m}]}{k!} \right| \\
& \leq \sum_{k=0}^K \left| \frac{\mathbb{E}_1[\ell^{k+m}]}{k!} \right| +
\sum_{k=K + 1}^\infty \left| \frac{\mathbb{E}_1[\ell^{k+m}]}{(k+m)!} \right| (1 + \epsilon)^{k+m} < \infty ,
\end{align*}
where the finiteness of the final series follows from $\mathbb{E}_1[e^{(1 + \epsilon)|\ell|}] < \infty$ and \Cref{lem: mgfs}.
Thus \eqref{eq: f assum} is satisfied.
\end{proof}

\begin{proof}[Proof of \Cref{cor: sufficient}]
From assumption we can express $f = f_1 + f_2$ where $|f_1| \leq C_1 |\ell|^0$ and $f_2 \leq C_2 |\ell|^m$.
Then $g(t) = \mathbb{E}_t[f] = \mathbb{E}_t[f_1] + \mathbb{E}_t[f_2]$, and since the sum of analytic functions is analytic it suffices to verify \eqref{eq: f assum} and \eqref{eq: sub exp} from \Cref{thm: main} for both $f_1$ and $f_2$.
In fact, we will verify these conditions for any function $\tilde{f}$ with $|\tilde{f}| \leq \tilde{C} |\ell|^{\tilde{m}}$ for some $\tilde{C} \in (0,\infty)$ and some $\tilde{m} \in \mathbb{N}_0$.
To this end, first we use \Cref{lem: finite info} to immediately establish \eqref{eq: sub exp} with $t = 1$.
Then we can establish \eqref{eq: f assum} using \Cref{lem: f growth}.
\end{proof}

\subsection{A Non-Elementary Argument}
\label{app: shortened proof}

The proofs that we present in \Cref{app: technical,app: proof main,app: proof corollary} are elementary and self-contained.
In particular, the explicit calculation of the derivatives enabled us to derive a computable formula for $g'(t)$ that was estimable from \ac{smc} output in \Cref{subsec: extrap as reg}.
On the other hand, if one simply wanted to deduce that $g$ was analytic, then the powerful result that complex differentiable functions are analytic can be used, as shown in \Cref{thm: alternative}.
As in \Cref{app: proof main}, we without loss of generality assume that $L_{\sup} < 1$.

\begin{theorem}[Regularity of tempered expectations II]
\label{thm: alternative}
    Assume there exists $\epsilon > 0$ such that
\begin{align}
    \int \max\{1,|f(x)|\} p_0(x) L(x)^{-\epsilon} \; \mathrm{d}x & < \infty . \label{eq: assumed moment}
\end{align}
Then $g$ is analytic on $[0,1]$.
\end{theorem}
\begin{proof}
Introduce the notation
\begin{align*}
    h_f(t) := \int f(x) p_0(x) L(x)^t \; \mathrm{d}x ,
\end{align*}
so that from \eqref{eq: tempered posterior} we may write
\begin{align*}
    g(t) = \int f(x) \frac{p_0(x) L(x)^t}{Z_t} \; \mathrm{d}x = \frac{h_f(t)}{h_\iota(t)}
\end{align*}
where $\iota$ is the identity function on $\mathbb{R}^d$.
\Cref{ass: stand} implies that $h_\iota \in (0,\infty)$ and thus to deduce $g$ is analytic on $[0,1]$ it is sufficient to show that the functions $h_f$ and $h_\iota$ are both analytic on $[0,1]$.
For simplicity we present the argument for $h_f$ being analytic, with the argument for $h_\iota$ being a special case of this general argument.
To do this, consider the complex extension
\begin{align*}
    h_f : B_{1/2}(\nicefrac{1}{2}) & \rightarrow \mathbb{C} \\
    z & \mapsto \int f(x) p_0(x) L(x)^z \; \mathrm{d}x.
\end{align*}
of $h_f$ to the ball of radius $\nicefrac{1}{2}$ centred at $\nicefrac{1}{2} \in \mathbb{C}$.
It suffices to shown that $h_f$ is complex differentiable, and hence analytic, on $B_{1/2}(\nicefrac{1}{2})$.

Fix $z \in B_{1/2}(\nicefrac{1}{2})$, noting that $|L(x)^z| = L(x)^{\mathrm{Re}(z)} \leq L(x)^1 \leq L_{\sup}$, since $z \in B_{1/2}(\nicefrac{1}{2})$.
Then, from the dominated convergence theorem,
\begin{align*}
    \frac{h_f(z + \delta) - h_f(z)}{\delta} = \int f(x) p_0(x) L(x)^z \left[ \frac{L(x)^\delta - 1}{\delta} \right]  \; \mathrm{d}x \rightarrow \int f(x) p_0(x) L(x)^z \ell(x)  \; \mathrm{d}x
\end{align*}
as $\delta \rightarrow 0$ in $\mathbb{C}$.
Indeed, we can appeal to the dominated convergence theorem since $\lim_{\delta \rightarrow 0} (L(x)^\delta - 1) / \delta = \partial_\delta (L(x)^\delta) |_{\delta = 0} = (L(x)^\delta \log L(x))|_{\delta = 0} = \ell(x)$ establishes pointwise convergence of the integrand, while the inequality
\begin{align*}
    \left| \frac{L(x)^\delta - 1}{\delta} \right| < \frac{L(x)^\epsilon + L(x)^{-\epsilon}}{\epsilon} , \qquad \forall \; |\delta| < \epsilon
\end{align*}
from Theorem 7.2 of \citet{barndorff2014information} implies 
\begin{align*}
    \left| f(x) p_0(x) L(x)^z \left[ \frac{L(x)^\delta - 1}{\delta} \right] \right| & <  |f(x)| p_0(x) |L(x)^z| \left| \frac{ L(x)^\delta - 1 }{\delta} \right|  \\
    & \leq |f(x)| p_0(x) L_{\sup} \frac{ L(x)^{\epsilon} + L(x)^{- \epsilon} }{\epsilon} , 
\end{align*}
yielding a uniform upper bound which is integrable in $x$ due to \eqref{eq: assumed moment}.
An analogous argument holds for $h_\iota$, and the claim is established.
\end{proof}

\section{Sequential Monte Carlo}
\label{app: SMC}

Historically, SMC algorithms were developed to infer the distribution of hidden states in state space models, where observations arrive sequentially. 
The application of SMC methods was then generalized to encompass sampling from a static distribution $p  = p_{t_n}$, achieved through tempering and sampling from the sequence of intermediate distributions $(p_{t_i})_{i=0}^{n-1}$, where $0=t_0 < t_1 ,\ldots <t_n = 1$ 
\citep{chopin2002sequential}.
 In \Cref{subsection:tempered_smc} we review details of traditional \ac{smc} for sampling from static distributions, including the adaptive selection of the temperature ladder. In \Cref{subsection:WF_SMC} we then focus on waste-free SMC, and explain how this framework allows to estimate the tempered expectations and their asymptotic variance which were used in the experiments.  

\subsection{Tempered SMC}\label{subsection:tempered_smc}

Assume initially that the temperature ladder $t_i,\, i=0,\ldots,n$ is given. Sequential Importance Sampling (SIS) offers a naive way to produce samples from $p_{t_i}$. In SIS, initially $N$ particles $\{x_j^{(0)}\}_{j=1}^N$ are sampled from the distribution $p_0 = p_{t_0}$, and an equal  unnormalized weight
$\tilde{w}_j^{(0)} = 1$ is assigned 
to each particle.
For each temperature $i\geq 1$
the  potential (or weight) function
\begin{align}
G_{i}(x)=\frac{L(x)^{t_i}}{L(x)^{t_{i-1}}} \label{eq:pot_funct}
\end{align}
is used to compute the (unnormalized) importance weights 
$
\tilde{w}_j^{(i)} = \tilde{w}_j^{(i-1)} \times G_{i}( x_j^{(i)}),
$
of 
the $N$ particles $x_j^{(i)} = x_j^{(0)}$, $j=1,\ldots,N$, 
which are then normalized to 
\begin{align}
{w}_j^{(i)} = {\tilde{w}_j^{({i})}}/{\sum_{j=1}^N \tilde{w}_j^{({i})}}.\label{eq:weigh_norm}
\end{align}
The empirical approximation of $p_{t_i}$ based on SIS is then of the form \eqref{eq:empirical_approximation_pti}.
In SIS, progressing over the temperature ladder, the importance weights of most particles become extremely small or negligible, while only a few particles retain significant weights.

Tempered SMC mitigates this weight degeneracy by $(i)$ resampling and $(ii)$ applying   
$P$ Markov transition kernels (here assumed to be MCMC kernels) to each particle, at each temperature, before computing the weights. 
Resampling provides a new set of particles by 
drawing, for each particle, the `ancestry variable'  $a_{j}^{(i-1)} 
\in\{1,\ldots, N\},$
for example through multinomial resampling based on the normalized weights 
\begin{align*}
a_{j}^{(i-1)} &\sim \text{Categorical}(N, w_1^{(i-1)}, \ldots, w_N^{(i-1)}).
\end{align*}
Then the resampled particles are defined as  $\tilde{x}_{j,1}^{(i-1)} = {x}_{a_j^{(i-1)}}^{(i-1)}$, that is they are copies of their `parent' (or `ancestor') particles, as indexed by the ancestry variable.  To summarize these steps, we use the shorthand notation $$\tilde{x}_{j,1}^{(i-1)}\sim\text{resample}(a_j^{(i-1)},w_{1:N}^{(i-1)}, x_{1:N}^{(i-1)}).$$
Resampling  effectively replicates particles with higher weights and eliminates those with lower weights, thereby 
reallocating computational resources to regions of high 
probability under $p_{t_{i-1}}$. This step alone might not suffice to eliminate particle degeneracy, but it proves effective when combined with additional MCMC moves, often called `particle rejuvenation'. 
Let \(M_t(x, \mathrm{d}x_t) \) denote a Markov kernel, that is a transition probability density from the state \( x\) to the state \( x_t \), and let \( M_t \) be designed to leave the distribution \( p_{t} \) invariant. 
Let also  \( M_t^P \) denote the composition of \( M_t \) applied \( P \) times.
Then, for each temperature, after resampling, tempered SMC applies $P$ Markov transition kernels that leave $p_{t_{i-1}}$ invariant to each of the resampled particles. After $P$ Markov transition steps, this produces the rejuvenated
set of particles $$\tilde{x}_{j,P}^{(i-1)} \sim M_{t_{i-1}}^P(\tilde{x}_{j,1}^{(i-1)},\mathrm{d}x_{t_{i-1}}).$$ 
with approximate distribution $p_{t_{i-1}}$. 
Due to resampling and rejuvenation, the re-weighting, or importance sampling, step in SMC requires only evaluating the potential function \eqref{eq:pot_funct} at the current temperature and not multiplying with the previous weights, so that the unnormalized weights in tempered SMC are 
\begin{align}
\tilde{w}_j^{(i)} = G_{i}( x_j^{(i)}).
\label{eq:smc_weights}
\end{align}
After weight normalization \eqref{eq:weigh_norm}, the weighted particles effectively give an empirical approximation to~$p_{t_i}$~\eqref{eq:empirical_approximation_pti}, such that the derived Monte Carlo estimators are asymptotically normal~\eqref{eq:AN_SMC} 
and the asymptotic variance depends on the whole particle genealogy. Asymptotic variance estimators are available in the standard SMC literature, but such methods degenerate if the set of ancestors collapses to one particle only. 

The specification of the temperature ladder, the Markov kernel, the number of particles, and the number of Markov iterations influence the overall
performance of SMC methods. In practice, these quantities can be selected adaptively, based on the weighted empirical measure available at each iteration. For example, the MCMC kernel and number of steps can be tuned using results from literature on adaptive MCMC methods and convergence diagnostics for MCMC. 
If not pre-specified, it is common practice to set the temperature ladder at run time as follows.

\subsubsection{Adaptive Selection of the Temperature Ladder}
\label{app: t selection}
The temperature  is initialized at $t_0=0$. 
For $i\geq 1$, given the current temperature $t_{i-1}$, in the reweighting step~\eqref{eq:smc_weights}
the unnormalized weights  are a function of  the subsequent temperature~$t_i$.  
This is set so that the average effective sample size (ESS) of the current particles
does not decrease below a predetermined minimum threshold $\text{ESS}_{\text{min}}\in (0,1)$ \cite{chopin2020introduction,dai2022invitation}. In practice,  the equation
\begin{align}
\frac{1}{N}\text{ESS}(t_i) := \frac{1}{N} \frac{{\left(\sum_{j=1}^N \tilde{w}_j^{(i)}\right)^2} }{\sum_{j=1}^N (\tilde{w}_j^{(i)})^2} 
= \text{ESS}_{\text{min}}. \label{eq:adaptive_temp_selection}
\end{align}
is solved for $t_i\in(t_{i-1},1]$. 
The effective sample size $\text{ESS}(t_i) \in [1,N]$ indicates the number of  particles that can be regarded as effectively independent samples from the target distribution~$p_{t_i}$. It serves as a measure of the quality of the weighted empirical distribution~\eqref{eq:empirical_approximation_pti}, and it is therefore reasonable to impose a lower bound on it.
Additionally, one can show that the average ESS is a sample approximation to $(1+\chi^2(p_{t_i}|p_{t_{i-1}}))^{-1},$
where 
$$\chi^2(p_{t_i}|p_{t_{i-1}}) 
:= \int\left(\frac{p_{t_i}(x)}{p_{t_{i-1}}(x)} - 1\right)^2 \mathrm{d}x,$$
is the $\chi^2$-divergence between the current and the following tempered distribution, bearing in mind that the latter is to be determined. Keeping the ESS above a certain  threshold is equivalent to finding the following temperature $t_i$ such that  $p_{t_i}$ is not too dissimilar from $p_{t_{i-1}}$, ensuring that the method performs well in practice, for a finite computing budget (number of particles and number of Markov steps). 
In practice, the temperature schedule often follows a geometric progression, with the spacing between successive temperatures increasing. This pattern arises because, at lower temperatures, the data exert a stronger influence when transitioning from the prior $p_0$ to the intermediate distribution $p_{t_i}$. As the temperature increases and $t_i$ approaches 1, the marginal impact of the data decreases, making it reasonable to use larger temperature steps.

\subsection{Waste-Free SMC}\label{subsection:WF_SMC}

\begin{algorithm}[t!]
\caption{Waste-free SMC}\label{alg:wf-smc}
\begin{algorithmic}[1]
  \Inputs{
    $M \geq 1$ \hfill \Comment{Number of resampled ancestors} \\
    $P \geq 2$ \hfill \Comment{Number of Markov steps per ancestor} \\  
    $\text{Adaptive}$ \hfill \Comment{Boolean flag for adaptive tempering} \\
    $T$ \hfill \Comment{Predefined temperature list (if $\text{Adaptive} = \textsc{false}$)} \\
    $M_t(x,\mathrm{d}x)$ \hfill \Comment{Markov transition kernel}
  }
  \Initialize{
  \strut
  $N = M \times P$ \hfill \Comment{Total number of particles} \\
    $i \gets 0$ \hfill \Comment{Initialize temperature step counter} \\
    $x_j^{(0)} \sim p_0, \quad w_j^{(0)} \gets 1/N, \quad j=1,\dots,N$ \hfill \Comment{Draw and initialize particle ensemble}
  }
  
  \While{$t_{i} < 1$}
    \State $i \gets i+1$
    \State $\tilde{x}_{m,1}^{(i-1)} \sim \text{resample}\left(a_m^{(i-1)}, w_{1:N}^{(i-1)}, x_{1:N}^{(i-1)}\right), \quad m=1,\dots,M$ \hfill \Comment{Resampling step} 

    \For{$p = 2$ \textbf{to} $P$}    
      \State $\tilde{x}_{m,p}^{(i-1)} \sim M_{t_{i-1}}\left(\tilde{x}_{m,p-1}^{(i-1)}, \mathrm{d}x\right), \quad m=1,\dots,M$ \hfill \Comment{Rejuvenation steps} 
    \EndFor

    \State $x_{1:N}^{(i)} \gets \tilde{x}_{1:M,1:P}^{(i-1)}$ \hfill \Comment{Collect updated joint particle locations}

    \If{$\text{Adaptive} = \textsc{true}$} 
      \State Find temperature $t_i$ by solving \eqref{eq:adaptive_temp_selection} \hfill \Comment{Adaptive selection}  
    \Else 
      \State $t_i \gets T[i]$ \hfill \Comment{Extract deterministic temperature}
    \EndIf
   
    \State $\tilde{w}_{j}^{(i)} \gets G_{i}\left(x_j^{(i)}\right), \quad j=1,\dots,N$ \hfill \Comment{Evaluate unnormalized weights}
    \State $w_j^{(i)} \gets \tilde{w}_j^{(i)} \,/\, \sum_{l=1}^N \tilde{w}_l^{(i)}, \quad j=1,\dots,N$ \hfill \Comment{Normalize weights}
    \State Approximate $p_{t_i}$ via \eqref{eq:empirical_approximation_pti} \hfill \Comment{Construct empirical distribution}
  \EndWhile
  
  \State \textbf{Output:} $x_{1:N}^{(i)}, w_{1:N}^{(i)}$ \hfill \Comment{Weighted particle approximation of $p_1$}
\end{algorithmic}
\end{algorithm}

In standard tempered SMC algorithms, only the final states of the $P$ Markov transition kernels are retained for propagation to the next iteration, and the intermediate samples are \textit{wastefully} discarded.  
\cite{dau2022waste} introduced waste-free SMC, aiming to improve the efficiency of standard SMC algorithms by maximizing the use of all generated samples. At the resampling step of each iteration, waste-free SMC resamples a subset of $M\ll N$ particles from the current particle set of size~$N$. Each resampled particle is then independently propagated through $P-1$ steps of a Markov transition kernel invariant to the current target distribution $p_{t_{i-1}}$, where  $P=N/M$. This process generates a set of $N$ new particles by retaining the ancestors and the Markov transition steps. After re-weighting, the new particles  serve for the empirical approximation of $p_{t_i}$ before being  re-sampled at the next iteration. \Cref{alg:wf-smc} summarizes the waste-free SMC sampler, with optional selection of the temperatures at run time. 

Besides not wasting the computation performed at the MCMC rejuvenation steps, waste-free SMC improves the approximation of $p_{t_i}$ compared to standard SMC, especially if the MCMC has slow mixing. 
The intuition for this is that, because $M\ll N$, in waste-free SMC particles with large weight are selected less often to be ancestors of new particles, compared to standard SMC.   
In fact, in standard SMC, an ancestor with a large weight will generate many nearly identical variables by the end of the rejuvenation step, which are then used to select ancestors in the next iteration. In contrast, waste-free SMC selects such ancestors less frequently from the start, and, even if rejuvenation yields similar samples, these are subsequently subjected to significant sub-sampling to form new ancestors.

The observation that waste-free SMC tends to improve the empirical approximation of the target compared to standard SMC is supported by the \textit{propagation of chaos theory} for SMC \cite[Chapter~8]{moral2004feynman}. 
According to this, the measure of dependency between particles introduced by resampling becomes negligible in the limit of an infinite number of particles $N\rightarrow \infty$. In practice, resampling  $M \ll N$ particles produces nearly independent samples from $p_{t_{i-1}}$, that are then rejuvenated behaving like $M$ independent Markov chains of length~$P$. 

In mathematical terms,  propagation of chaos is observed in the asymptotic variance derived by \cite{dau2022waste} for the central limit theorem \eqref{eq:AN_SMC} that holds for waste-free SMC.
For $P\rightarrow \infty$, with $M$ growing as $P^\alpha$, $\alpha \geq 0$ (thus the including the case of fixed $M$), when the  Markov kernel $M_t$ is uniformly ergodic, and the weight function $G_i$ is upper bounded, the asymptotic variance for every $p_{t_i}$, $i=0,\ldots, n$, is 
\begin{align}
    \sigma^2_i[f] = s^2_i\left[\bar{G}_{i}\times(f-g(t_i))\right],   \label{eq:asympt_var_full}
\end{align}
where: 
$\bar{G}_{i}:=G_{i}/\mathbb{E}_{t_i}[G_i]$;  
$g(t_i) = \mathbb{E}_{t_i}[f]$ is the tempered expectation of interest; and 
\begin{align}
    s^2_i[f] = 
    \begin{cases}
         \text{Var}(f(Y_0)), & \text{if $i=0$,}
        \\
        \text{Var}(f(Y_0)) + 2\sum_{p=1}^\infty \text{Cov}(f(Y_0), f(Y_p)), & \text{if $i\geq 1$,}
    \end{cases}
    \label{eq:asympt_var}
\end{align}
is the usual asymptotic variance, as $P\rightarrow\infty$, of $\sum_{i=1}^Pf(Y_p)/P$, of a Monte Carlo estimator based on $P$ samples from  
$(Y_p)_{p\geq 0}$, a single stationary Markov chain with transition kernel $M_{t_i}$ and stationary distribution $p_{t_i}$. 
Therefore, the $N = M \times P$ waste-free SMC samples can be viewed as $M$ independent Markov chains of length $P$, because the asymptotic variance~\eqref{eq:asympt_var_full} depends only on the current and previous temperatures $t_i$ and $t_{i-1}$, but not on the particles genealogy at the temperatures~$\{t_j\}_{j=0}^{i-2}$. This is in contrast to standard SMC, where dependency between particles at all the previous temperatures needs to be tracked to compute  the asymptotic variance. 

\subsubsection{Estimation of the Asymptotic Variance}
Given that the $N = M \times P $ waste-free SMC samples behave as $M$ independent Markov chains of length $P$, it is possible to estimate the asymptotic variance \eqref{eq:asympt_var}, and thus \eqref{eq:asympt_var_full}, using the MCMC literature.
Most of this literature is based on computing the sample covariances of order $q \in \{0,\ldots, P-1\}$, based on one  chain at temperature~$t_i$, so that   dealing with $M$ independent chains simply requires  additional averaging across chains
\begin{align}
\tilde{\gamma}_q^{(i)}:=
\frac{1}{MP}
\sum_{m=1}^M\sum_{p=1}^{P-q}
\left[f\left(\tilde{x}^{(i-1)}_{m,p}\right)-\tilde{g}(t_i)\right]
\left[f\left(\tilde{x}^{(i-1)}_{m,p+q}\right)-\tilde{g}(t_i)\right], \label{eq:sample_cov_wf_SMC}
\end{align}
where $\tilde{g}(t_i)= \sum_{m=1}^M\sum_{p=1}^{P} f\left(\tilde{x}^{(i-1)}_{m,p}\right)/(MP)$ is the overall empirical mean, based on equally weighting all the waste-free SMC samples $\tilde{x}^{(i-1)}_{1:M,1:P}$. 
Then several MCMC estimators of \eqref{eq:asympt_var}
are of the form 
$$\hat{s}_i^2[f] = \psi_P(\tilde{\gamma}_0^{(i)}, \ldots, \tilde{\gamma}_{P-1}^{(i)}),$$ where $\psi_P:\mathbb{R}^P\rightarrow \mathbb{R}^+$ is a function that maps the vector of $P$ sample covariances to the asymptotic variance estimator.  \cite{dau2022waste} advocate in favour of the initial monotone sequence estimator of \cite{geyer1992practical}, which estimates the asymptotic variance by summing sample covariances of order $q\in{0,\ldots,L}$, where $L\leq P-1$ is the last index for which $\tilde{\gamma}_q^{(i)}$ are positive and monotonically decreasing. In fact, this is known to be a property of the exact covariances of a stationary, irreducible, and reversible Markov chain. Given that the estimate \eqref{eq:sample_cov_wf_SMC} is ultimately affected by sampling noise, the variance estimator sums only the terms where the noise is not prevalent. 

Finally, based on \eqref{eq:asympt_var_full}, the asymptotic variance of $\hat{g}(t_i)$, the waste-free SMC estimator of the tempered expectation of interest \eqref{eq:AN_SMC}, can be estimated as 
\begin{align}
\hat\sigma_i^2[f]=\hat{s}_i^2[\tilde{G}_{i}\times(f-\hat{g}(t_i))], \label{eq:WF_SMC_AV_est}
\end{align}
where $$\tilde{G}_{i}:= \frac{{G}_{i}}{\frac{1}{N}\sum_{j=1}^N \tilde{w}_j^{(i)}},$$
and  the intractable quantities $g(t_i)$ and 
$\mathbb{E}_{t_i}[G_i]$
are replaced by their Monte Carlo estimates, based on the current set of weighted particles.

\subsubsection{Waste-Free \ac{smc} Implementation}
To better satisfy the propagation of chaos theory behind the waste-free SMC asymptotic analysis, \cite{dau2022waste} suggest choosing the input variables in favour of small $M$ and large $P$. The open-source software \texttt{Particles} \cite{particles} implements $M_t$ as a 1-fold adaptive Metropolis kernel, whereby the proposal is adapted using the total set of current particles. 
This default choice is guided by asymptotic reasoning on fixed-lag thinning in the MCMC literature and it provides a reversible Markov kernel, which supports the validity of the asymptotic variance estimator \eqref{eq:WF_SMC_AV_est}. 

\section{Computation for the \ac{name} Regression Model}
\label{ap: gp computation}

The aim of this appendix is to explain how the conditional \ac{gp}, based on the function value and gradient data described in \Cref{subsec: extrap as reg}, can be explicitly computed  \citep[Chapter 9]{rasmussen2006gaussian}, and to provide practical details for the optimization of the \ac{gp} prior hyperparameters. 
To simplify the notation, in what follows we define the prior mean function 
$$m_{{\theta}}(t) =\dfrac{\sum_{j=0}^{r} a_j t^j}{1 + \sum_{k=1}^{s} b_k t^k}, $$
with ${\theta} = (a_{0:r}, b_{1:s})$, and 
we introduce the linear operator
\begin{align*}
    (\mathcal{L} h)(t) & = [h(t_0) , \dots , h(t_n) , h'(t_0) , \dots , h'(t_n) ]^\top
\end{align*}
that acts on functions $h : [0,1] \rightarrow \mathbb{R}$.
For a bivariate function $h : [0,1] \times [0,1] \rightarrow \mathbb{R}$ let $\mathcal{L}_1h$ and $\mathcal{L}_2h$ denote, respectively, the action of $\mathcal{L}$ on the first and second argument.
In particular,
\begin{align*}
    (\mathcal{L}_1 k_\phi)(t) = \left[ \begin{array}{c} k_\phi(t_0,t) \\ \vdots \\ k_\phi(t_n,t) \\ \hline \partial_1 k_\phi(t_0,t) \\ \vdots \\ \partial_1 k_\phi(t_n,t)  \end{array} \right] , \qquad
    (\mathcal{L}_2 k_\phi)(t) = \left[ \begin{array}{c} k_\phi(t,t_0) \\ \vdots \\ k_\phi(t,t_n) \\ \hline \partial_2 k_\phi(t,t_0) \\ \vdots \\ \partial_2 k_\phi(t,t_n)  \end{array} \right]^\top,
\end{align*}
and
\begin{align*}
    \mathcal{L}_1 \mathcal{L}_2 k_\phi = 
\begin{pNiceArray}{ccc|ccc}
  k_\phi(t_0,t_0) & \cdots & k_\phi(t_0,t_n) & \partial_2 k_\phi(t_0,t_0) & \cdots & \partial_2 k_\phi(t_0,t_n) \\
  \vdots & \ddots & \vdots & \vdots & \ddots & \vdots \\
  k_\phi(t_n,t_0) & \cdots & k_\phi(t_n,t_n) & \partial_2 k_\phi(t_n,t_0) & \cdots & \partial_2 k_\phi(t_n,t_n) \\
  \hline
  \partial_1 k_\phi(t_0,t_0) & \cdots & \partial_1 k_\phi(t_0,t_n) & \partial_1 \partial_2 k_\phi(t_0,t_0) & \cdots & \partial_1 \partial_2 k_\phi(t_0,t_n) \\
  \vdots & \ddots & \vdots & \vdots & \ddots & \vdots \\
  \partial_1 k_\phi(t_n,t_0) & \cdots & \partial_1 k_\phi(t_n,t_n) & \partial_1 \partial_2 k_\phi(t_n,t_0) & \cdots & \partial_1 \partial_2 k_\phi(t_n,t_n) 
\end{pNiceArray}.
\end{align*}
In addition, let
\begin{align}
y := \left[ \begin{array}{c} \hat{g}(t_0) \\ \vdots \\ \hat{g}(t_n) \\ \hline \hat{g}'(t_0) \\ \vdots \\ \hat{g}'(t_n) \end{array} \right] , \qquad
    \Sigma := 
\begin{pNiceArray}{ccc|ccc}
  \hat{\sigma}_0^2 & &  & & & \\
  & \ddots &  &  &  & \\
   &  & \hat{\sigma}_n^2 &  & &  \\
  \hline
  &  & & \hat{\gamma}_0^2  & & \\
   &  & & & \ddots & \\
   & &  & & & \hat{\gamma}_n^2 
\end{pNiceArray}
\end{align}
denote our combined function value and gradient training data and its associated error covariance matrix.
The conditional process can then be expressed compactly as a \ac{gp} $g | y  \sim \mathcal{GP}(m_{\theta,\phi,n} , k_{\phi,n})$ with posterior mean and covariance
\begin{align}
    m_{\theta,\phi,n}(t) & := m_\theta(t) + [ \mathcal{L}_2 k_\phi(t) ] [ \mathcal{L}_1 \mathcal{L}_2 k_\phi + \Sigma ]^{-1} [ y - \mathcal{L} m_\theta ], \label{eq: post mean}  \\
    k_{\phi,n}(t,t') & := k_\phi(t,t') - [ \mathcal{L}_2 k_\phi(t) ] [\mathcal{L}_1 \mathcal{L}_2 k_\phi + \Sigma]^{-1} [\mathcal{L}_1 k_\phi(t')].  \label{eq: post var}
\end{align}

\paragraph{\ac{gp} Hyperparameter Optimization}
The log-marginal likelihood of the posterior \ac{gp} is given, up to an additive $(\theta,\phi)$-independent constant, by\begin{align}\log p(y \mid \theta , \phi) & \stackrel{+C}{=} - \frac{1}{2} \log \text{det} ( \mathcal{L}_1 \mathcal{L}_2 k_\phi + \Sigma ) - \frac{1}{2} (y - \mathcal{L} m_\theta)^\top ( \mathcal{L}_1 \mathcal{L}_2 k_\phi + \Sigma )^{-1} (y - \mathcal{L} m_\theta). \label{eq:log_marginal_lik}\end{align}The degrees $r$ and $s$, the parameters $\theta$ of the prior mean function, and the parameters $\phi = \{\lambda , \ell\}$ of the prior kernel function are jointly selected to maximize \eqref{eq:log_marginal_lik}. This is operationalized by enumerating the rational function orders $(r,s) \in \mathcal{R}$, where $\mathcal{R} = \{0,1,\ldots, k\}^2$. For each combination, \eqref{eq:log_marginal_lik} is optimized with respect to $\theta$ and $\phi$ via Newton's method, with the relevant gradients computed using automatic differentiation. The final degrees $(r,s)$ are chosen as those yielding the maximum value of \eqref{eq:log_marginal_lik} across all candidates. To promote numerical stability while maintaining a highly flexible prior mean function over the support $[0, 1]$, the maximum degrees of the numerator and denominator polynomials are constrained to $k=2$. Even with this degree constraint, numerical instability occurs if the rational prior mean function $m_{\theta}$ exhibits a pole within the domain $[0,1]$.
To prevent the denominator polynomial from vanishing on this interval, which would introduce singularities and render trajectory extrapolation ill-posed, we introduce a regularization penalty to the optimization framework. Specifically, for a denominator polynomial defined by $Q_{\theta}(t) = 1 + b_1 t + \dots + b_s t^s$, the penalty function $\mathcal{P}(\theta)$ is formulated as\begin{align}\label{eq:pole-penalty}\mathcal{P}(\theta) := \int_{\mathcal{T}} \frac{1}{Q_{\theta}(t)^2} , \mathrm{d}t, \qquad \mathcal{T} := [t_{\min}, t_{\max}],\end{align}where $t_{\min}=-0.1$ and $t_{\max}=1.1$ are fixed constants chosen to provide a safety buffer around the \ac{gp} training interval. If there exists a root $t_0 \in \mathcal{T}$ such that $Q_{\theta}(t_0)=0$, the integrand in \eqref{eq:pole-penalty} becomes non-integrable in any neighborhood of $t_0$, causing the penalty to diverge to infinity. Consequently, the parameters $\theta$ and $\phi$ are robustly estimated by solving the penalized optimization problem
\begin{align*}\max_{\theta,\phi}  \log p(y \mid \theta,\phi) - c \mathcal{P}(\theta) ,
\end{align*}
where the penalty weight controlling the enforcement strength is set to $c=10^{-11}$ across all experiments. The rational function coefficients are initialized by performing a weighted least-squares fit to the data, using weights inversely proportional to the estimated \ac{smc} variances:
\begin{align*}
\theta_{\text{init}}:=\arg\min_{\theta}  -\mathfrak{L}(g) - c \mathcal{P}(\theta) ,
\end{align*}
where the likelihood $\mathfrak{L}(g)$ 
is defined in \eqref{eq: heur log like}, and the penalty term is added to avoid poles in the denominator at initialization. The \ac{gp} prior kernel parameters $\phi$ are  initialized as follows:  the log-amplitude is set to the log-empirical variance of the residuals between the raw data and this initial trend line, $\sigma_{\text{res}}^2$, yielding the starting value $\lambda_{\text{init}} = \text{softplus}\big(\log(\sigma_{\text{res}}^2)\big) + 10^{-4}$. Finally, the initial length-scale parameter is fixed to $\ell_{\text{init}} = \text{softplus}(0) \approx 0.693$. Pseudocode detailing the steps of the \ac{gp} regression involved in \ac{name} smoothing is given in \Cref{alg:elate}. \ac{name} extrapolation is implemented analogously, by restricting the input data to a subset of initial tempered expectations and derivatives. Similarly, one can choose to condition the \ac{gp} prior solely on function values, discarding the derivatives, by adjusting the observation vector, likelihood, and corresponding \ac{gp} expressions accordingly.

\begin{algorithm}[t!]
\caption{ELATE smoothing (function value and gradient data)}\label{alg:elate}
\begin{algorithmic}[1]
  \Inputs{
    $\{\hat{g}(t_i)\}_{i=0}^n, \{\hat{\sigma}_i^2\}_{i=0}^n$ \hfill \Comment{Tempered expectations  and variances} \\
    $\{\hat{g}'(t_i)\}_{i=0}^n, \{\hat{\gamma}_i^2\}_{i=0}^n$ \hfill \Comment{Tempered derivatives  and variances} \\
    $k$ \hfill \Comment{Maximum degree of  rational function} \\
    $c$ \hfill \Comment{Penalty weight} 
  }
  \Initialize{\strut
    $\mathcal{R} \gets \{0, 1, \dots, k\}^2$ \hfill \Comment{Candidate degree search space} \\
    ${\Sigma} \gets \text{diag}\left(\{\hat{\sigma}_i^2\}_{i=0}^n, \{\hat{\gamma}_i^2\}_{i=0}^n\right)$ \hfill \Comment{Data covariance matrix} \\
    $y \gets \left[\hat{g}(t_0), \dots, \hat{g}(t_n), \hat{g}'(t_0), \dots, \hat{g}'(t_n)\right]^T$ \hfill \Comment{Joint observation vector}
  }

  \ForAll{$(r,s) \in \mathcal{R}$}
    \State $\theta \gets \theta_{\text{init}}$ \hfill \Comment{Initialize prior mean parameters}
    \State $\phi \gets \phi_{\text{init}}$ \hfill \Comment{Initialize kernel hyperparameters}
    \State $(\theta^*_{r,s}, \phi^*_{r,s}) \gets \arg\max_{\theta, \phi} \left[ \log p(y \mid \theta, \phi) - c \mathcal{P}(\theta) \right]$ \hfill \Comment{Maximize log-marginal likelihood}
    \State $\mathscr{L}_{r,s} \gets \log p(y \mid \theta^*_{r,s}, \phi^*_{r,s}) - c \mathcal{P}(\theta^*_{r,s})$ \hfill \Comment{Store objective  for model selection}
  \EndFor

 \State Select $(r^*, s^*) \gets \arg\max_{(r,s) \in \mathcal{R}} \mathscr{L}_{r,s}$ \hfill \Comment{Identify optimal rational degrees}
  \State $\theta^* \gets \theta_{r^*,s^*}^*, \quad \phi^* \gets \phi_{r^*,s^*}^*$ \hfill \Comment{Extract optimal parameters}
  
  \State \textbf{Output:} $m_{\theta^*,\phi^*,n}(1)$ \hfill \Comment{\ac{gp} posterior conditional mean via \eqref{eq: post mean}}
\end{algorithmic}
\end{algorithm}

\section{Importance Tempering}
\label{app: IT}
In this appendix, we summarize the \ac{it} method of \citet{gramacy2010importance} and we describe how we approximate the variance of \ac{it} data, and the gradient \ac{it} data and its variance, used in \ac{name}. At the outset, \ac{it}  constructs self-normalised importance sampling estimators of the tempered expectation $g(t_k)$ using samples $\{x_j^{(i)}\}_{j=1}^N$ from $p_{t_i}$, for $t_i \leq t_k$, obtained for instance using \ac{smc}.
These can be expressed as  
\begin{align}
    \tilde{g}_i(t_k) := \frac{\sum_{j=1}^N \omega_i(x_j^{(i)}) f(x_j^{(i)}) }{ \sum_{j=1}^N \omega_i(x_j^{(i)}) } , \qquad \omega_i(x) := \frac{p_{t_k}(x)}{p_{t_i}(x)}, \qquad i = 0, \ldots, k, \label{eq:self_normilized_is}
\end{align}
and, under appropriate regularity assumptions, they are consistent estimators of the tempered expectation $g(t_k)$. Strictly, only estimation of the original posterior expectation was considered in \citealt{gramacy2010importance}, corresponding to $k=n$, but we  make use also of estimates of tempered expectations in our methodological development, letting $k=0,\ldots, n$.
The issue with using any of the $\tilde{g}_i(t_j)$ individually is that their variance can be substantial.
To address this, \citet{gramacy2010importance} proposed to consider convex combinations of the form
\begin{align}
    \hat{g}_{\mathrm{IT}}(t_k) := \sum_{i=0}^k \lambda_i \tilde{g}_i(t_k), \qquad \lambda_i \propto \frac{ ( \sum_{j=1}^N \omega_i(x_j^{(i)}) )^2 }{ \sum_{j=1}^N \omega_i(x_j^{(i)})^2 } , \label{eq:it_estimator}
 \end{align}
where the weights $\lambda_i$ are selected to maximise the effective sample size associated with the estimator $\hat{g}_{\mathrm{IT}}(t_j)$ \citep[][Proposition 2.1]{gramacy2010importance}.

\paragraph{Bootstrap Variance of \ac{it} Data}
Quantifying the variability of the \ac{it} estimators is not straightforward because the weights $\lambda_i$ depend on the samples. Furthermore, when the samples are obtained via \ac{smc}, both the self-normalized estimators $\tilde{g}_i$ and the weights exhibit correlation across temperatures.
Since the role of the error model in \eqref{eq: heur log like} is limited to guiding the choice of a suitable regression function, 
in our experiments we employ bootstrapping to derive a crude estimator of the variance of the \ac{it} estimators. 
For each temperature $t_k$,
we sample with replacement $B=100$ times from the original \ac{smc} samples 
$\{x^{(k)}_{j}\}_{j=1}^{N}$, obtaining the bootstrap sample sets
$
\{x_{k,j}^{*(b)}\}_{j=1}^{N}, b=1, \ldots, B. 
$
Using these resampled particles, we compute $B$ importance tempering estimators  $
\hat g_{\mathrm{IT}}^{*(b)}(t_k)
$ at the temperature $t_k$,  following~\eqref{eq:it_estimator}. The variance of the IT estimator at temperature $t_k$ is then approximated by
\begin{equation*}
\widehat{\mathrm{Var}}_{boots}\!\left(
\hat g_{\mathrm{IT}}(t_k)
\right)
:=
\frac{1}{B-1}
\sum_{b=1}^B
\left(
\hat g_{\mathrm{IT}}^{*(b)}(t_k)
-
\overline g_{\mathrm{IT}}^{*}(t_k)
\right)^2,
\label{eq:it-bootstrap-var}
\end{equation*}
where
\[
\overline g_{\mathrm{IT}}^{*}(t_k)
=
\frac{1}{B}
\sum_{b=1}^B
\hat g_{\mathrm{IT}}^{*(b)}(t_k).
\]
By design, in our setting the \ac{it} estimators depend directly on the \ac{smc} output. 
However, because the \ac{it} framework prioritizes unbiasedness, its bootstrap variance estimates are drastically smaller than those of the raw \ac{smc} estimators, regardless of whether the number of \ac{smc} particles $M$ is large or small. 

\paragraph{Gradient \ac{it} Data and its Bootstrap Variance}
Finally, when performing \ac{name} with gradient information, the \ac{it} gradient data are obtained as follows. Let
\[
h(t) := \mathbb{E}_t[f\ell],
\qquad
q(t) := \mathbb{E}_t[\ell],
\]
where $f$ is the integrand of interest, and $\ell$ the log-likelihood, 
and denote their \ac{it} estimators at temperature $t_k$ by
$\hat{h}_{\mathrm{IT}}(t_k)$ and $\hat{q}_{\mathrm{IT}}(t_k)$, respectively.
The corresponding \ac{it} gradient estimator is then
\[
\hat{g}'_{\mathrm{IT}}(t_k)
=
\hat{h}_{\mathrm{IT}}(t_k)
-
\hat{g}_{\mathrm{IT}}(t_k) \hat{q}_{\mathrm{IT}}(t_k),
\]
which mirrors the structure of \eqref{eq: derivative estimator}. Its variance is estimated by applying the same delta-method approximation that we use for \ac{smc} data
\[
\widehat{\mathrm{Var}}_{boots}\!\left(
\hat g'_{\mathrm{IT}}(t_k)
\right):=\widehat{\mathrm{Var}}_{boots}\!\left(
\hat h_{\mathrm{IT}}(t_k)
\right)+\hat{g}_{\mathrm{IT}}(t_k)^2\widehat{\mathrm{Var}}_{boots}\!\left(
\hat q_{\mathrm{IT}}(t_k)
\right)+\hat{q}_{\mathrm{IT}}(t_k)^2\widehat{\mathrm{Var}}_{boots}\!\left(
\hat g_{\mathrm{IT}}(t_k)
\right).
\]
The estimated variances 
$\widehat{\mathrm{Var}}_{boots}\!\left(
\hat g_{\mathrm{IT}}(t_k)
\right)$
and $\widehat{\mathrm{Var}}_{boots}\!\left(
\hat g'_{\mathrm{IT}}(t_k)
\right)$ are  used in place of $\hat\sigma^2_k$
 and $\hat\gamma^2_k$, respectively,  when applying \ac{name} to smooth tempered expectations obtained via importance tempering.

\section{Details for the Gaussian Mixture Model
}\label{app: illus}

In this section we  elaborate on the two-dimensional Gaussian mixture model introduced in \Cref{subsec: illust} and provide the comprehensive simulation results that support the discussion in \Cref{subsec: gmm}. 
In this model, the prior is $p_0 = \mathcal{N}(0 , \sigma_0^2 I)$ where $\sigma_0^2 =10$, the likelihood is $L(x) \propto \sum_{i=1}^K
\mathcal{N}(x; \mu_i , v_i^2 I)$ where $v_i^2 =0.5$, and 
the locations of the  $K=9$  mixture components in the likelihood are $\mu_i \in \{-4,\, 0,\, 4\}^2$.  The posterior is then $p_1(x) \propto \sum_{i=1}^K \alpha_i \mathcal{N}(x; \tilde\mu_i , \tilde\sigma_i^2 I)$,  where $\tilde\sigma_i^2 =10/21$, the locations are  
$\tilde\mu_i \in 20/21\times \{-4, \, 0, \, 4 \}^2$, and unnormalized weights are proportional to $\exp\left(-\frac{1}{21}\|\mu_i\|^2 \right)$, leading to the set of weights $\alpha_i \in \{0.25, \, 0.125, \, 0.063\}$, with the largest weight given to the central mode, second largest to modes aligned with the cartesian axes, and smallest to the diagonal modes. We consider $f(x) =x_1^2 $ to be the function of interest.  The posterior expectation  can be computed in closed form 
\begin{align*}
g(1) = \sum_{i=1}^K \alpha_i \left(\tilde\sigma^2 + (\tilde\mu_{i1})^2\right)\approx 7.495, 
\end{align*}
where $\tilde\mu_{i1}$ denotes the first component of $\tilde\mu_{i}$. 
Whilst in this toy example the posterior can be directly sampled and the expectation $g(1)$ explicitly computed, 
the expectations under tempered posteriors do not admit closed-form expressions. 
Therefore, the ground truth tempered expectations displayed in \Cref{fig:illustration:ELATE_SMC}  and the associated illustrations in this appendix are computed via numerical integration.

\begin{figure}[t!]
    \centering
    \includegraphics[width=\textwidth]{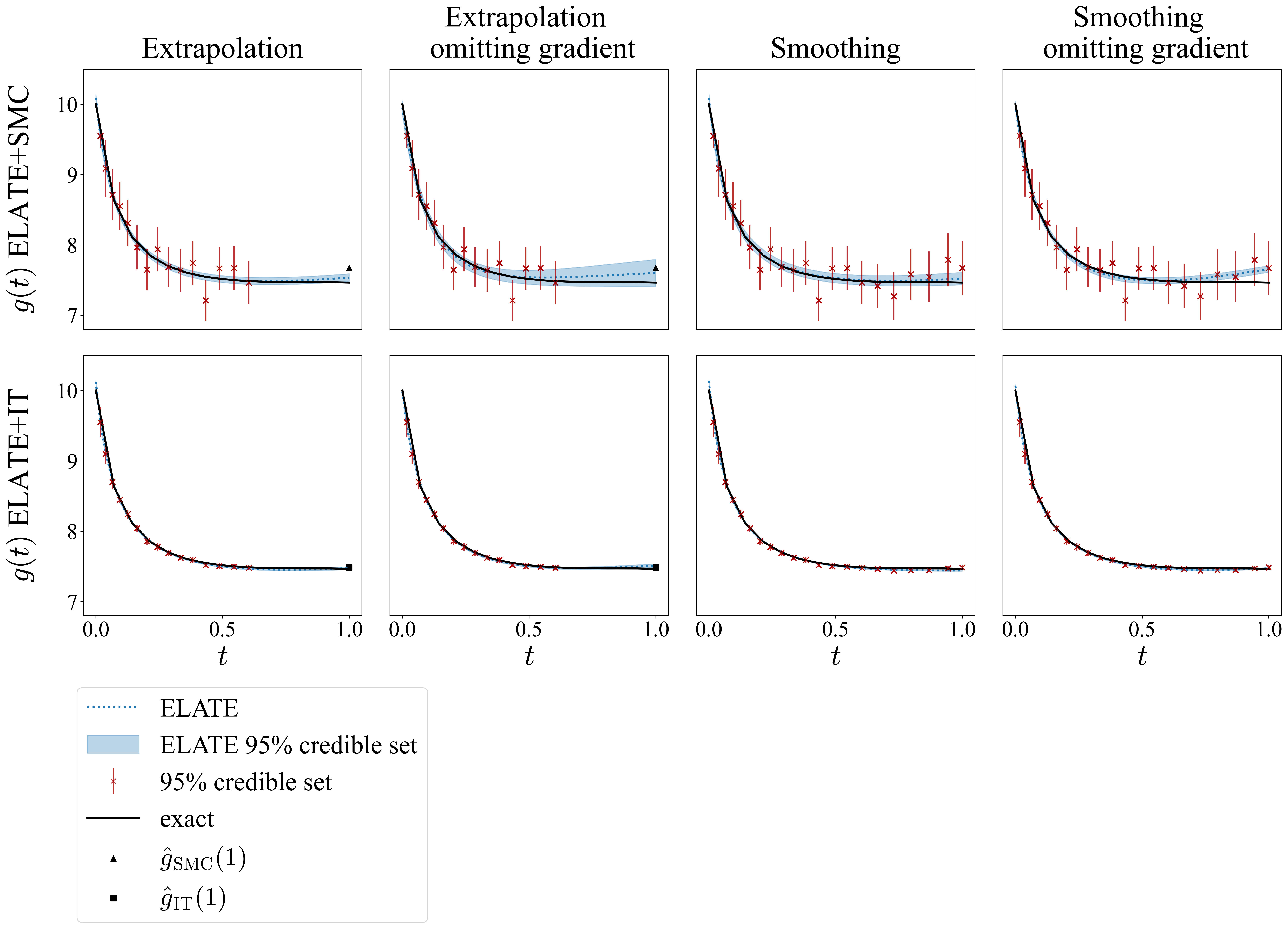}
    \caption{Illustration of \ac{name} on the \textbf{Gaussian mixture} model, based on the same \ac{smc} samples as in \Cref{fig:illustration:ELATE_SMC}, with $\mathbf{M=200}$
    resampled particles and $\mathbf{P=100}$ MCMC steps. The first row shows estimators obtained directly from \ac{smc}, and the second row shows those obtained via \ac{it}. From left to right, the four  panels for each row correspond to:  \ac{name} extrapolation using $t<0.6$  with gradient data included;  extrapolation  using $t<0.6$ without gradient data;  smoothing using the full dataset with gradient data; and smoothing using the full dataset without gradient data. 
Line styles and symbols follow the same convention as in \Cref{fig:illustration:ELATE_SMC}, with an additional black square to represent the reference \ac{it} estimator $\hat{g}_{\text{IT}}(1)$, when extrapolation is performed. 
    	}
    \label{fig:illustration:ELATE_SMC_large_M}
\end{figure}
\begin{figure}[t!]
    \centering
    \includegraphics[width=\textwidth]{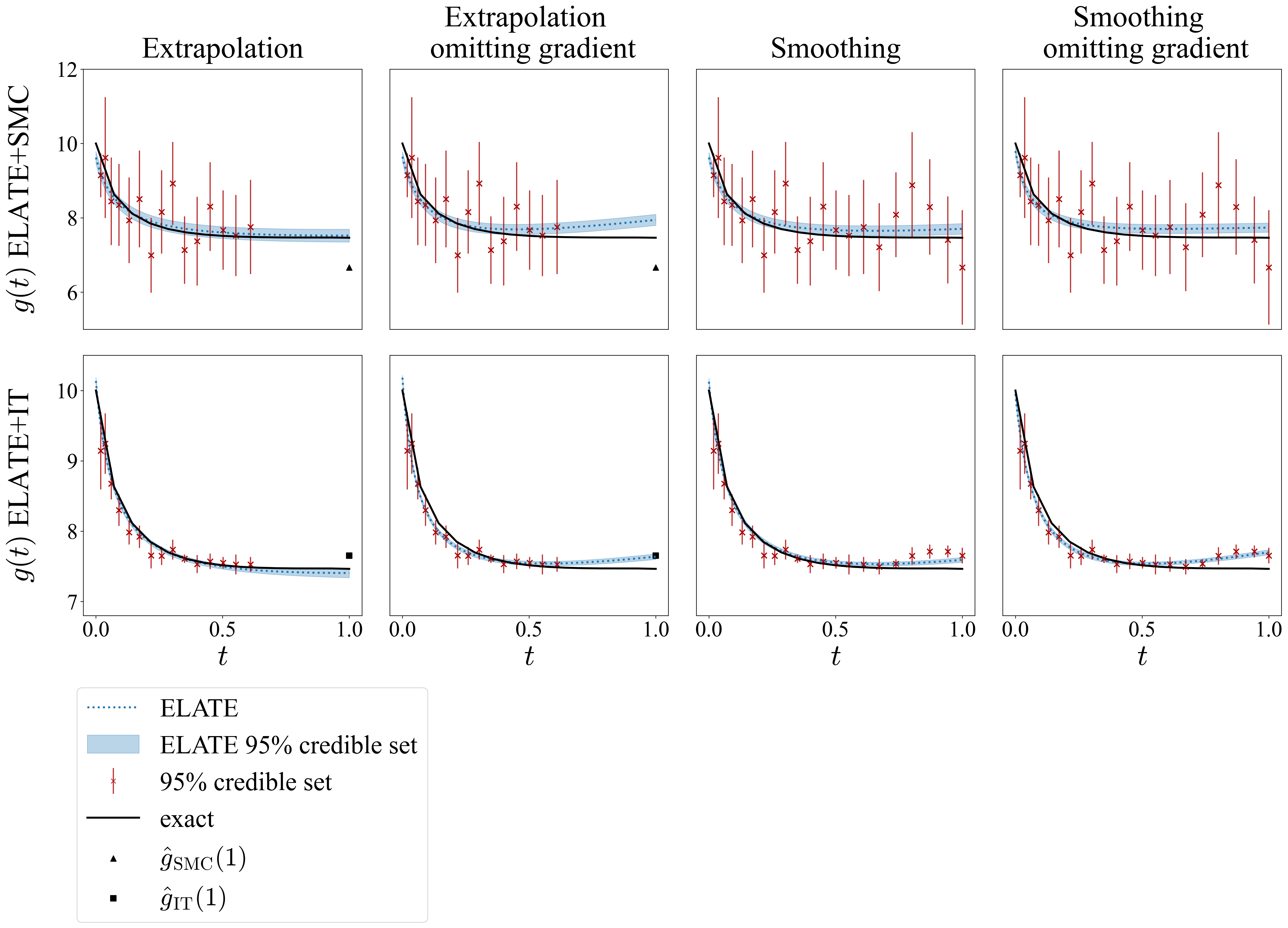}
    \caption{Illustration of \ac{name} on the \textbf{Gaussian mixture} model, based on \ac{smc} samples obtained with 
    	with $\mathbf{M=15}$ resampled particles and $\mathbf{P=100}$ MCMC steps.  The first row shows estimators obtained directly from \ac{smc}, and the second row shows those obtained via \ac{it}. From left to right, the four  panels for each row correspond to:  \ac{name} extrapolation using $t<0.6$  with gradient data included;  extrapolation using $t<0.6$ without gradient data;  smoothing using the full dataset with gradient data; and  smoothing using the full dataset without gradient data. 
Line styles and symbols follow the same convention as in \Cref{fig:illustration:ELATE_SMC}, with an additional black square to represent the reference  \ac{it} estimator $\hat{g}_{\text{IT}}(1)$, when extrapolation is performed. 
    }
 \label{fig:illustration:ELATE_SMC_small_M}
\end{figure}
\begin{figure}[t!]
    \centering
    \includegraphics[width=\textwidth
    ]{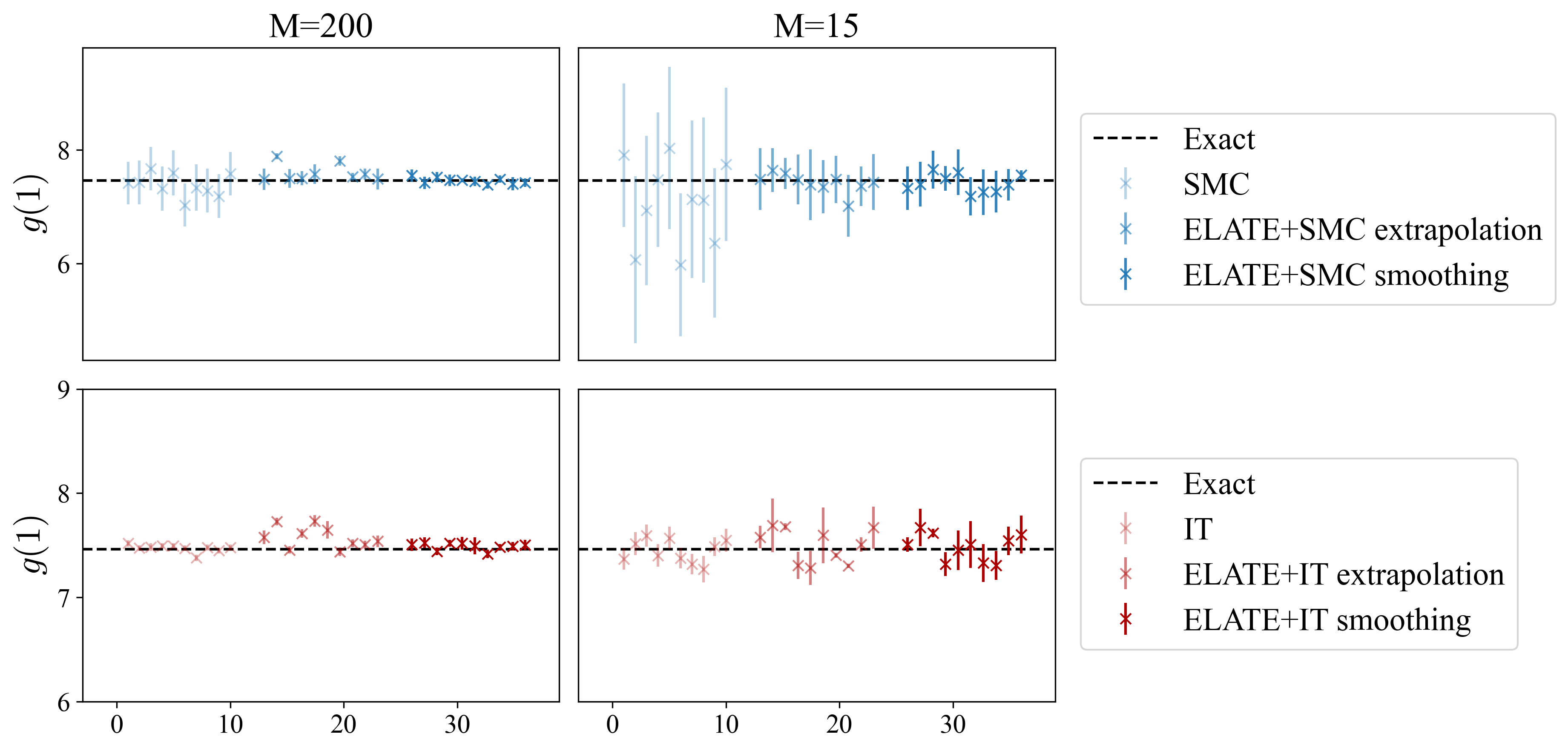}
    \caption{Illustration of \ac{name} on the \textbf{Gaussian mixture} model, showing the \ac{gp} predictive outcome across 
10 experiments for each design choice (panel). In each panel, the  GP is conditioned on both function value and gradient data. 
  Results in the left panel are based on \ac{smc} samples obtained with a resample size of $M = 200$, while the right panel uses $M = 15$; both were rejuvenated using $P = 100$ MCMC steps. Each cross represents an estimate of $g(1)$ obtained in a single experiment using each of the methods: 
 \ac{name} extrapolation and smoothing based on \ac{smc} in the top row;   \ac{name} extrapolation and smoothing based on \ac{it} data in the bottom row. 
  Error bars indicate the uncertainty quantified by each method. 
  } 
    \label{fig:illust_MC}
\end{figure}

We perform two sets of experiments. We first
employ a large number of \ac{smc} samples ($M=200$, $P=100$, for a total of $N=20 \times 10^3$ samples), whereby the  variances of the estimators are moderate; see \Cref{fig:illustration:ELATE_SMC_large_M}. We then
run an experiment where fewer \ac{smc} samples are used ($M=15$, $P=100$), leading to higher  variances across temperatures; see   \Cref{fig:illustration:ELATE_SMC_small_M}. 
 The minimum threshold for the effective sample size is set to $\text{ESS}_{\text{min}} = 0.995$, corresponding to an average  of 21 temperatures selected according to the procedure specified in \Cref{app: t selection}. 
Repeated simulations for these settings are presented in \Cref{fig:illust_MC}, showing that the performance of \ac{name} (evaluated via the \ac{gp} predictive posterior mean) strongly depends on the quality of the point estimators and variance estimates used to construct the \ac{gp} regression model for both \ac{smc} and \ac{it} data. Focusing first on the case when gradient data is included in the regression, we make the following observations: 

\paragraph{\ac{smc} Data} When the baseline \ac{smc} estimator variance is small (\Cref{fig:illustration:ELATE_SMC_large_M,fig:illust_MC}), \ac{name} smoothing consistently improves the estimate of $g(1)$ relative to $\hat{g}_{\mathrm{SMC}}(1)$. In contrast, extrapolation exhibits mixed performance; it occasionally outperforms the baseline but can also slightly degrade the estimate. This behavior aligns with our intuition that in low-variance regimes, ``outlier" estimators exert a stronger influence on \ac{name} extrapolation than on smoothing, as the extrapolation model is trained on a smaller subset of the data. Conversely, when the \ac{smc} estimators are highly variable (\Cref{fig:illustration:ELATE_SMC_small_M,fig:illust_MC}), both smoothing and extrapolation yield systematic improvements over the baseline, a stabilization that is likely governed by the regularizing effect of the smooth \ac{gp} prior.

\paragraph{\ac{it} Data} 
In low-variance regimes (\Cref{fig:illustration:ELATE_SMC_large_M,fig:illust_MC}), applying \ac{name} smoothing to \ac{it} data can outperform plain \ac{it}, effectively combining two distinct sources of variance reduction relative to vanilla \ac{smc}. Nonetheless, this additional gain is modest compared to the substantial variance reduction already achieved by plain \ac{it}, and it is not consistently observed across all experiments, though no meaningful degradation occurs. In this setting, extrapolation performs slightly worse than smoothing, a result consistent with our findings for the \ac{smc} data. On the other hand, when the underlying \ac{smc} estimators exhibit large variance (\Cref{fig:illustration:ELATE_SMC_small_M,fig:illust_MC}), the \ac{it} estimators are also more variable; yet, they remain highly accurate with remarkably small bootstrap variance estimates, thereby dominating the likelihood in the \ac{gp} regression model. As a consequence, \ac{name} smoothing and extrapolation yield similar performance, neither substantially improving nor degrading the baseline \ac{it} estimates. 

\begin{figure}[t!]
    \centering
    \includegraphics[width=\textwidth]{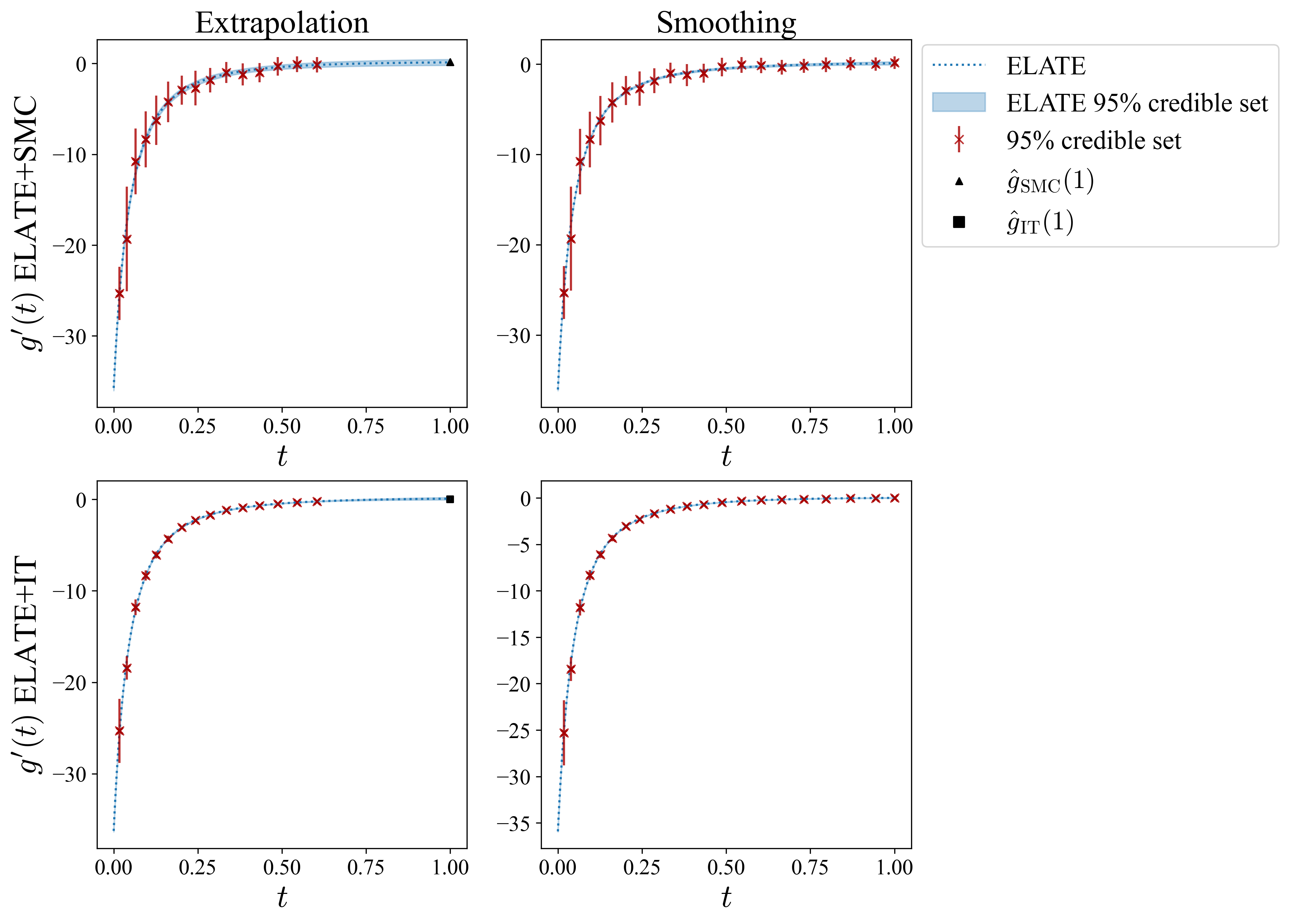}
    \caption{ Illustration of the GP fit to gradient data, for the \textbf{Gaussian mixture} model, based on the same \ac{smc} samples (and corresponding \ac{it} samples) as in \Cref{fig:illustration:ELATE_SMC}. The first row shows gradient estimators obtained directly from \ac{smc}, while the second row shows those obtained via \ac{it}. For each row, the left panel displays extrapolation results using  $t < 0.6$, and the right panel shows \ac{name} fitted using the full  dataset.
Line styles and symbols follow the same convention as in \Cref{fig:illustration:ELATE_SMC}, with an additional black square to represent the reference standard \ac{it} estimator $\hat{g}_{\text{IT}}(1)$, when extrapolation is performed. 
}
\label{fig:illustration:ELATE_SMC_derivatives}
\end{figure}

\paragraph{Incorporating the gradient data}  \Cref{fig:illustration:ELATE_SMC_large_M,fig:illustration:ELATE_SMC_small_M}
 show that omitting gradient information consistently degrades performance,
supporting the inclusion of both terms in the heuristic
likelihood~\eqref{eq: heur log like}. 
To confirm valid behavior, \Cref{fig:illustration:ELATE_SMC_derivatives} shows the GP fit to the \ac{smc} and \ac{it} gradient data corresponding to the same setting as 
in \Cref{fig:illustration:ELATE_SMC}; 
in both cases, the gradient GP fits well the gradient data, as expected.

\clearpage

\section{Details for the mRNA Transfection Model}\label{appendix:mrna}

\Cref{app: LVM spec} provides the comprehensive implementation details necessary to reproduce the \ac{mrna} experiments described in \Cref{subsec: first}.  \Cref{app: LVM vary integrand,app: failure modes} present supplementary empirical results referenced throughout the main text, demonstrating the performance of the proposed methodology across a variety of test functions and outlining  considerations regarding potential failure modes.

\subsection{Model Specification and Tempered Distributions}
\label{app: LVM spec}
We consider the  \ac{ode} model of \citep{leonhardt2014single}, that describes the transfection process of cells, that is the dynamics of \ac{mrna} delivery $M(\tau)$, $\tau\geq 0$, 
and the expression of the  
 enhanced green fluorescent protein (eGFP)  $G(\tau)$, a 
 commonly used  fluorescence reporter sequence in \ac{mrna} therapeutics 
\begin{align*}
                \frac{\mathrm{d}}{\mathrm{d}\tau}G(\tau)&=k_{TL} M(\tau)-\beta  G(\tau),\\
                \frac{\mathrm{d}}{\mathrm{d}\tau}M(\tau)&=-\delta  M(\tau). 
            \end{align*}
The parameter $k_{TL}>0$ refers to the translation rate, and $\delta>0$ and $\beta>0$ correspond to the degradation rates of \ac{mrna} and eGFP, respectively.
Let $\tau_0$ denote the initial time of the transfection response; the state variables at $\tau_0$ are given by $G(\tau_0)=0$ and $M(\tau_0)=M_0$.          
Following the literature, we treat the product \( \psi:=k_{TL}  M_0 \) as a single parameter,  and define the parameter vector of interest as  \( \vartheta := \{ \psi, \delta, \beta, \tau_0 \} \), upon which we set uniform priors: 
$\psi \sim \text{Uniform}(0,6)$, $\delta\sim \text{Uniform}(0,1)$, $\beta \sim \text{Uniform}(0,1)$, and $t_0\sim \text{Uniform}(0,3)$.
Observations of the model  $O_\tau$ are simulated at   the discrete times $\tau = 1,2,\dots 50$, to 
represent noisy observations of eGFP concentration, using a Gaussian additive model with fixed noise level $v=1$, in correspondence of the parameter value $\vartheta = (5, 0.1, 0.8, 2)$. Given that the \ac{ode} model is linear, it admits closed-form solution, and the observations $O_\tau$ have likelihood $O_\tau|\vartheta\sim \mathcal{N}(\mu(\tau),v^2)$, where 
\begin{align*}
    \mu(\tau)=\frac{\psi}{\delta-\beta}(1-e^{-(\delta-\beta)(\tau-\tau_0)}) e^{-\beta(\tau-\tau_0)}.
\end{align*}

\begin{figure}[t!]
    \centering
    \includegraphics[width=0.9\textwidth
    ]{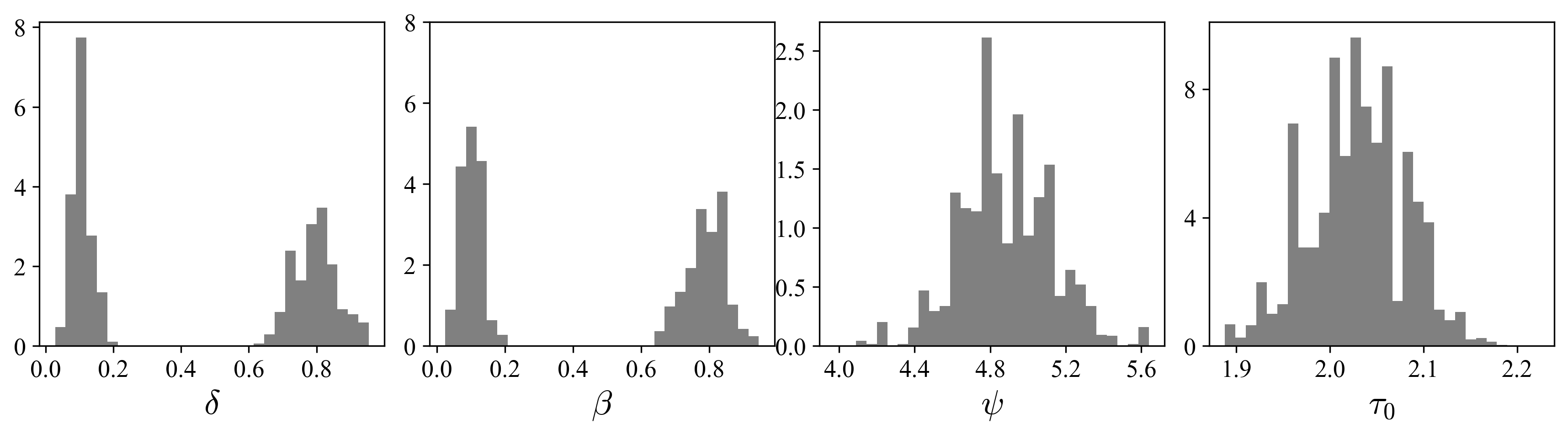}
    \caption{Histograms of posterior \ac{smc} samples in the \textbf{\ac{mrna}} model. 
    } 
    \label{fig: mrna samples}
\end{figure}
\begin{figure}[t!]
    \centering
    \includegraphics[width=0.65\textwidth
    ]{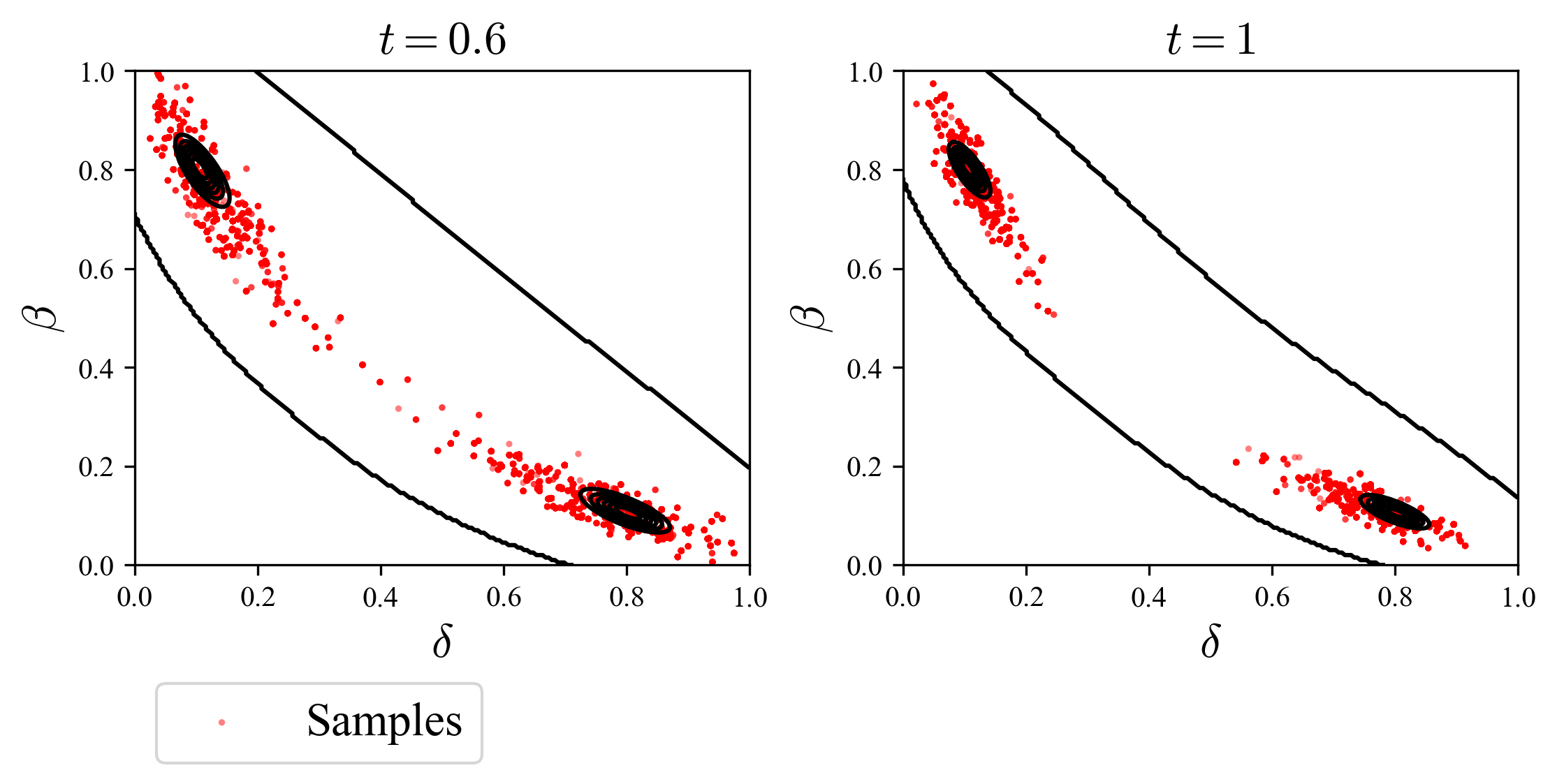}
    \caption{Contour plots of $(\beta, \,\delta)$-slices of the  tempered posteriors in the \textbf{\ac{mrna}} model, with the other parameters set to their ground truth value, and scatterplot of \ac{smc} particles. 
    } 
    \label{fig: posterior mrna}
\end{figure}
The sample marginal posterior distributions at temperature $t=1$ 
obtained from an \ac{smc} run with $M=10$ resampled particles, $P=1000$ MCMC steps, and  $\text{ESS}_{\text{min}}=0.97$, are shown in \Cref{fig: mrna samples}. The posterior distributions of the parameters $\delta$ and $\beta$ are bimodal, due to the exchangeability of the two parameters in the \ac{ode} model. Additionally, in \Cref{fig: posterior mrna} we show the $(\beta, \, \delta)$-marginal tempered posteriors, for the same \ac{smc} setting, and in correspondence of the true value of the remaining parameters. When $t=0.6$, \ac{smc} samples are spread in the two modes and in the low probability region between the modes, but they are concentrated only around the modes when $t=1$. Overall, \ac{smc}  seems to have performed well without any post-processing, but we wish to investigate if \ac{name} could be used to improve posterior expectations derived from tempered \ac{smc} estimators.

\subsection{\ac{name} Estimate of  Moments}\label{app: LVM vary integrand}
\begin{figure}[t!]
    \centering
        \includegraphics[width=\textwidth]{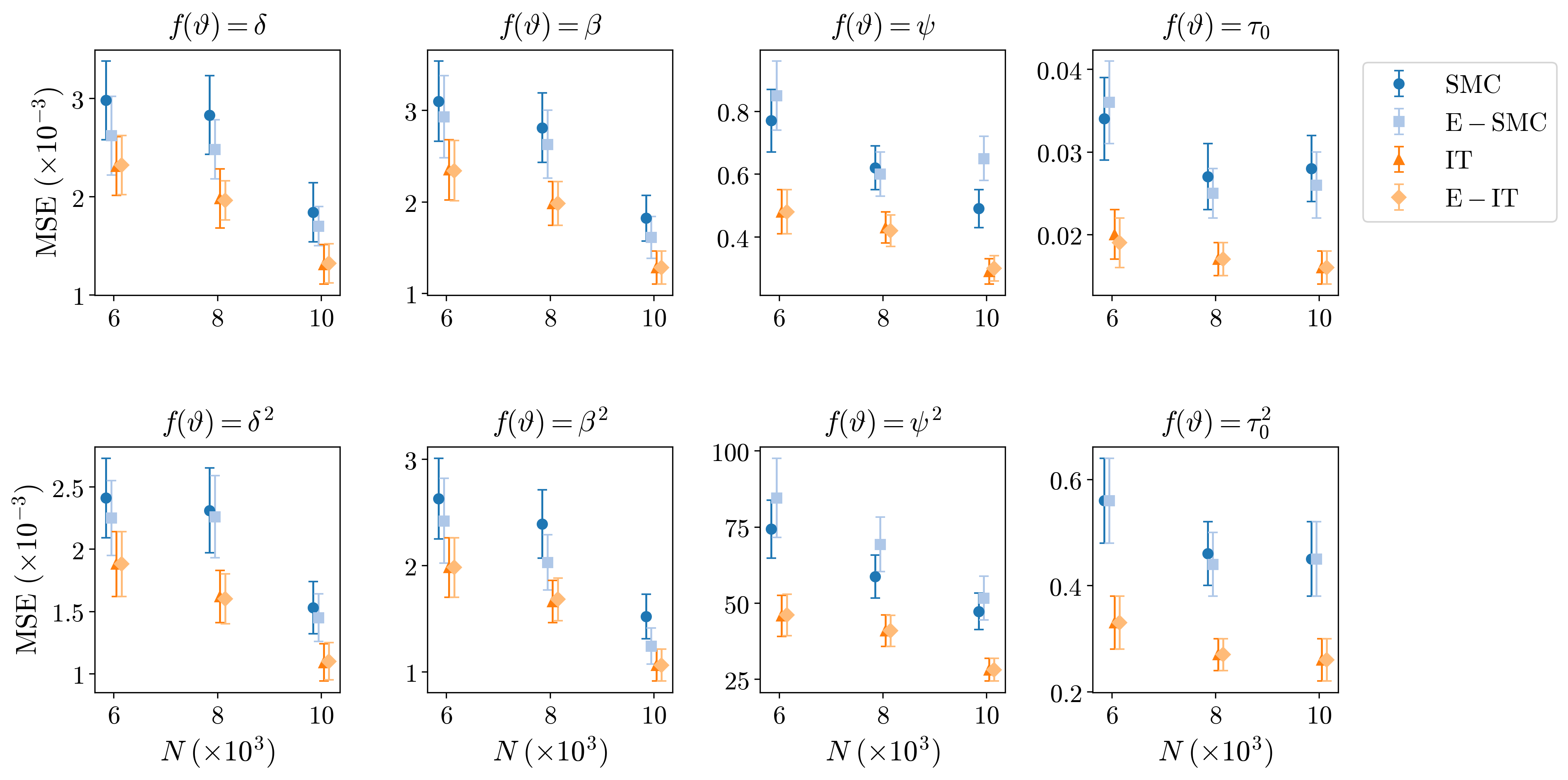}
    \caption{Parameter estimation for \acp{ode}. Estimator \ac{mse} and associated standard error for the \textbf{mRNA} model, when varying   the test function 
$f(\vartheta)$, as indicated in the title of each subplot, computed over 100 independent realizations of SMC. Marker styles follow the same convention as in \Cref{fig: lvm mse}.
We fix the resample size to  $M =50$ across experiments, and standalone markers display performance for increasing particle sizes $N$. \ac{mse} values are presented in units of $10^{-3}$.  
   }
\label{fig: lvm mse 2}
\end{figure}

We report the results of a Monte Carlo experiment similar to that in \Cref{subsec: first}, where we now vary the integrand test function to showcase the estimation of means and second moments of all the parameters of the \ac{mrna} model. 
Performance is measured in terms of the \ac{mse} relative to a gold standard obtained averaging 100 brute force extended \ac{smc} runs, each $M=200$ resampled particles, 
 chain length $P=2500$
 for a total $N=500\times10^3$ total particles,
and  $\text{ESS}_{\text{min}} = 0.7$.
\Cref{fig: lvm mse 2} compares the effectiveness of 
\ac{name} smoothing, using function and gradient data.
When predicting the mean and second moment of the exchangeable parameters $\beta$ and $\delta$, \ac{name} consistently outperforms standard \ac{smc} across all sample and resample sizes. In our setup, the standard \ac{smc} particle approximation at $t=1$ slightly misrepresents the posterior modes, whereas earlier tempered states capture them more effectively due to their more diffuse targets. This setting presents an ideal scenario for leveraging the smoothing capability of \ac{name} to reduce the \ac{mse}. Conversely, the marginal distributions of $\psi$ and $\tau_0$ are robust to inadequate representation of a mode, ensuring that baseline \ac{smc} estimators for expectations of these parameters remain accurate. In such instances, \ac{name} achieves comparable performance without introducing numerical degradation.
Analogous to the results for $\psi$ and $\tau_0$, combining \ac{name} with \ac{it} yields consistent outcomes across all test functions, given that \ac{it} already performs exceptionally well in this low-dimensional regime.

\subsection{\ac{name} Failure Modes}
\label{app: failure modes}
\begin{figure}[t!]
    \centering
    \includegraphics[width=0.9\textwidth]{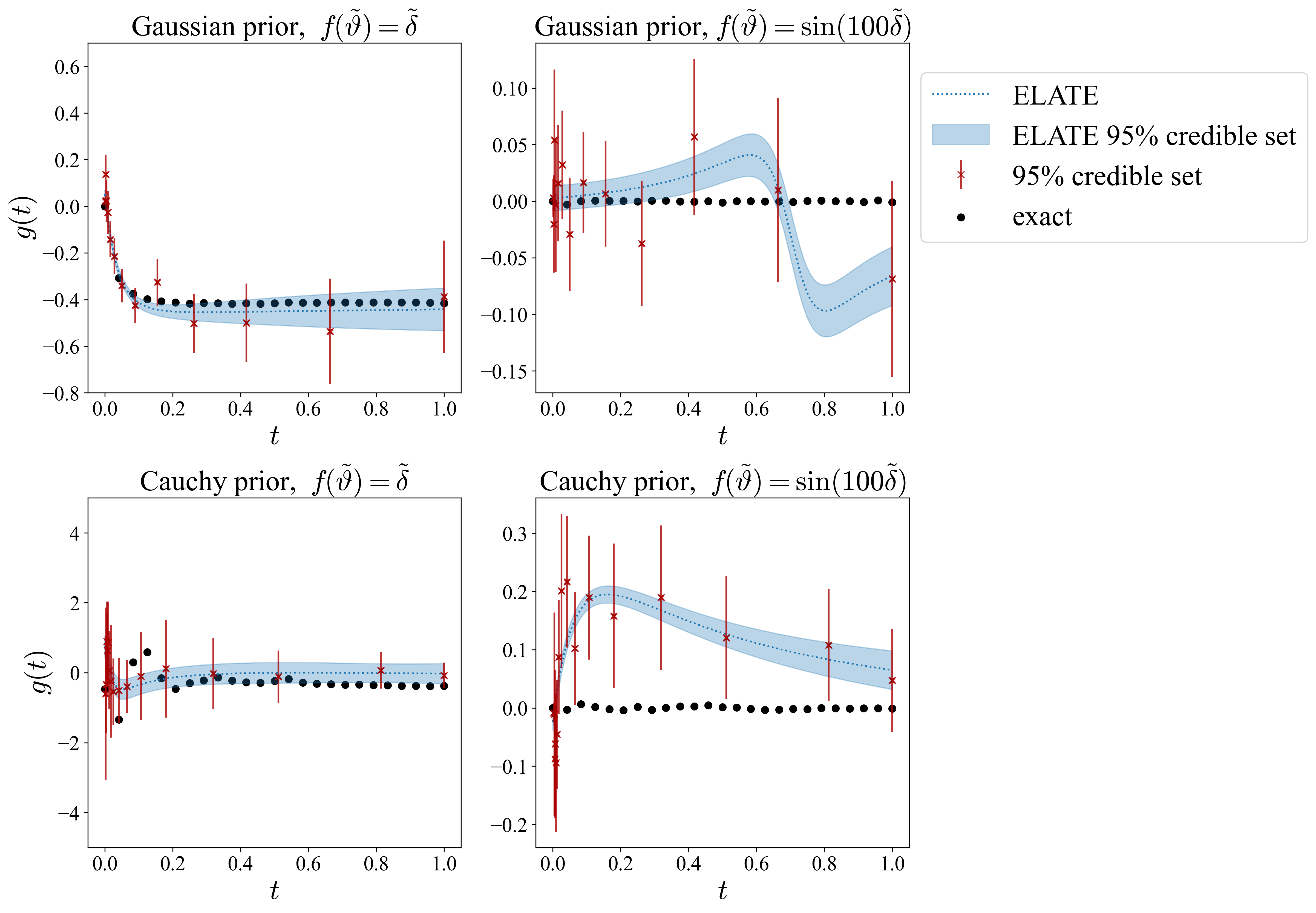}
    \caption{\ac{name} failure modes in the \textbf{\ac{mrna}} model. 
    Black dots aim to represent the ground truth values,
    red crosses and vertical lines represent the baseline  \ac{smc} estimators and their  variances, and \ac{name} smoothing is applied to \ac{smc} data and their gradients. The blue dashed lines correspond to the \ac{gp}  posterior
mean, and the blue shaded areas indicate the predictive credible intervals.
    The prior $p_0$ and test function $f$ are indicated in the title of each subplot. }
    \label{fig:failure modes}
\end{figure}

We present three distinct scenarios where \ac{name} either breaks down or exhibits a lack of robustness: $(a)$ heavy-tailed Cauchy prior $p_0$, $(b)$  highly oscillatory integrand $f$, and $(c)$ a combination of both challenges. To accommodate for heavy-tailed priors, the \ac{smc} sampler is run on the log-transformed parameter space $\tilde{\vartheta} := \log \vartheta = (\tilde{\psi}, \tilde{\delta}, \tilde{\beta}, \tilde{\tau}_0)$. 
Ground truth values  are obtained on an equally spaced temperature ladder, averaging
100 \ac{smc} runs with $M = 50$ and $P = 10\times 10^3$;   the baseline \ac{smc} estimators are obtained with $M = 100$,
$P = 100$ and $\text{ESS}_{\text{min}}  = 0.7$. 
The corresponding results are displayed in \Cref{fig:failure modes}. For a baseline comparison, the top-left panel illustrates the  performance obtained under a standard Gaussian prior paired with the identity test function, where the sufficient conditions for  analyticity of $g(t)$ are  satisfied. In this  setting,  \ac{name} smoothing, constructed using both function values and gradient data, behaves as expected and  outperforms the plain \ac{smc} estimator at $t=1$.

Scenario $(a)$, bottom-left panel,  showcases the effect of the non-informative priors  $\tilde{\psi} \sim \text{Cauchy}(-2,1)$, $\tilde{\delta}\sim \text{Cauchy}(0,1)$, $\tilde{\beta} \sim \text{Cauchy}(0,1)$, and $\tilde{\tau}_0\sim \text{Cauchy}(-2,0.5)$, where $\text{Cauchy}(x_0,\gamma)$ denotes the Cauchy distribution with location $x_0$ and scale $\gamma$. The \ac{smc} estimators exhibit large variance in the early tempering regime ($t \to 0$), where the tempered posterior is dominated by the heavy-tailed prior. As $t$ increases, the tempered posterior concentrates around the data-driven modes, but numerical instabilities introduced at the early stages propagates forward. 
The variability of the \ac{smc} estimators decreases, leading  \ac{name} to be driven primarily by the data at higher temperatures, and therefore making the method highly susceptible to local  fluctuations and outliers close to $t=1$.
In scenario~$(b)$, top-right panel, the integrand is set to the highly oscillatory function $f(\tilde{\vartheta}) = \sin(100 \tilde{\delta})$. Although the growth condition in \Cref{def: growth} is satisfied,  the high-frequency oscillations of the integrand cause the \ac{smc} estimators oscillate and exhibit high variability across temperatures. 
Therefore, \ac{name} is unable to infer a coherent smoothing trajectory, leaving it incapable of correcting the baseline estimator bias at $t=1$.
Finally, in scenario $(c)$, bottom right, the two failure modes occur simultaneously, making \ac{smc} perform poorly, fitting the given data, and the 
\ac{name} predictions  far from the ground true values.

\section{Details for the Logistic Regression Model}\label{app: sonar_model}
In this appendix we provide the implementation details for the logistic regression model fitted to the \textit{Sonar} dataset from the UCI Machine Learning Repository \citep{Dua:2019}.  As noted by \cite{chopin2017leave}, this dataset serves as a challenging benchmark for Bayesian inference algorithms due to its high-dimensional and strongly correlated covariates. Given the observations and covariates $(y_i, z_i) \in \{-1,1\} \times \mathbb{R}^p$ for $i=1, \ldots, n$, and the parameter vector $x \in \mathbb{R}^p$, the likelihood function is given by
\begin{align}L(x) = \prod_{i=1}^{n} F(y_i x^\top z_i), \qquad F(a) = \frac{1}{1 + e^{-a}}.
\end{align}
The dimension of the parameter vector, including the intercept,  is $p=63$. 
Each predictor is rescaled to have a mean of 0 and a standard deviation of 0.5, and we assign a $\mathcal{N}(0,20^2)$ prior to the intercept and independent $\mathcal{N}(0,5^2)$ priors to the remaining coefficients.

\begin{figure}[t!]
    \centering
\includegraphics[width=\textwidth]{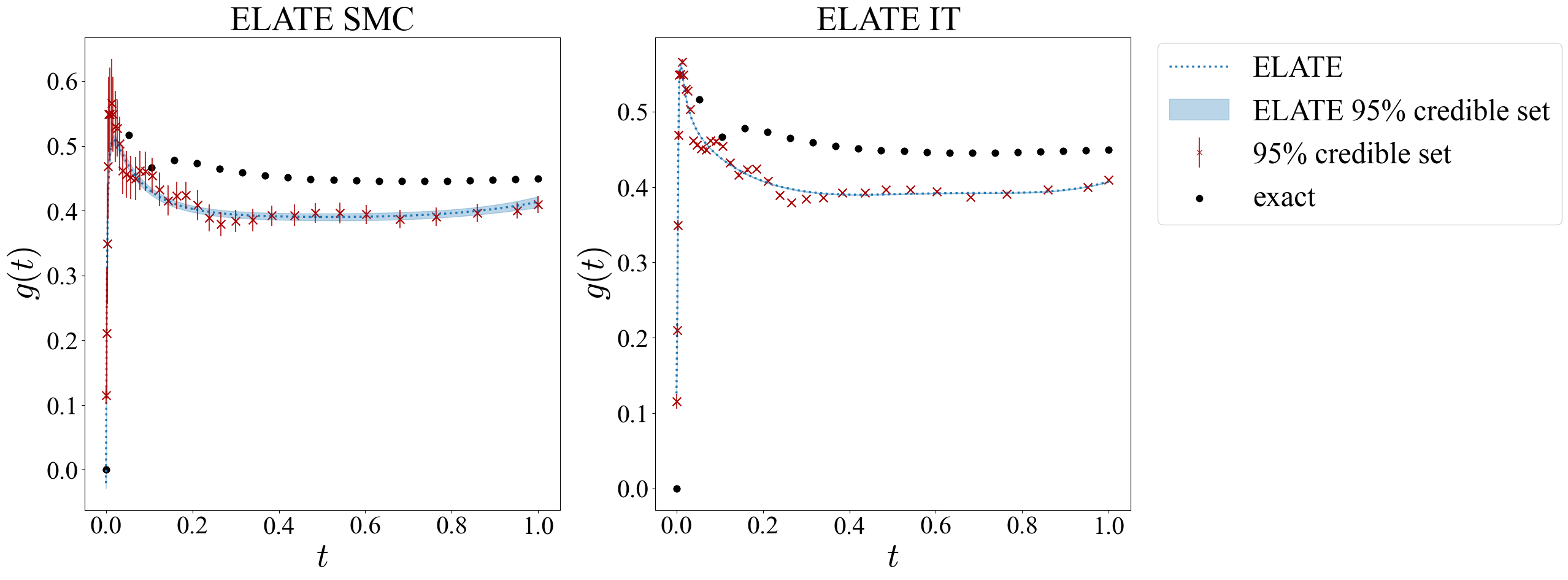}
    \caption{Illustration of ELATE for the posterior expectation of 
$f(x)=\sum_{i=1}^{p} x_i$ in the \textbf{Sonar} logistic regression model. The left panel shows estimators obtained directly from SMC, while the right panel shows those obtained via IT. In both panels, ELATE is fitted using the full set of tempered expectation estimates, together with gradient data. Line styles and symbols follow the same convention as in \Cref{fig:failure modes}, with the GP posterior predictive mean and corresponding predictive credible interval shown alongside the baseline estimator and its associated error bars.}
    \label{fig:sonar_GP}
\end{figure}

\begin{figure}[t!]
    \centering
        \includegraphics[width=0.7\textwidth]{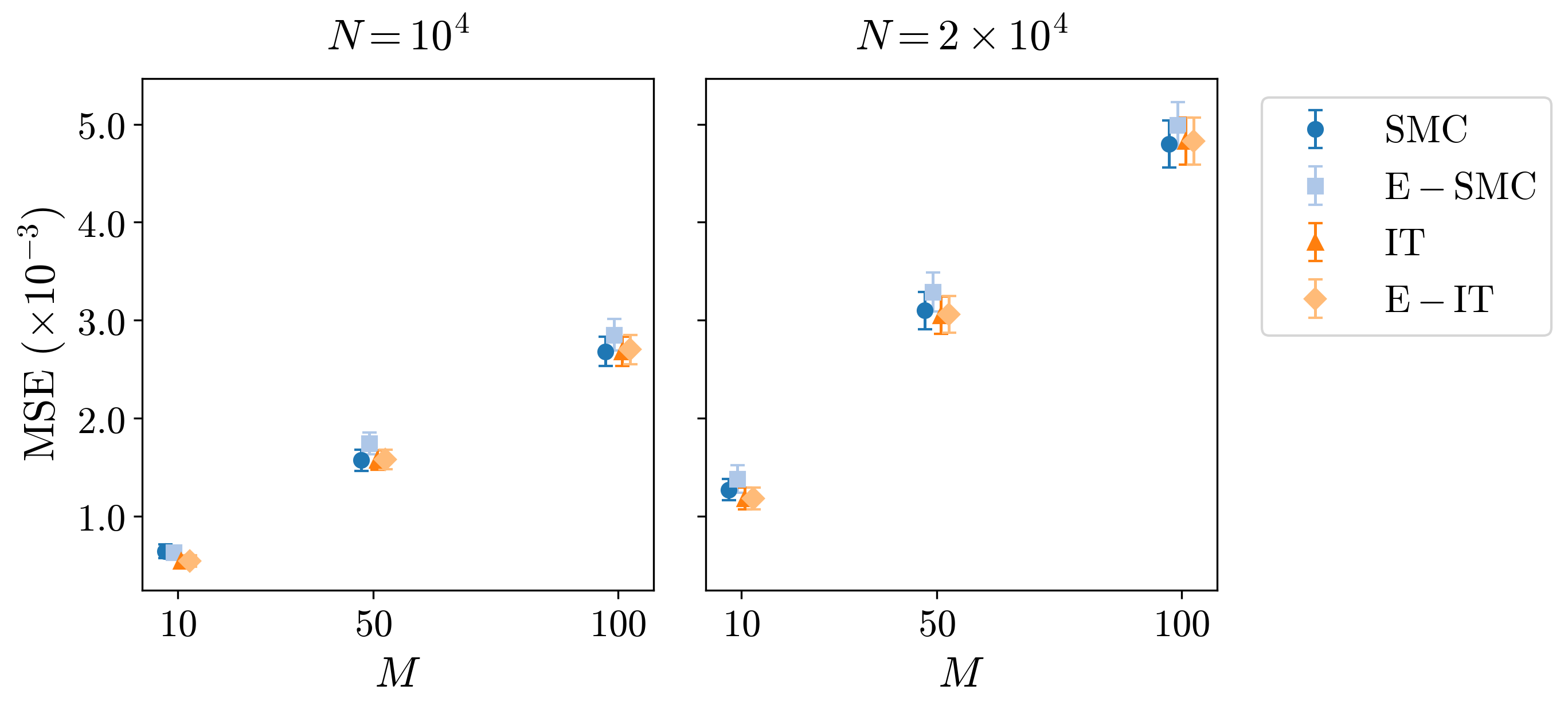}
    \caption{Estimation of the posterior expectation of $f(x) = \sum_{i=1}^px_i$ in the \textbf{Sonar} logistic regression model. Estimator \ac{mse} and associated standard error computed over 100 independent realizations of \ac{smc}. Marker styles follow the same convention as in \Cref{fig: lvm mse}. Panels display increasing particle sizes $N$ from left to right, with standalone markers displaying performance across distinct resample sizes $M$. \ac{mse} values are presented in units of $10^{-3}$.  
   }
\label{fig:sonar1}
\end{figure}

\begin{figure}[t!]
    \centering
        \includegraphics[width=0.8\textwidth]{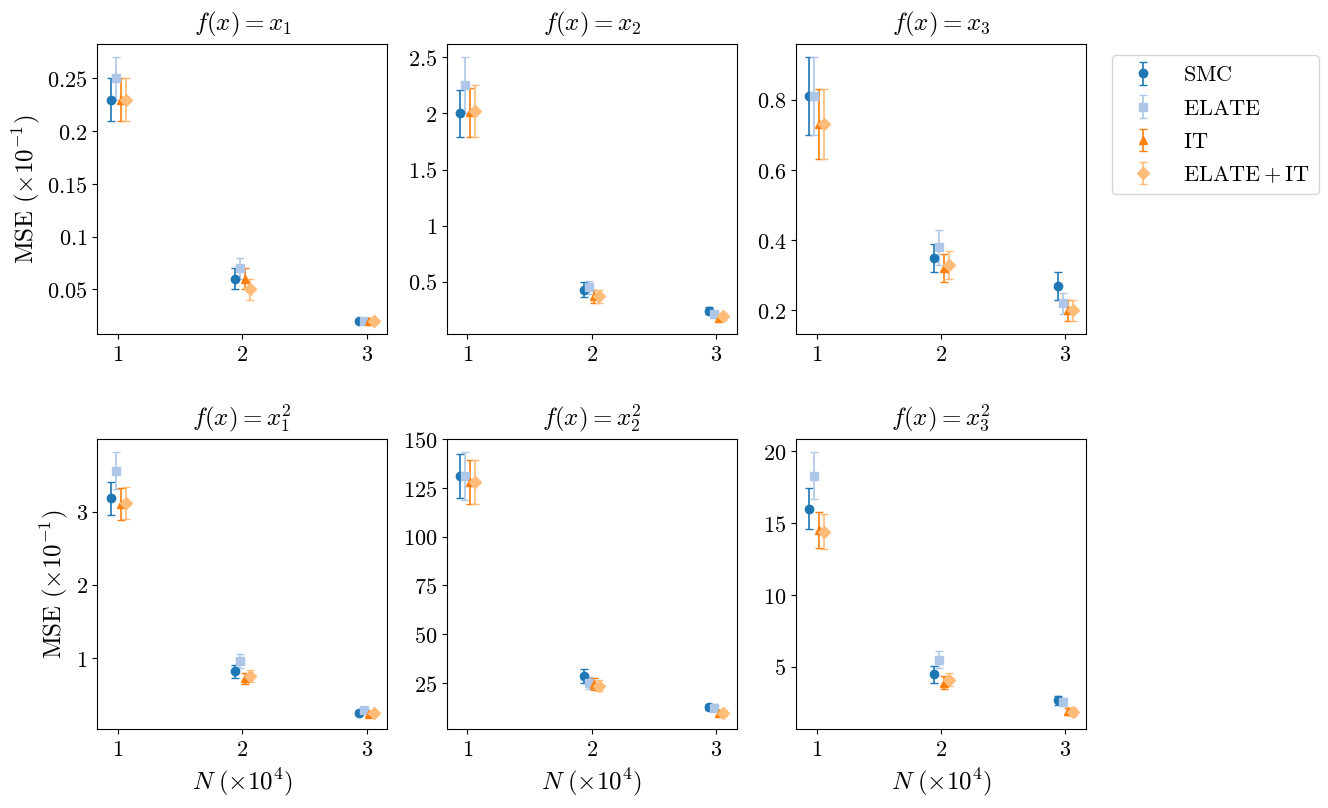}
    \caption{Parameter estimation for  the \textbf{Sonar} logistic regression model.  Estimator \ac{mse} and associated standard error, when varying   the test function $f(x)$, as indicated in the title of each subplot, computed over 100 independent realizations of SMC. Marker styles follow the same convention as in \Cref{fig: lvm mse}. We fix the resample size to  $M =50$ across experiments, and standalone markers display performance for increasing particle sizes $N$. \ac{mse} values are presented in units of $10^{-1}$.  
   }
\label{fig:sonar2}
\end{figure}

For completeness, we evaluate the performance of \ac{name} in estimating the moments of selected elements of the parameter vector. Following \cite{dau2022waste}, we first target the posterior expectation of the test function $f(x) = \sum_{i=1}^p x_i$. The ground truth value is established by averaging 100 high-precision baseline \ac{smc} runs,  with $M=100$, $P=10^4$, and a pre-set linear tempering schedule of 20 equally spaced steps. \Cref{fig:sonar_GP} displays a typical realization of the \ac{smc} and \ac{it} estimators ($M = 100$, $P = 100$, and $\text{ESS}_{\text{min}} = 0.80$) along with their corresponding \ac{name}-smoothed trajectories. The baseline \ac{smc} and \ac{it} estimates exhibit pronounced bias and oscillations, signaling that the underlying sampler struggles within this high-dimensional setting.  
To statistically evaluate the performance of \ac{name}, \Cref{fig:sonar1} illustrates the \ac{mse} computed across 100 independent replications under varying sample and resample sizes, and  $\text{ESS}_{\text{min}} = 0.5$. 
In this example, similar to observations for certain integrands in \Cref{app: LVM vary integrand}, the \ac{name} estimator yields negligible accuracy gains, but crucially introduces no numerical degradation to the underlying inference. These findings are further corroborated by additional experiments targeting the posterior expectations of a wider variety of test functions, as reported in \Cref{fig:sonar2}.

\section{Details for Thermodynamic Integration}
\label{app: thermo}

In \Cref{app: BQ} we provide the details needed to compute the \ac{name} estimate when inferring the marginal log-likelihood using  Bayesian quadrature. 
Additionally, to support the findings shown for the logistic regression example, in \Cref{app:TI_additional_results} we  report results for 
all the other examples considered in the paper. 

\subsection{Bayesian Quadrature}\label{app: BQ}

In our setting, Bayesian quadrature amounts to using the posterior \ac{gp} as a surrogate for the exact integrand.
That is, epistemic uncertainty in the value of the log-marginal likelihood $\log Z_1$ is modelled as a Gaussian random variable with mean
\begin{align}
    \int_0^1 m_{\theta,n}(t) \; \mathrm{d}t & = \int_0^1 m_\theta(t) \; \mathrm{d}t + \left[ \int_0^1 \mathcal{L}_2 k_\phi(t) \; \mathrm{d}t \right] [ \mathcal{L}_1 \mathcal{L}_2 k_\phi + \Sigma ]^{-1} [ y - \mathcal{L} m_\theta ]  . \label{eq: kernel quad}
\end{align}
and variance
\begin{align}
\int_0^1 \int_0^1 k_{\phi,n}(t,t') \; \mathrm{d}t \mathrm{d}t' & = \int_0^1 \int_0^1  k_\phi(t,t') \; \mathrm{d}t \mathrm{d}t' \nonumber \\
& \qquad - \left[ \int_0^1 \mathcal{L}_2 k_\phi(t) \; \mathrm{d}t \right] [\mathcal{L}_1 \mathcal{L}_2 k_\phi + \Sigma]^{-1} \left[ \int_0^1 \mathcal{L}_1 k_\phi(t') \mathrm{d}t' \right] . \label{eq: BQ var}
\end{align}
For the \ac{gp} model we set out in \Cref{subsec: extrap as reg} and \Cref{ap: gp computation}, the kernel integrals appearing on the right hand side of \eqref{eq: kernel quad} and \eqref{eq: BQ var} can be exactly computed, see \cite{briol2025dictionary}. On the other hand, the integral of the \ac{gp} prior mean in \eqref{eq: kernel quad} is computed numerically, because the  rational functions we use  do not always admit closed-form integrals.

\subsection{Normalizing Constant Simulation Results}\label{app:TI_additional_results}

\begin{figure}[t!]
    \centering
    \includegraphics[width=0.75\textwidth]{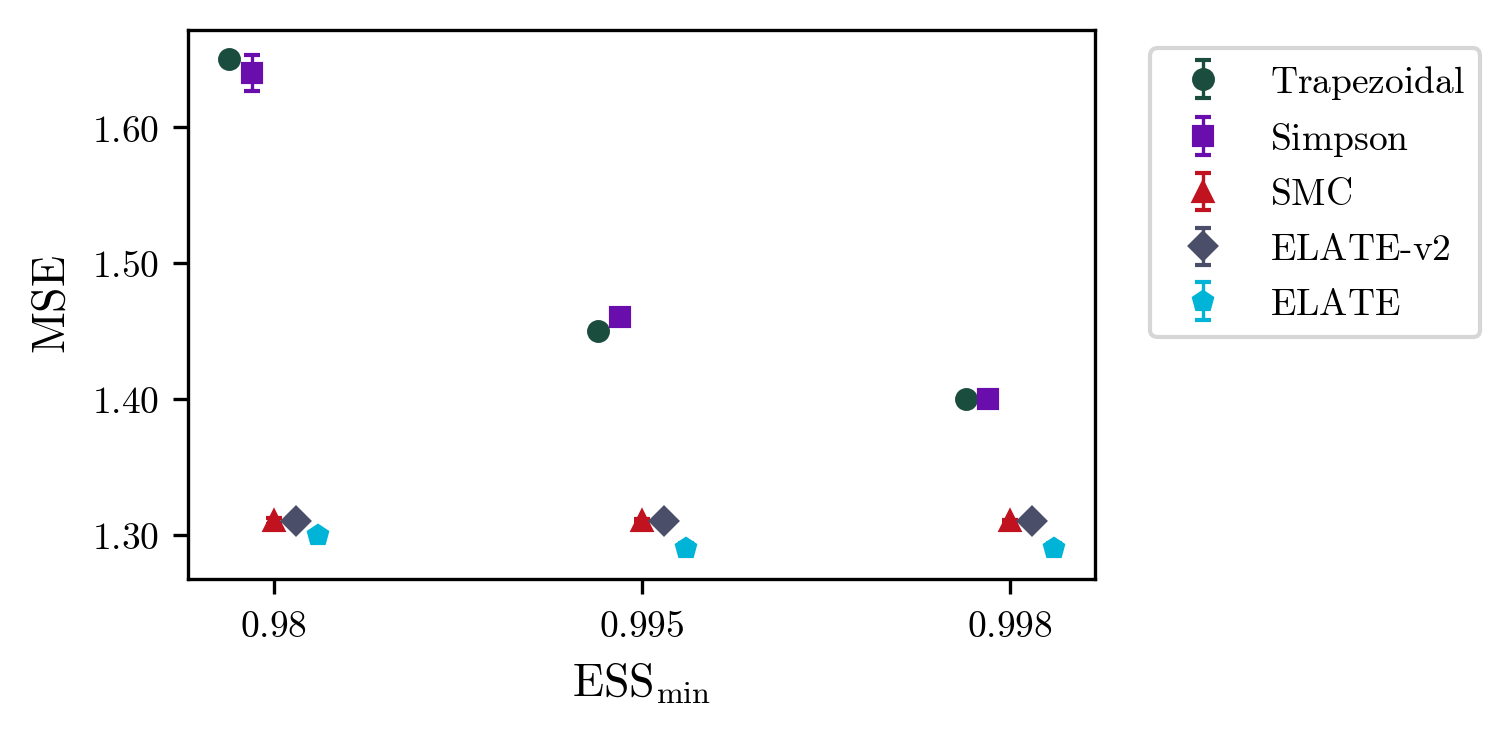}
    \caption{Thermodynamic integration.
Estimator \ac{mse} and associated standard error for the marginal log-likelihood of the \textbf{Gaussian mixture model}, computed over 100 independent realizations of \ac{smc}, varying the $\text{ESS}_{\text{min}}$ threshold. The gold standard can be obtained in closed form.  \ac{smc} was run with {$N=20\times 10^3$}, and {$M=50$}. 
Marker styles follow the same convention as in \Cref{fig: thermo}.}
\label{fig: normconst_egg}
\end{figure}

\begin{figure}[t!]
    \centering
    \includegraphics[width=0.75\textwidth]{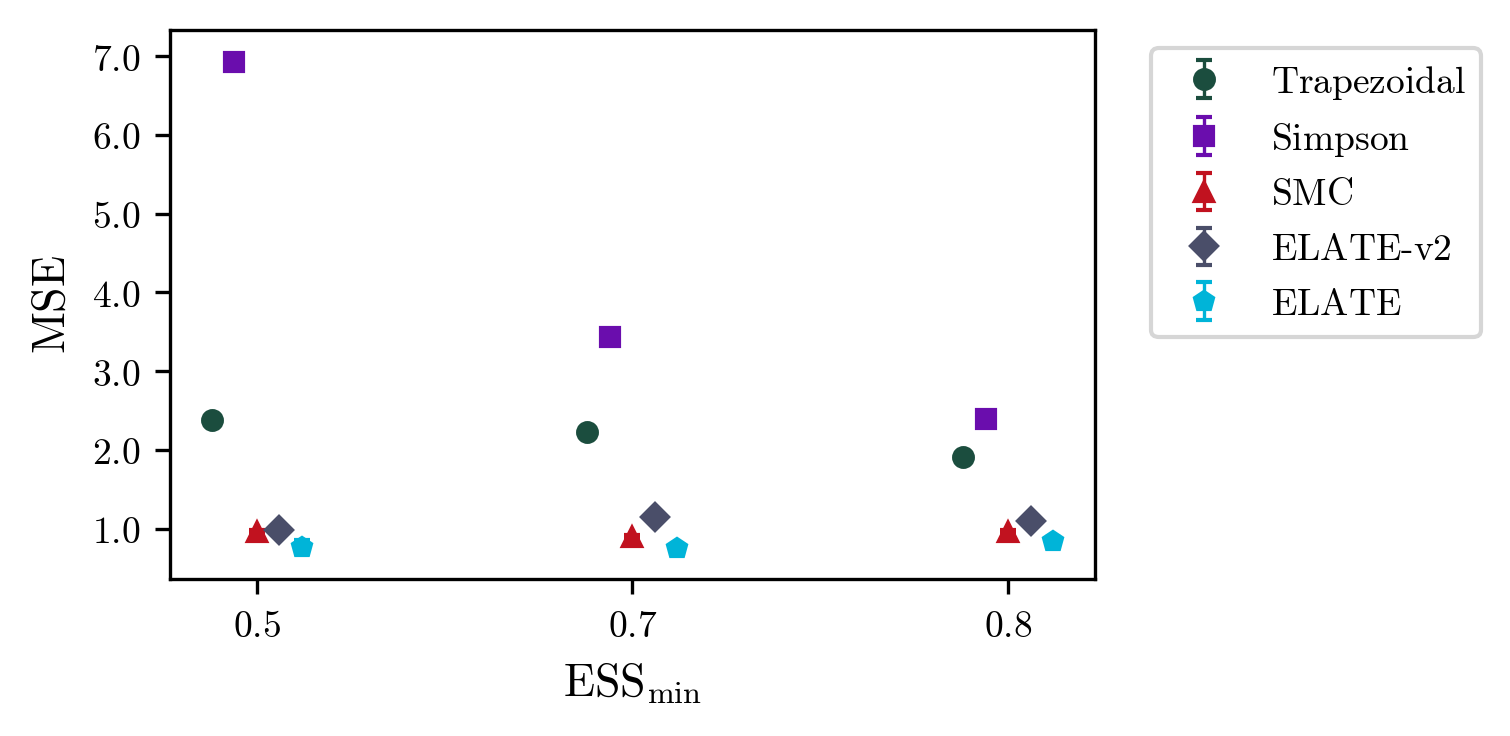}
    \caption{Thermodynamic integration.
Estimator \ac{mse} and associated standard error for the marginal log-likelihood of the  \textbf{\ac{mrna}} model, computed over 100 independent realizations of \ac{smc}, varying the $\text{ESS}_{\text{min}}$ threshold.     The gold standard was obtained using Simpson's rule based on Monte Carlo estimates from \ac{smc} samples, using 130 equally spaced temperatures, and half a million samples for each temperature.  \ac{smc} was run to produce 
 $N = 10^4$ samples, with fixed resample size $M=50$.  
Marker styles follow the same convention as in \Cref{fig: thermo}.}
\label{fig:normconst_mrna}
\end{figure}

To compare the effectiveness of the methods described in \Cref{subsec: thermo} in estimating the log-marginal  likelihood, we carried out 100 independent experiments on various test beds, using the average MSE as performance metric. \Cref{fig: normconst_egg} presents  
results  for the Gaussian
mixture model, and  \Cref{fig:normconst_mrna} for the \ac{mrna} model, showing 
that  \ac{name}  consistently outperforms the other methods.

\end{document}